\documentclass[aip,pof,graphicx]{revtex4-1}

\draft 

\usepackage{amsthm}
\usepackage{amsmath,amssymb}
\usepackage{here}
\usepackage{multirow}
\usepackage{comment}
\usepackage{enumerate}
\usepackage{ulem}

\usepackage{bm}
\usepackage{mathrsfs}
\usepackage{mathtools}
\usepackage{physics}
\usepackage{siunitx}

\newcommand{\rmL}{\mathrm{L}}
\newcommand{\rmR}{\mathrm{R}}
\newcommand{\rmK}{\mathrm{K}}
\newcommand{\ic}{i\mathrm{c}}
\theoremstyle{remark}

\newcommand{\reviseEditor}[1]{{\color{black}{#1}}}

\newif\iffigure 
\figuretrue 

\begin{document}


\title{A low-dissipation numerical method based on boundary variation diminishing principle for compressible gas-liquid two-phase flows with phase change on unstructured grid} 



\author{Hiro Wakimura}
\email[]{wakimura.h.aa@m.titech.ac.jp}
\affiliation{Department of Mechanical Engineering, Institute of Science Tokyo, 2-12-1 i6-29 Ookayama, Meguro-ku, Tokyo, 152-8550, Japan}

\author{Takayuki Aoki}
\affiliation{Center for Information Infrastructure, Institute of Science Tokyo, 2-12-1 i7-3 Ookayama, Meguro-ku, Tokyo 152-8550, Japan}

\author{Feng Xiao}
\email[]{xiao.f.aa@m.titech.ac.jp}
\affiliation{Department of Mechanical Engineering, Institute of Science Tokyo, 2-12-1 i6-29 Ookayama, Meguro-ku, Tokyo, 152-8550, Japan}


\date{\today}

\begin{abstract}
A low-dissipation numerical method for compressible gas-liquid two-phase flow with phase change on unstructured grids is proposed. The governing equations adopt the six-equation model. The non-conservative terms included in the volume fraction and total energy equations of the six-equation model are defined on cell boundaries using second-order accurate approximations and calculated without interpolating the spatial derivatives. To capture discontinuities such as contact discontinuities and gas-liquid interfaces with low dissipation, the MUSCL-THINC/QQ-BVD scheme, which combines the Monotone Upstream-centered Schemes for Conservation Laws (MUSCL) method and the \reviseEditor{Tangent Hyperbola for INterface Capturing} method with \reviseEditor{Q}uadratic surface representation and Gaussian 
\reviseEditor{Q}uadrature (THINC/QQ) method, is employed. The MUSCL method is one of the mainstream numerical solvers for compressible flows, achieving second-order accuracy for smooth solutions, but it introduces excessive numerical dissipation errors near discontinuous solutions. The THINC/QQ method uses a reconstruction function developed for interface capturing on unstructured grids, making use of a sigmoidal function with a quadratic surface. By combining these reconstruction functions according to the Boundary Variation Diminishing (BVD) principle, the MUSCL method is selected for smooth solutions, while the THINC/QQ method is chosen for discontinuous solutions, preserving the solution structure accurately. Several benchmark tests are solved, demonstrating that the MUSCL-THINC/QQ-BVD scheme not only captures contact discontinuities with low dissipation but also resolves dynamically generated gas-liquid interfaces due to phase changes clearly.
\end{abstract}

\pacs{}

\maketitle 



%
%

%



\section{Introduction}

Compressible gas-liquid two-phase flows with phase change are observed in various engineering situations, such as the collapse of cavitation bubbles near propeller blades, the fuel flow within jet engines or rocket engines, and so on. In the compressible gas-liquid two-phase flows, flow structures such as shock waves, contact discontinuities, and expansion waves coexist with gas-liquid interfaces. Moreover, due to phase change, dynamic generation of gas-liquid interfaces occurs within the flow. Accurately capturing these fluid phenomena presents one of the most challenging tasks in computational fluid dynamics.

In the numerical simulation of compressible gas-liquid two-phase flows, the most critical factor is the numerical treatment of the gas-liquid interface. The Sharp Interface Model (SIM) treats the gas-liquid interface as a distinct discontinuity, assuming that only pure gas and liquid phases exist in the computational domain. Representative methods belonging to the SIM include the Volume of Fluid (VOF) method \cite{Hirt1981, Kothe1998ReconstructingTracking, Welch2000}, the level-set method \cite{Mulder1992, Sussman1994}, and the front-tracking method \cite{Unverdi1992AFlows, Tryggvason2001AFlow}. These methods are widely used in the simulation of multiphase flows; however, when the interface undergoes complex deformation, or when gas-liquid interfaces are dynamically generated due to phase changes, it is generally challenging to accurately reproduce the gas-liquid interface. On the other hand, the Diffuse Interface Model (DIM) represents the gas-liquid interface as an artificial mixture region of gas and liquid, allowing for more flexible computation of the interface compared to the SIM. Additionally, the DIM enables the direct application of numerical techniques developed for compressible single-phase flows to the two-phase flow simulations. Considering these advantages, the DIM has become the dominant approach in the numerical simulation of compressible gas-liquid two-phase flow simulations.

Development of governing equation models for the gas-liquid two-phase flows within the DIM framework has advanced over time. These models can be classified into four-equation to seven-equation models, depending on the number of equations they include. The seven-equation model, originally proposed by Baer and Nunziato \cite{Baer1986}, as well as Saurel and Abgrall \cite{Saurel1999}, is a two-velocity two-pressure model consisting of equations for the mass, momentum, and total energy of each phase, along with a transport equation of the volume fraction. Mechanical equilibrium between the two phases, specifically pressure and velocity equilibrium, is achieved through corresponding relaxation terms in the seven-equation model. The five-equation model, derived from the seven-equation model by Kapila et al. \cite{Kapila2001}, simplifies to a one-velocity one-pressure model. It consists of the mass conservation laws for the gas and liquid phases, the conservation of mixture momentum and total energy, and a transport equation of the volume fraction. Because of its simplicity, the five-equation model is the most widely used in compressible gas-liquid two-phase flow simulations. However, it has been reported that the non-monotonic behavior with respect to the volume fraction of the mixture sound speed, known as Wood's sound speed, can lead to numerical instability within the mixed-phase region. The six-equation model, proposed by Saurel et al. \cite{RichardSaurel2009}, is a one-velocity two-pressure model that includes equations for the mass and internal energy of each phase, the conservation of mixture momentum, and a transport equation of the volume fraction. The mixture sound speed in the six-equation model is referred to as the frozen sound speed \cite{Picard1987, Flatten2011}, which exhibits a monotonic distribution with respect to the volume fraction, offering superior numerical stability. Pelanti and Shyue \cite{Pelanti2014} further proposed a variant of Saurel's six-equation model that employs the total energy equation instead of the internal energy equation, ensuring mixture total energy conservation at the discrete level.

In practical applications, numerical methods on unstructured grids are important. For example, when analyzing flows around propellers or within engines, it is necessary to employ meshes that can conform to the complex geometries of the computational domain. However, achieving high accuracy in numerical computations on unstructured grids is generally more challenging compared to structured grids. On structured grids, a simple one-dimensional approach can be applied to each dimensional direction in a multi-dimensional space. In contrast, unstructured grids require multi-dimensional interpolation, which increases the complexity of the computational algorithms. Representative high-order numerical methods on unstructured grids include the Discontinuous Galerkin (DG) method \cite{Cockburn1998TheV, Krivodonova2004ShockLaws, Dumbser2008AMeshes, Zhu2013Runge-KuttaMeshes}, the spectral volume (SV) method \cite{Wang2002SpectralFormulation}, the spectal difference (SD) method \cite{Wang2007SpectralEquations}, the flux reconstruction (FR) method \cite{Huynh2007AMethods, Witherden2014PyFR:Approach, Haga2019OnMethod, Abe2020}, and the multi-moment constrained finite volume (MCV) method \cite{Ii2009HighFormulation, Xiao2013ASchemes}. In this study, numerical computations are conducted within the framework of the Godunov-type finite volume method (FVM), which is highly compatible with the DIM approach.

The computational procedure of the FVM primarily involves spatial reconstruction, numerical flux evaluation, and time evolution. Since spatial reconstruction significantly influences the accuracy and stability of the numerical solution, many reconstruction methods have been proposed over several decades. The most widely used methods are the Monotone Upstream-centered Schemes for Conservation Laws (MUSCL) scheme \cite{VanLeer1977} and the Weighted Essentially Non-Oscillatory (WENO) scheme \cite{Liu1994, Jiang1996}. The MUSCL scheme employs piecewise-linear interpolation functions, with gradients limited to satisfy the Total Variation Diminishing (TVD) property \cite{Harten1983}, ensuring second-order accuracy for smooth and monotonic solutions. The WENO scheme achieves arbitrarily high-order accuracy by weighting reconstructed values from sub-stencils according to the smoothness of the solution, and essentially avoiding numerical oscillations. Both the MUSCL and WENO schemes have been extended to unstructured grids, with research efforts demonstrated in works such as \cite{Venkatakrishnan1995ConvergenceLimiters, Hubbard1999MultidimensionalGrids, Darwish2003TVDGrids, Li2008AnSchemes, CastroDiaz2009Two-dimensionalMeshes, Park2010Multi-dimensionalGrids, Park2012Multi-dimensionalGrids, Zhang2016AMeshes, Li2021AnGrids} for MUSCL-type scheme and \cite{Hu1999WeightedMeshes, Dumbser2007Quadrature-freeSystems, Dumbser2007ArbitrarySystems, Zhang2009ThirdMeshes, Li2012HighorderGrids, Liu2013ASchemes, Christlieb2015HighMeshes, Ji2022AMeshes, Ji2023High-OrderMeshes} for WENO-type scheme.

While these schemes effectively suppress numerical oscillations near discontinuities and maintain high-order accuracy for smooth solutions, it is well-known that they exhibit excessive numerical dissipation errors at discontinuities. This dissipation error can lead to non-physical diffusion of discontinuous solutions with time evolution. There have been proposals \cite{Chiapolino2017b} to sharpen interfaces using the MUSCL scheme with an Overbee limiter, which enhances interface resolution. However, this approach still exhibits slight numerical diffusion on the downstream side of discontinuity, and the geometric shape of the interface is approximated as a straight line. The methods such as the Central WENO (CWENO) \cite{Tsoutsanis2021CWENOMeshes} and hybrid approaches combining DG and FV \cite{Maltsev2024High-orderMeshes} have been suggested for interface capturing. Nevertheless, the performance of these schemes on long-time advection in capturing discontinuous solutions remains unclear.

The Boundary Variation Diminishing (BVD) principle \cite{Sun2016, Xie2017a, Deng2017, Deng2018, Deng2018a, Deng2019, Xie2019High-orderFlow, Deng2020, Abe2020, Jiang2021HybridFlows, Cheng2021, Wakimura2021a, Majima2023AFlows, Li2023AFlows, Wakimura2024High-resolutionEvaporation} has recently been proposed and studied as an approach to capture both smooth and discontinuous solutions in the compressible flows with low dissipation errors and without oscillation errors. The BVD scheme employs multiple spatial reconstruction methods as candidates for interpolation functions, and selects the interpolation function that minimizes the difference between the reconstructed left and right cell boundary values (boundary variation). Since the artificial viscosity contained in the numerical flux, computed by an approximate Riemann solver, is proportional to the value of the boundary variation, choosing an interpolation function with less boundary variation value effectively suppresses numerical dissipation errors in the numerical solution. For example, the MUSCL-THINC-BVD scheme proposed by Deng et al. \cite{Deng2018a} uses the MUSCL method and the Tangent of Hyperbola for INterface Capturing (THINC) method \cite{Xiao2005, Xiao2011} as interpolation candidates, which are well-suited for interpolating smooth solutions and discontinuous solutions, respectively. By selecting the interpolation function for each cell according to the BVD selection algorithm, this scheme successfully captures both smooth and discontinuous solutions with high fidelity.

On unstructured grids, the MUSCL-THINC/QQ-BVD scheme proposed by Cheng et al. \cite{Cheng2021} combines the MUSCL method with the MLP limiter \cite{Park2010Multi-dimensionalGrids, Park2012Multi-dimensionalGrids} and the THINC method with \reviseEditor{Q}uadratic surface representation and Gaussian \reviseEditor{Q}uadrature (THINC/QQ) method \cite{Xie2017} as interpolation candidates. The THINC/QQ method, which was originally developed for capturing moving interfaces, represents interfaces using a quadratic surface and interpolates discontinuous distributions with a sigmoid function. The MUSCL-THINC/QQ-BVD scheme has been successful in capturing both smooth and discontinuous solutions with high fidelity on triangular and quadrilateral grids.

In this study, we perform simulations of compressible gas-liquid two-phase flows with phase change using the MUSCL-THINC/QQ-BVD scheme, demonstrating that the BVD scheme effectively suppresses numerical dissipation errors on unstructured grids and accurately captures phase change phenomena. Furthermore, as there are no known references of computations on unstructured grids using the six-equation model proposed by Pelanti and Shyue, to the best of our knowledge, this study clarifies a simple and effective procedure for implementing the six-equation model on unstructured grids. Specifically, we propose a method for calculating the non-conservative terms without directly interpolating the spatial derivative values, which would introduce additional computational costs and errors. We solve several benchmark tests and compare the numerical results with those obtained using the MUSCL scheme, a representative numerical scheme on unstructured grids, to demonstrate that the BVD scheme can capture discontinuous solutions with low dissipation errors.

\section{Governing equation}

\subsection{Six-equation model with heat and mass transfer}

The six-equation two-phase model proposed by Pelanti and Shyue \cite{Pelanti2014} including heat and mass transfer is,
\begin{subequations}
    \begin{alignat}{1}
        &\pdv{\alpha_1}{t} + \bm{u} \vdot \grad \alpha_1=\mu (p_1-p_2)+\frac{\theta(T_2-T_1)}{\kappa}+\frac{\nu(G_2-G_1)}{\rho_{\mathrm{I}}}, \label{eq:six-eq_alpha1} \\
        &\pdv{(\alpha_1 \rho_1)}{t}+\div(\alpha_1 \rho_1 \bm{u})=\nu(G_2-G_1), \label{eq:six-eq_alpha1rho1} \\
        &\pdv{(\alpha_2 \rho_2)}{t}+\div(\alpha_2 \rho_2 \bm{u})=-\nu(G_2-G_1), \label{eq:six-eq_alpha2rho2} \\
        &\pdv{(\rho \bm{u})}{t}+\div(\rho \bm{u} \otimes \bm{u}+(\alpha_1 p_1 +\alpha_2 p_2)\mathbb{I})=\bm{0}, \label{eq:six-eq_rhou} \\
        &\pdv{(\alpha_1 \rho_1 E_1)}{t}+\div(\alpha_1 (\rho_1 E_1 +p_1)\bm{u})+\mathit{\Upsilon} =-\mu p_\mathrm{I} (p_1-p_2)+\theta(T_2-T_1)+\nu e_{\mathrm{I}}(G_2-G_1), \label{eq:six-eq_alpha1E1} \\
        &\pdv{(\alpha_2 \rho_2 E_2)}{t}+\div(\alpha_2 (\rho_2 E_2 +p_2)\bm{u})-\mathit{\Upsilon} =\mu p_\mathrm{I} (p_1-p_2)-\theta(T_2-T_1)-\nu e_{\mathrm{I}}(G_2-G_1), \label{eq:six-eq_alpha2E2}
    \end{alignat}
    \label{eq:six-eq}
\end{subequations}
where $\alpha_k$ is a volume fraction, $\rho_k$ is a phasic density, $\bm{u}=(u,v,w)^\mathsf{T}$ is a velocity vector, $E_k$ is a phasic total energy, $p_k$ is a phasic pressure, $T_k$ is a phasic temperature, and $G_k$ is a phasic Gibbs energy. The relaxation parameters $\mu,~\theta,~\nu$ denote relaxation speeds of pressure, temperature, and Gibbs energy equilibrium respectively, and are assumed to be infinite here. The parameters $\kappa,~\rho_\mathrm{I},~p_\mathrm{I},~e_\mathrm{I}$ are defined on a phase interface. The non-conservative term $\mathit{\Upsilon}$ reads,
\begin{align}
    \mathit{\Upsilon} =-\bm{u} \vdot (Y_2\grad{(\alpha_1 p_1)}-Y_1\grad{(\alpha_2 p_2)}),
\end{align}
where $Y_k=\frac{\alpha_k \rho_k}{\rho}$ is a mass fraction. The mixture rules are given as,
\begin{subequations}
    \begin{alignat}{1}
        &\alpha_1 + \alpha_2 = 1, \\
        &\alpha_1 \rho_1 + \alpha_2 \rho_2 = \rho, \\
        &\alpha_1 \rho_1 e_1 + \alpha_2 \rho_2 e_2 = \rho e, \\
        &\alpha_1 \rho_1 E_1 + \alpha_2 \rho_2 E_2 = \rho E, 
    \end{alignat}
\end{subequations}
where $e_k=E_k-\frac{1}{2}\norm{\bm{u}}^2$ is a phasic internal energy.

The six-equation model in a compact form reads,
\begin{align}
    \pdv{\bm{q}}{t} + \div \bm{f}(\bm{q}) + \bm{\varsigma}(\bm{q},\grad \bm{q}) = \bm{\psi}_{\mathrm{p}}(\bm{q}) + \bm{\psi}_{\mathrm{T}}(\bm{q}) + \bm{\psi}_{\mathrm{G}}(\bm{q}),
    \label{eq:six-eq_compact}
\end{align}
where,
\begin{align}
    \begin{aligned}
        &
        \bm{q}=\mqty[\alpha_1 \\ \alpha_1 \rho_1 \\ \alpha_2 \rho_2 \\ \rho \bm{u} \\ \alpha_1 \rho_1 E_1 \\ \alpha_2 \rho_2 E_2],~
        \bm{f}(\bm{q})=\mqty[\alpha_1 \bm{u} \\ \alpha_1 \rho_1 \bm{u} \\ \alpha_2 \rho_2 \bm{u} \\ \rho \bm{u} \otimes \bm{u}+(\alpha_1 p_1 +\alpha_2 p_2)\mathbb{I} \\ \alpha_1 (\rho_1 E_1 +p_1)\bm{u} \\ \alpha_2 (\rho_2 E_2 +p_2)\bm{u}],~
        \bm{\varsigma}(\bm{q},\grad \bm{q})=\mqty[-\alpha_1 \div{\bm{u}} \\ 0 \\ 0 \\ \bm{0} \\ \mathit{\Upsilon}  \\ -\mathit{\Upsilon} ],~ \\
        &
        \bm{\psi}_{\mathrm{p}}(\bm{q})=\mqty[\mu(p_1-p_2) \\ 0 \\ 0 \\ \bm{0} \\ -\mu p_{\mathrm{I}} (p_1-p_2) \\ \mu p_{\mathrm{I}} (p_1-p_2)],~
        \bm{\psi}_{\mathrm{T}}(\bm{q})=\mqty[\frac{\theta(T_2-T_1)}{\kappa} \\ 0 \\ 0 \\ \bm{0} \\ \theta(T_2-T_1) \\ -\theta(T_2-T_1)],~
        \bm{\psi}_{\mathrm{G}}(\bm{q})=\mqty[\frac{\nu(G_2-G_1)}{\rho_{\mathrm{I}}} \\ \nu(G_2-G_1) \\ -\nu(G_2-G_1) \\ \bm{0} \\ \nu e_{\mathrm{I}}(G_2-G_1) \\ -\nu e_{\mathrm{I}}(G_2-G_1)].
    \end{aligned}
    \label{eq:six-eq_compact_variable}
\end{align}
We note that the advection term in volume fraction equation \eqref{eq:six-eq_alpha1} is transformed using the chain rule:
\begin{align}
    \bm{u} \vdot \grad \alpha_1 = \div(\alpha_1 \bm{u}) - \alpha_1 \div{\bm{u}},
\end{align}
for the computation on the unstructured grid, explained in section \ref{sec:non-conservative_terms}.

\subsection{Stiffened gas equation of state}

To close the system \eqref{eq:six-eq}, we add a stiffened gas equation of state. For each phase, the phasic pressure, temperature, and Gibbs energy are determined as,
\begin{subequations}
    \begin{alignat}{1}
        &p_k(e_k,\rho_k)=(\gamma_k-1)\rho_k(e_k-\eta_k)-\gamma_k\pi_k, \label{eq:sgeos_p} \\
        &T_k(p_k,\rho_k)=\frac{p_k+\pi_k}{(\gamma_k-1)C_{\mathrm{v}k}\rho_k}, \label{eq:sgeos_T} \\
        &G_k(p_k,T_k)=(\gamma_kC_{\mathrm{v}k}-\eta_k')T_k-C_{\mathrm{v}k}T_k\ln{\frac{T_k^{\gamma_k}}{(p_k+\pi_k)^{\gamma_k-1}}}+\eta_k, \label{eq:sgeos_g}
    \end{alignat}
    \label{eq:sgeos}
\end{subequations}
where $\gamma_k,~\pi_k,~\eta_k,~C_{\mathrm{v}k},~\eta_k'$ are material-dependent constant parameters. In this study, we follow \cite{Saurel2008, Pelanti2014} and 
\begin{table}[htbp]
    \centering
    \begin{tabular}{ccccccc}\hline
        & \ & $\gamma$ & $\pi \ \mathrm{[Pa]}$ & $\eta \ \mathrm{[J/kg]}$ & $\eta' \ \mathrm{[J/(kg\cdot K)]}$ & $C_\mathrm{v} \ \mathrm{[J/(kg\cdot K)]}$ \\ \hline
        \multirow{2}{*}{water} & liquid & $2.35$ & $10^9$ & $-1167\times10^3$ & $0$ & $1816$ \\
        & vapor & $1.43$ & $0$ & $2030\times10^3$ & $-23.4\times10^3$ & $1040$ \\ \hline
        \multirow{2}{*}{dodecane} & liquid & $2.35$ & $4\times10^8$ & $-775.269\times10^3$ & $0$ & $1077.7$ \\
        & vapor & $1.025$ & $0$ & $-237.547\times10^3$ & $-24.4\times10^3$ & $1956.45$ \\ \hline
    \end{tabular}
    \caption{SGEOS parameters of water and dodecane for 300-500K.}
    \label{tab:SGEOS_param}
\end{table}

The associated phasic sound speed $c_k$ is obtained from SGEOS \eqref{eq:sgeos} as,
\begin{align}
    c_k = \sqrt{\frac{\gamma_k (p_k + \pi_k)}{\rho_k}},
\end{align}
and the mixture sound speed corresponding to the homogeneous part of the six-equation model: $\partial_t\bm{q} + \div \bm{f}(\bm{q}) + \bm{\varsigma}(\bm{q},\grad \bm{q}) = \bm{0}$ can be written as,
\begin{align}
    c_\mathrm{hom} = \sqrt{Y_1 c_1^2 + Y_2 c_2^2},
    \label{eq:frozen_sound_speed}
\end{align}
which is called frozen sound speed \cite{Picard1987, Flatten2011}. The frozen sound speed has a monotonic distribution with respect to the volume fraction, resulting in superior numerical stability in the interface region \cite{RichardSaurel2009}. It can be confirmed that the homogeneous part of the six-equation model is hyperbolic as long as $\rho_k,p_k,Y_k > 0$. The mixture sound speed for the overall six-equation model is known to follow the typical sound speed in gas-liquid two-phase flows, which is called Wood's sound speed, 
\begin{align}
    c_\mathrm{Wood}=\frac{1}{\rho \sqrt{\frac{Y_1}{\rho_1^2 c_1^2} + \frac{Y_2}{\rho_2^2 c_2^2}}}.
\end{align}

\section{Computation procedure on unstructured grid}

In this section, the computational procedure for solving the six-equation model on unstructured grids is described. The six-equation model is solved using a fractional step time-integration method, which separates the hyperbolic homogeneous part and the relaxation part as,
\begin{subequations}
    \begin{alignat}{1}
        &\pdv{\bm{q}}{t} = - \div \bm{f}(\bm{q}) - \bm{\varsigma}(\bm{q},\grad \bm{q}), \quad \text{(homogeneous part)} \label{eq:six-eq_hom}\\
        &\pdv{\bm{q}}{t} = \bm{\psi}_{\mathrm{p}}(\bm{q}) + \bm{\psi}_{\mathrm{T}}(\bm{q}) + \bm{\psi}_{\mathrm{G}}(\bm{q}). \quad \text{(relaxation part)} \label{eq:six-eq_relax}
    \end{alignat}
\end{subequations}
Specifically, the update of the solution to the next time step can be expressed as follows,
\begin{align}
    \bm{q}^{n+1}=\mathcal{L}_\mathrm{relax} \mathcal{L}_\mathrm{hyp}^{\Delta t} \bm{q}^n,
\end{align}
where $ \mathcal{L}_{\text{hyp}}^{\Delta t} $ and $ \mathcal{L}_{\text{relax}} $ denote the computational operators for the hyperbolic homogeneous part \eqref{eq:six-eq_hom} and the relaxation part \eqref{eq:six-eq_relax}, respectively.

For the computation of the homogeneous part \eqref{eq:six-eq_hom}, we propose a high-fidelity calculation method within the framework of the Godunov-type FVM. We assume that the 2D computational domain is divided into non-overlapping triangular meshes (see Fig. \ref{fig:triangle_mesh}).
\iffigure
\begin{figure}[htbp]
    \centering
    \includegraphics[width=8cm]{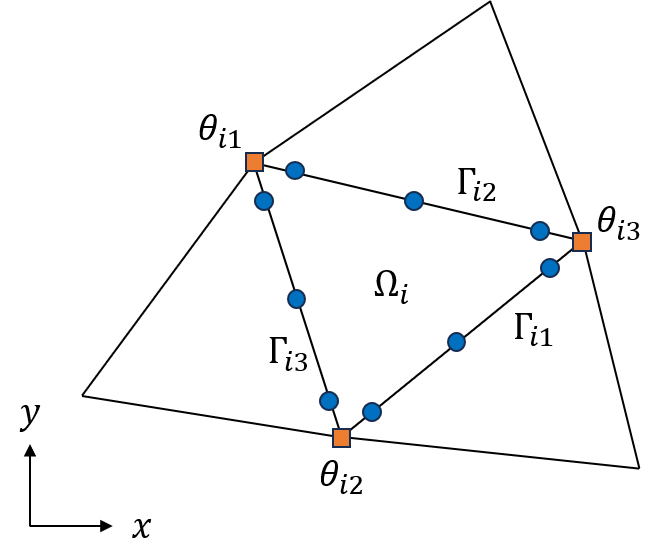}
    \caption{Conceptual diagram of 2D triangle mesh. The blue dots indicate the location of the Gaussian quadrature points where the numerical fluxes and non-conservative terms are defined. The orange squares indicate the location of cell vertices, where gradients are calculated using the least square method.}
    \label{fig:triangle_mesh}
\end{figure}
\fi
Then, the semi-discrete form of \eqref{eq:six-eq_hom} in the framework of the FVM reads,
\begin{align}
    \dv{\overline{\bm{q}}_i}{t}
    &=-\frac{1}{\qty|\Omega_i|}\iint_{\Omega_i}\div{\bm{f}(\bm{q})}\dd \Omega_i - \frac{1}{\qty|\Omega_i|}\iint_{\Omega_i}\bm{\varsigma}(\bm{q},\grad{\bm{q}})\dd \Omega_i,
    \label{eq:six-eq_semi-discrete}
\end{align}
where the bar above the variable indicates the volume-integrated average in a computational cell. 
The volume integral of the divergence of flux in \eqref{eq:six-eq_semi-discrete} is calculated as,
\begin{align}
    \frac{1}{\qty|\Omega_i|}\iint_{\Omega_i}\div{\bm{f}(\bm{q})}\dd \Omega_i
    &= \frac{1}{\qty|\Omega_i|}\oint_{\Gamma_i}\bm{f}(\bm{q})\cdot\bm{n}_{\Gamma_i}\dd \Gamma_i \nonumber\\
    &= \frac{1}{\qty|\Omega_i|} \Sigma_{j} \qty(\int_{\Gamma_{ij}}\bm{f}(\bm{q})\cdot\bm{n}_{\Gamma_{ij}}\dd \Gamma_{ij}) \nonumber\\
    &= \frac{1}{\qty|\Omega_i|} \Sigma_{j} \qty(\Sigma_{g} \omega_g \bm{f}(\bm{q}_{ijg})\cdot\bm{n}_{\Gamma_{ij}} |\Gamma_{ij}|),
\end{align}
where $\bm{f}(\bm{q}_{ijg})$ is a flux at $g$-th Gaussian quadrature point on $j$-th cell-boundary segment of cell $\Omega_i$, and $\bm{n}_{\Gamma_{ij}}$ is a unit normal vector of cell boundary $\Gamma_{ij}$. The volume integration of the non-conservative terms in \eqref{eq:six-eq_semi-discrete} is approximated by integration over the cell boundaries (explained in section \ref{sec:non-conservative_terms}). Consequently, both the fluxes and non-conservative terms are defined at the Gaussian quadrature points on the cell boundaries as Fig. \ref{fig:triangle_mesh}.

The main differences from structured grids in the computational procedure are the need for 1) multidimensional reconstruction functions and 2) the evaluation of non-conservative terms. For the difference 1), while a one-dimensional reconstruction function can be applied for each direction in structured grids, the unstructured grids require multidimensional reconstruction functions, which are obviously more complex compared with the case of structured grids. We propose the MUSCL-THINC/QQ-BVD scheme \cite{Cheng2021} as a high-fidelity reconstruction method for unstructured grids, which suppresses numerical dissipation error near the discontinuous solutions. For the difference 2), in the case of structured grids, the wave-propagation method \cite{LeVeque1997, Ketcheson2013High-OrderSystems} or the path-conservative method \cite{CastroDiaz2009Two-dimensionalMeshes, Dumbser2010FORCESystems, Tokareva2010, Nguyen2015ATension} can be used to avoid direct calculation of the non-conservative terms as expressed in \eqref{eq:six-eq_semi-discrete}. However, for unstructured grids, the applications of these methods are not straightforward, and fluxes and non-conservative terms are generally computed separately. This process requires interpolations of spatial derivatives included in the non-conservative terms, leading to additional computational costs and computational errors. In this work, we propose a second-order accurate method for calculating non-conservative terms that avoids the interpolation of spatial derivatives.

Therefore, the computational procedure proposed in this work consists of a) reconstruction, b) calculation of numerical flux, c) evaluation of non-conservative terms, d) time evolution, and e) relaxation. Below, each of these specific computational methods will be explained.

\subsection{High-fidelity reconstruction method}

We propose a high-fidelity reconstruction method that simultaneously captures both smooth and discontinuous solutions. The proposed method incorporates the MUSCL scheme with the MLP limiter and the THINC/QQ method as candidate reconstruction functions. For selecting the appropriate reconstruction function, the selection algorithm based on the Boundary Variation Diminishing (BVD) principle is presented. The reconstruction is performed for each quasi-primitive variable $\bm{q}^{\mathrm{qp}}=(\alpha_1,\alpha_1\rho_1,\alpha_2\rho_2,\bm{u},p)^\mathsf{T}$, but for simply describing the reconstruction method, the reconstructed variables will be denoted as $ q $.

\subsubsection{MUSCL reconstruction}

The MUSCL reconstruction method is an interpolation technique based on a first-degree polynomial with a limited gradient, maintaining second-order accuracy for smooth solutions while avoiding numerical oscillations near discontinuous solutions. Extensions of the MUSCL scheme to unstructured grids have been studied, and the MLP limiter \cite{Park2010Multi-dimensionalGrids, Park2012Multi-dimensionalGrids} is one of the representative successful limiters for the MUSCL method on unstructured grids. In this study, the MUSCL reconstruction function with the MLP limiter is employed as a candidate interpolation function within the MUSCL-THINC/QQ-BVD method.

A general reconstruction function of the MUSCL scheme on two-dimensional space is expressed as,
\begin{align}
    \mathcal{Q}^{\mathrm{MUSCL}}_i(x,y)=\overline{q}_i+\phi_i\qty(q_{x\ic}(x-x_{\ic})+q_{y\ic}(y-y_{\ic})),
    \label{eq:MUSCL_reconst_func}
\end{align}
where $(\grad{q})_{\ic}=(q_{x\ic},q_{y\ic})$ is a gradient vector at a mass center of cell $\Omega_i$, estimated using least square method.

The slope limiter $\phi_i$ in \eqref{eq:MUSCL_reconst_func} is introduced to avoid numerical oscillation. Park et al. \cite{Park2010Multi-dimensionalGrids, Park2012Multi-dimensionalGrids} demonstrated that, on unstructured grids, when the interpolated value $\hat{q}_{ik}$ at each cell vertex $\theta_{ik}~(k=1,2,3)$ lies between the minimum and maximum of the cell-averaged values of the cells sharing that vertex, i.e. $\overline{q}_{ik}^{\mathrm{min}} \leq \hat{q}_{ik} \leq \overline{q}_{ik}^{\mathrm{max}}$, the following maximum principle will be satisfied under an appropriate Courant–Friedrichs–Lewy (CFL) condition,
\begin{align}
    \overline{q}_{i,\mathrm{neighbor}}^{\mathrm{min},n} \leq \overline{q}_i^{n+1} \leq \overline{q}_{i,\mathrm{neighbor}}^{\mathrm{max},n},
    \label{eq:maximum_principle}
\end{align}
where $\overline{q}_{i,\mathrm{neighbor}}^{\mathrm{min/max},n}$ means the minimum/maximum of the cell-averaged values of the cells that share at least one vertex with cell $\Omega_i$. Based on this concept, the MLP limiter on unstructured grids is designed as,
\begin{align}
    \phi_i=\underset{k}{\min}
    \begin{dcases}
        \Phi(R_{ik}), &\text{if} \quad (\grad{q})_i \cdot \bm{r}_{ik} \neq 0, \\
        1, &\text{otherwise},
    \end{dcases}
\end{align}
where $\bm{r}_{ik}$ is a vector from the mass center of cell $\Omega_i$ to the vertex $\theta_{ik}$, and $R_{ik}$ indicates a slope ratio, calculated as,
\begin{align}
    R_{ik}=\max\qty(\frac{\overline{q}_{ik}^{\mathrm{min}}-\overline{q}_i}{(\grad{q})_i \cdot \bm{r}_{ik}}, \frac{\overline{q}_{ik}^{\mathrm{max}}-\overline{q}_i}{(\grad{q})_i \cdot \bm{r}_{ik}}).
\end{align}
We apply the MLP-u2 limiter \cite{Park2010Multi-dimensionalGrids} for the limiter function $\Phi(R)$, which can be expressed as,
\begin{align}
    \Phi_{\mathrm{MLPu2}}(R)=\frac{R^2+2R+\epsilon}{R^2+R+2+\epsilon}, \quad \epsilon=10^{-15}.
    \label{eq:MLP-u2}
\end{align}
The MLP-u2 limiter \eqref{eq:MLP-u2} ensures the interpolated value at each cell vertex lies between the minimum and maximum of the cell-averaged values of the cells sharing that vertex, and the numerical solution satisfies the maximum principle \eqref{eq:maximum_principle}. The MUSCL reconstruction with the MLP limiter does not generate numerical oscillation near the discontinuity, however, the numerical dissipation error is too large to capture the discontinuity. Next, we introduce the THINC/QQ reconstruction method for capturing discontinuous solutions.

\subsubsection{THINC/QQ reconstruction}

The THINC/QQ method \cite{Xie2017, Chen2022RevisitRobustness} is derived from the original THINC method \cite{Xiao2005} and was proposed as an interface-capturing method for unstructured grids. The distinguishing feature of the THINC/QQ method is its quadratic surface representation. This allows it to maintain the geometric shape of the interface with higher accuracy compared to the conventional THINC method, which uses linear functions to represent the interface. The THINC/QQ method has also been extended to the simulation of compressible flows \cite{Xie2019High-orderFlow, Cheng2021}, where it has been reported to capture discontinuities, such as shock waves and contact discontinuities, with suppressing numerical dissipation compared to existing numerical methods.

The reconstruction function of the THINC/QQ method can be written as,
\begin{align}
    \mathcal{Q}_i^{\mathrm{THINC/QQ}}(x,y)=\frac{\overline{q}_i^{\mathrm{min}}+\overline{q}_i^{\mathrm{max}}}{2}+\frac{\overline{q}_i^{\mathrm{max}}-\overline{q}_i^{\mathrm{min}}}{2}\tanh{\qty(\frac{\beta}{L_i}(P_i(x,y)+d_i))},
    \label{eq:THINC/QQ_func}
\end{align}
where $\overline{q}_i^{\mathrm{min/max}}$ means the minimum/maximum of the cell-averaged values of the cells that share at least one vertex with cell $\Omega_i$. The parameter $L_i$ is a typical cell length such as the diameter of a circumcircle of a triangle cell. The polynomial function $P_i(x,y)$ indicates the geometrical shape of the discontinuity. Specifically, $P_i(x,y)+d_i=0$ shows the location of the discontinuity. In the THINC/QQ formulation, $P_i(x,y)$ is set to be a second-degree polynomial function as,
\begin{align}
    P_i(x,y)&=\sum_{\substack{s,t \geq 0, \\ 1 \leq s+t \leq 2}} a_{st}(x-x_{\ic})^s (y-y_{\ic})^t \nonumber \\
    &=a_{10}(x-x_{\ic}) + a_{01}(y-y_{\ic}) + a_{20}(x-x_{\ic})^2 + a_{11}(x-x_{\ic})(y-y_{\ic}) + a_{02}(y-y_{\ic})^2,
\end{align}
where $(x_{\ic},y_{\ic})$ is the coordinate of center of cell $\Omega_i$. The coefficients $a_{st}$ are determined so that the polynomial function $P_i(x,y)$ is consistent with the unit normal vector and curvature tensor of the discontinuity. More specifically, the following constraint conditions should be satisfied,
\begin{align}
    \begin{dcases}
        \eval{\pdv{P_i(x,y)}{x}}_{\ic}=\varphi_{x,\ic}, \quad 
        \eval{\pdv{P_i(x,y)}{y}}_{\ic}=\varphi_{y,\ic}, \quad \\
        \eval{\pdv[2]{P_i(x,y)}{x}}_{\ic}=\varphi_{xx,\ic}, \quad 
        \eval{\pdv{P_i(x,y)}{x}{y}}_{\ic}=\frac{1}{2}\qty(\varphi_{xy,\ic}+\varphi_{yx,\ic}), \quad 
        \eval{\pdv[2]{P_i(x,y)}{y}}_{\ic}=\varphi_{yy,\ic}, \quad 
    \end{dcases}
    \label{eq:THINC/QQ_Pi_cond}
\end{align}
where $(\varphi_x,\varphi_y)_{\ic}=(\grad{q})_{\ic}/|(\grad{q})_{\ic}|$ is a unit normal vector, and $\varphi_{ab,\ic}~((a,b) \in (x,y))$ is a curvature tensor at cell center. Solving the constraint conditions \eqref{eq:THINC/QQ_Pi_cond}, we obtain the coefficients $a_{st}$ as,
\begin{align}
    \begin{dcases}
        a_{10}=\varphi_{x,\ic}, \quad
        a_{01}=\varphi_{y,\ic}, \quad \\
        a_{20}=\frac{1}{2}\varphi_{xx,\ic}, \quad
        a_{11}=\frac{1}{2}\qty(\varphi_{xy,\ic}+\varphi_{yx,\ic}), \quad
        a_{02}=\frac{1}{2}\varphi_{yy,\ic}.
    \end{dcases}
\end{align}
The unit normal vector $(\varphi_x,\varphi_y)_{\ic}$ and the curvature tensor $\varphi_{ab,\ic}=\eval{\partial_a \varphi_b} _{\ic} ~((a,b) \in (x,y))$ at cell center are calculated by using the least square method.

Except for the steepness parameter $\beta$, the only unknown parameter in \eqref{eq:THINC/QQ_func} is $d_i$, which indicates a jump location of the THINC/QQ reconstruction function. The value of $d_i$ is determined so that a volume integral average of the reconstruction function \eqref{eq:THINC/QQ_func} agrees with the numerical solution,
\begin{align}
    \frac{1}{\qty|\Omega_i|}\iint_{\Omega_i}\mathcal{Q}_i^{\mathrm{THINC/QQ}}(x,y)\dd x \dd y=\overline{q}_i.
    \label{eq:THINC/QQ_cond}
\end{align}
We conduct the Gaussian quadrature to evaluate the integral in \eqref{eq:THINC/QQ_cond} and utilize the Newton-Raphson method to solve the jump location of the reconstruction function. We can uniquely determine the value of $d_i$ with a few iteration steps. For more information, see the details in \cite{Xie2017, Kumar2021THINCSchemes}.

Finally, giving a certain value for parameter $\beta$, the THINC/QQ reconstruction function is determined. We set the $\beta$ value to 1.8 to effectively capture the discontinuous solution and maintain computational stability.

\subsubsection{BVD selection algorithm}

The MUSCL and THINC/QQ reconstruction methods introduced above have distinct characteristics. The MUSCL method achieves second-order accuracy in smooth solutions and avoids numerical oscillations. On the other hand, the THINC/QQ method can represent the geometrical shape of discontinuous solutions using quadratic curves, and capture discontinuities with small numerical dissipation error. Therefore, by selecting the MUSCL method for smooth solutions and the THINC/QQ method for discontinuous solutions, it is anticipated that both smooth and discontinuous solutions can be captured with high fidelity. The following section explains the algorithm for selecting the reconstruction function based on the BVD principle, which proposes minimizing the difference in interpolated values at cell boundaries.

\begin{enumerate}[(i)]
    \item First, the cell boundary values are computed using each of the candidate reconstruction methods in all cells,
    \begin{align}
        q_{ijg}^{\Xi}=\mathcal{Q}_i^\Xi(x_{ijg},y_{ijg}),
    \end{align}
    where the candidate reconstruction method $\Xi$ is the MUSCL or THINC/QQ method, and the subscript $ ijg $ represents the $ g $-th Gaussian quadrature point at the $ j $-th cell boundary of cell $ i $.

    \item Next, the difference in interpolated values at the cell boundary, referred to as the boundary variation (BV), is defined as follows,
    \begin{align}
        BV_{ij}^\Xi
        &\equiv \qty|\int_{\Gamma_{ij}} \qty(\mathcal{Q}_{ij\mathrm{R}}^{\Xi}(x,y)-\mathcal{Q}_{ij\mathrm{L}}^{\Xi}(x,y)) \dd \Gamma_{ij}| \nonumber\\
        &\approx \qty|\sum_g \omega_g \qty(q_{ijg\mathrm{R}}^{\Xi}-q_{ijg\mathrm{L}}^{\Xi}) ||\Gamma_{ij}|,
    \end{align}
    where $\mathcal{Q}_{ij\mathrm{R/L}}^{\Xi}(x,y)$ means the reconstruction function on the right/left side of the cell boundary $\Gamma_{ij}$, and $\omega_g$ is Gaussian quadrature weight. Then, the value of the total of $BV$ (TBV) on the all cell boundaries of cell $\Omega_i$ is calculated as,
    \begin{align}
        TBV_i^\Xi = \sum_j BV_{ij}^\Xi.
    \end{align}

    \item Finally, according to the BVD principle, the candidate reconstruction function with a small value of the $TBV_i$ is selected for cell $\Omega_i$,
    \begin{align}
        \mathcal{Q}_i(x,y)
        =
        \begin{dcases}
            \mathcal{Q}_i^{\mathrm{MUSCL}}(x,y) & \text{if} \quad TBV_i^{\mathrm{MUSCL}}<TBV_i^{\mathrm{THINC/QQ}}, \\
            \mathcal{Q}_i^{\mathrm{THINC/QQ}}(x,y) & \text{otherwise}.
        \end{dcases}
    \end{align}
    
\end{enumerate}
The reconstruction method selected through this procedure is expected to have the smallest BV value among the candidate reconstruction functions, thereby minimizing numerical dissipation errors.

\subsection{Numerical flux}
\label{sec:numerical_flux}

After selecting the reconstruction function and obtaining the reconstructed values at each Gaussian quadrature point on each cell boundary, the numerical fluxes are calculated using an approximate Riemann solver at each quadrature point. The numerical flux along the normal direction of the cell boundary is computed as follows \cite{Toro2009},
\begin{align}
    \bm{f}_{\mathrm{n}}(\bm{q})=\bm{f}(\bm{q}) \cdot \bm{n}=T^{-1}\hat{\bm{f}}(T\bm{q}_\rmL,T\bm{q}_\rmR),
\end{align}
where $\hat{\bm{f}}(\cdot,\cdot)$ indicates an operator of Riemann solver, the rotation matrix $T$ is
\begin{align}
    T=\mqty[1 & 0 & 0 & 0 & 0 & 0 & 0 \\
            0 & 1 & 0 & 0 & 0 & 0 & 0 \\
            0 & 0 & 1 & 0 & 0 & 0 & 0 \\
            0 & 0 & 0 & n_x & -n_y & 0 & 0 \\
            0 & 0 & 0 & n_y & n_x & 0 & 0 \\
            0 & 0 & 0 & 0 & 0 & 1 & 0 \\
            0 & 0 & 0 & 0 & 0 & 0 & 1 ],
\end{align}
and $\bm{n}=(n_x,n_y)$ is a unit normal vector of the cell boundary. In this section, a velocity vector $\bm{u}=(u, v)$ is represented in a coordinate system along the normal vector of the cell boundary.

While some Riemann solvers have been proposed, we here introduce the Harten-Lax-van Leer-Contact (HLLC) Riemann solver \cite{Zein2010, Pelanti2014, DeLorenzo2018, Wakimura2024High-resolutionEvaporation} for better accuracy and robustness. In this framework, three waves $(S_\rmL, S_\star, S_\rmR)$ and four constant states $(\bm{q}_\rmL, \bm{q}_{\star\rmL}, \bm{q}_{\star\rmR}, \bm{q}_\rmR)$ are considered. The following equations like the Rankine-Hugoniot conditions (neglecting non-conservative terms without significant numerical errors \cite{Pelanti2014, DeLorenzo2018}) are solved,
\begin{subequations}
    \begin{alignat}{1}
        &\bm{f}_{\star\rmL}-\bm{f}_{\rmL} = S_\rmL (\bm{q}_{\star\rmL}-\bm{q}_{\rmL}), \\
        &\bm{f}_{\star\rmR}-\bm{f}_{\star\rmL} = S_\star (\bm{q}_{\star\rmR}-\bm{q}_{\star\rmL}), \\
        &\bm{f}_{\rmR}-\bm{f}_{\star\rmR} = S_\rmR (\bm{q}_{\rmR}-\bm{q}_{\star\rmR}). 
    \end{alignat}
    \label{eq:Rankine-Hugoniot}
\end{subequations}
The speeds of left- and right-waves are estimated as,
\begin{align}
    S_{\rmL}=\min(u_{\rmL}-c_{\rmL}, u_{\rmR}-c_{\rmR}), \quad S_{\rmR}=\max(u_{\rmL}+c_{\rmL}, u_{\rmR}+c_{\rmR}),
\end{align}
where $c$ is a mixture sound speed \eqref{eq:frozen_sound_speed}. Then, the speed of intermediate wave $S_{\star}$ and the intermediate states $\bm{q}_{\star\rmL}, \bm{q}_{\star\rmR}$ are obtained from solving \eqref{eq:Rankine-Hugoniot} as,
\begin{align}
    S_{\star}=\frac{(p_\rmL-p_\rmR)+(\rho_\rmL u_\rmL (S_\rmL-u_\rmL) - \rho_\rmR u_\rmR (S_\rmR-u_\rmR))}{\rho_\rmL (S_\rmL - u_\rmL) - \rho_\rmR (S_\rmR - u_\rmR)},
\end{align}
\begin{align}
    \begin{aligned}
        \bm{q}_{\star\rmK}=\mqty[\alpha_{1,\rmK} \chi_{\rmK} \\ 
        (\alpha_1 \rho_1)_{\rmK} \chi_{\rmK} \\ 
        (\alpha_2 \rho_2)_{\rmK} \chi_{\rmK} \\ 
        \rho_{\rmK} S_{\star} \chi_{\rmK} \\ 
        \rho_{\rmK} v_{\rmK} \chi_{\rmK} \\ 
        (\alpha_1 \rho_1)_{\rmK} \chi_{\rmK} \qty(E_{1,\rmK}+(S_{\star}-u_{\rmK})\qty(S_{\star}+\frac{p_{1,\rmK}}{\rho_{1,\rmK}(S_{\rmK}-u_{\rmK})})) \\ 
        (\alpha_2 \rho_2)_{\rmK} \chi_{\rmK} \qty(E_{2,\rmK}+(S_{\star}-u_{\rmK})\qty(S_{\star}+\frac{p_{2,\rmK}}{\rho_{2,\rmK}(S_{\rmK}-u_{\rmK})}))],
    \end{aligned}
    \quad \rmK=\rmL\ \mathrm{or}\ \rmR,
    \label{eq:HLLC_middle_state}
\end{align}
where $\chi_{\rmK}=\frac{S_{\rmK}-u_{\rmK}}{S_{\rmK}-S_{\star}}$. The numerical flux is determined according to the signs of wave speeds as,
\begin{align}
    \bm{f}^\mathrm{HLLC} = 
    \begin{dcases}
        \bm{f}_{\rmL} & \text{if} \quad 0 \leq S_{\rmL}, \\
        \bm{f}_{\star\rmL} & \text{if} \quad S_{\rmL} \leq 0 \leq S_{\star}, \\
        \bm{f}_{\star\rmR} & \text{if} \quad S_{\star} \leq 0 \leq S_{\rmR}, \\
        \bm{f}_{\rmR} & \text{if} \quad S_{\rmR} \leq 0, 
    \end{dcases}
    \label{eq:HLLC_flux}
\end{align}
or calculated by simplified formulation as,
\begin{align}
    \bm{f}^\mathrm{HLLC} = \frac{1 + \mathrm{sign}(S_\star)}{2} \qty(\bm{f}_\rmL + S_\mathrm{L-} (\bm{q}_{\star\rmL}-\bm{q}_{\rmL})) + \frac{1 - \mathrm{sign}(S_\star)}{2} \qty(\bm{f}_\rmR + S_\mathrm{R+} (\bm{q}_{\star\rmR}-\bm{q}_{\rmR})),
\end{align}
where $S_\mathrm{L-}=\min(S_\rmL, 0)$ and $S_\mathrm{R+}=\max(S_\rmR, 0)$.

In addition to the numerical flux, it is also necessary to obtain the values of the normal velocity $u$ and phasic pressure $\alpha_k p_k$ at the cell boundary to compute the non-conservative terms. These values can be computed based on the Riemann solution at the cell boundaries to maintain computational stability. For calculating the velocity value, the formulation proposed in \cite{Johnsen2006} is well-known, which is written as,
\begin{align}
    u^\mathrm{HLLC} = \frac{1 + \mathrm{sign}(S_\star)}{2} \qty(u_\rmL + S_\mathrm{L-} (\chi_\rmL - 1)) + \frac{1 - \mathrm{sign}(S_\star)}{2} \qty(u_\rmR + S_\mathrm{R+} (\chi_\rmR - 1)).
    \label{eq:u_HLLC}
\end{align}
About the phasic pressure values, following \eqref{eq:HLLC_flux}, we calculate them according to the signs of wave speeds as,
\begin{align}
    (\alpha_k p_k)^\mathrm{HLLC} = 
    \begin{dcases}
        (\alpha_k p_k)(\bm{q}_{\rmL}) & \text{if} \quad 0 \leq S_{\rmL}, \\
        (\alpha_k p_k)(\bm{q}_{\star\rmL}) & \text{if} \quad S_{\rmL} \leq 0 \leq S_{\star}, \\
        (\alpha_k p_k)(\bm{q}_{\star\rmR}) & \text{if} \quad S_{\star} \leq 0 \leq S_{\rmR}, \\
        (\alpha_k p_k)(\bm{q}_{\rmR}) & \text{if} \quad S_{\rmR} \leq 0, 
    \end{dcases}
    \label{eq:alpha_k_p_k_HLLC}
\end{align}
where $(\alpha_k p_k)(\cdot)$ is calculated as a function of state variables. For example, $(\alpha_k p_k)_{\star\rmL}$ is calculated as a function of $\bm{q}_{\star\rmL}$ as,
\begin{align}
    (\alpha_k p_k)_{\star\rmL}&=(\alpha_k p_k)(\bm{q}_{\star\rmL}) \nonumber\\
    &= (\gamma_k - 1)\qty((\alpha_k \rho_k E_k)_{\star\rmL}-(\alpha_k \rho_k)_{\star\rmL}\qty(\frac{1}{2}\frac{(\rho u)^2_{\star\rmL} + (\rho v)^2_{\star\rmL}}{((\alpha_1 \rho_1)_{\star\rmL}+(\alpha_2 \rho_2)_{\star\rmL})^2}+\eta_k)) - (\alpha_k)_{\star\rmL}\gamma_k\pi_k.
\end{align}
By using the velocity $u^\mathrm{HLLC}$ and phasic pressure $(\alpha_k p_k)^\mathrm{HLLC}$ values calculated based on the Riemann solution as described above, the non-conservative terms can be computed stably without the need for interpolation of derivative values, as will be explained in the next section.

\subsection{Non-conservative terms}
\label{sec:non-conservative_terms}

In the six-equation model, the non-conservative terms are included in the equations for the volume fraction and the total energy of each phase, as \eqref{eq:six-eq_compact_variable}. In the case of structured grids, using methods such as the wave-propagation method \cite{LeVeque1997, Ketcheson2013High-OrderSystems} or the path-conservative method \cite{CastroDiaz2009Two-dimensionalMeshes, Dumbser2010FORCESystems, Tokareva2010, Nguyen2015ATension} can eliminate the need to directly calculate the non-conservative terms. However, for unstructured grids, such formulations are challenging, necessitating separate calculations for fluxes and non-conservative terms. The non-conservative terms involve spatial gradients, and accurate interpolation of these gradients can lead to additional computational costs and potential errors. Therefore, this study proposes a method for calculating non-conservative terms that does not require the interpolation of a spatial gradient.

The equation for the volume fraction in the semi-discrete form \eqref{eq:six-eq_semi-discrete} can be expressed as follows,
\begin{align}
    \dv{\overline{\alpha_1}_i}{t}
    =-\frac{1}{\qty|\Omega_i|}\oint_{\Gamma_i}\alpha_1\bm{u}\vdot\bm{n}_{\Gamma_i}\dd \Gamma_i + \frac{1}{\qty|\Omega_i|}\iint_{\Omega_i}\alpha_1\div{\bm{u}}\dd \Omega_i.
    \label{eq:semi-discrete_volume_fraction}
\end{align}
The second term in \eqref{eq:semi-discrete_volume_fraction} represents the volume integral of the non-conservative term. The well-known approximate calculation method proposed in \cite{Johnsen2006} can be used, as shown below,
\begin{align}
    \frac{1}{\qty|\Omega_i|}\iint_{\Omega_i}\alpha_1\div{\bm{u}}\dd \Omega_i 
    &= \qty(\frac{1}{\qty|\Omega_i|}\iint_{\Omega_i}\alpha_1\dd \Omega_i) \cdot \qty(\frac{1}{\qty|\Omega_i|}\iint_{\Omega_i}\div{\bm{u}}\dd \Omega_i) + \mathcal{O}(h^2) \nonumber \\
    &= \frac{\overline{\alpha_1}_i}{\qty|\Omega_i|} \oint_{\Gamma_i}\bm{u}\vdot\bm{n}_{\Gamma_i}\dd \Gamma_i + \mathcal{O}(h^2) \nonumber \\
    &= \frac{\overline{\alpha_1}_i}{\qty|\Omega_i|} \Sigma_j \qty(\int_{\Gamma_{ij}} u_{\mathrm{n},ij} \dd \Gamma_{ij}) + \mathcal{O}(h^2) \nonumber \\
    &= \frac{\overline{\alpha_1}_i}{\qty|\Omega_i|} \Sigma_j \qty(\Sigma_{g} \omega_g u_{\mathrm{n},ijg} |\Gamma_{ij}|) + \mathcal{O}(h^2),
    \label{eq:non_conservative_alpha_calculation}
\end{align}
where $\Gamma_{ij}$ indicates a $j$-th cell-boundary segment belonging to the cell $\Omega_i$, and $u_{\mathrm{n},ijg}=\bm{u}_{ijg} \vdot \bm{n}_{\Gamma_{ij}}$ is a velocity component projected in the normal direction to the cell boundary $\Gamma_{ij}$ at $g$-th Gaussian quadrature point. This approximation follows the midpoint rule, giving it second-order accuracy. Since the reconstruction itself is at most second-order accurate, this does not significantly degrade the overall accuracy of the computation. By substituting $ u_{n,ijg} $ with the $ u^{\text{HLLC}} $ derived in the Riemann solver as \eqref{eq:u_HLLC}, the volume integral of the non-conservative term in the volume fraction equation can be obtained without the direct interpolation of the divergence of the velocity.

It is also necessary to calculate the non-conservative terms included in the equations for the phasic total energy. The semi-discrete forms of them are written as,
\begin{align}
    \dv{(\overline{\alpha_k \rho_k E_k})_i}{t}
    =-\frac{1}{\qty|\Omega_i|}\oint_{\Gamma_i}\alpha_k(\rho_k E_k + p_k)\bm{u}\vdot\bm{n}_{\Gamma_i}\dd \Gamma_i - \frac{(-1)^k}{\qty|\Omega_i|}\iint_{\Omega_i}\mathit{\Upsilon} \dd \Omega_i.
\end{align}
The volume integral of the non-conservative term $\mathit{\Upsilon}$ can be approximated using the midpoint rule, similar to \eqref{eq:non_conservative_alpha_calculation}, as follows,
\begin{align}
    \frac{1}{\qty|\Omega_i|}\iint_{\Omega_i}\mathit{\Upsilon} \dd \Omega_i
    &= \frac{1}{\qty|\Omega_i|}\iint_{\Omega_i}(-\bm{u} \vdot Y_2 \grad{(\alpha_1 p_1)} + \bm{u} \vdot Y_1 \grad{(\alpha_2 p_2)}) \dd \Omega_i \nonumber \\
    &= -\overline{\bm{u}}_i \vdot \frac{\overline{Y_2}_i}{\qty|\Omega_i|}\iint_{\Omega_i} \grad{(\alpha_1 p_1)} \dd \Omega_i + \overline{\bm{u}}_i \vdot \frac{\overline{Y_1}_i}{\qty|\Omega_i|}\iint_{\Omega_i} \grad{(\alpha_2 p_2)} \dd \Omega_i + \mathcal{O}(h^2) \nonumber \\
    &= -\overline{\bm{u}}_i \vdot \frac{\overline{Y_2}_i}{\qty|\Omega_i|}\oint_{\Gamma_i} (\alpha_1 p_1)\bm{n}_{\Gamma_i} \dd \Gamma_i + \overline{\bm{u}}_i \vdot \frac{\overline{Y_1}_i}{\qty|\Omega_i|}\oint_{\Gamma_i} (\alpha_2 p_2)\bm{n}_{\Gamma_i} \dd \Gamma_i + \mathcal{O}(h^2) \nonumber \\
    &= -\overline{\bm{u}}_i \vdot \frac{\overline{Y_2}_i}{\qty|\Omega_i|} \Sigma_j \qty(\int_{\Gamma_{ij}} (\alpha_1 p_1)_{ij} \dd \Gamma_{ij}) \bm{n}_{\Gamma_{ij}} \nonumber\\
    &\quad + \overline{\bm{u}}_i \vdot \frac{\overline{Y_1}_i}{\qty|\Omega_i|} \Sigma_j \qty(\int_{\Gamma_{ij}} (\alpha_2 p_2)_{ij} \dd \Gamma_{ij}) \bm{n}_{\Gamma_{ij}} + \mathcal{O}(h^2) \nonumber \\
    &= -\frac{1}{|\Omega_i|}\Sigma_j \qty(\qty(\overline{Y_2}_i \Sigma_g \omega_g (\alpha_1 p_1)_{ijg} - \overline{Y_1}_i \Sigma_g \omega_g (\alpha_2 p_2)_{ijg}) |\Gamma_{ij}| \qty(\overline{\bm{u}}_i \vdot \bm{n}_{\Gamma_{ij}})) + \mathcal{O}(h^2),
\end{align}
where $(\alpha_k p_k)_{ijg}$ is a phasic pressure at $g$-th Gaussian quadrature point on $j$-th cell-boundary segment of cell $\Omega_i$. Substituting $(\alpha_k p_k)_{ijg}$ with $(\alpha_k p_k)^{\text{HLLC}}$ calculated in the Riemann solver as \eqref{eq:alpha_k_p_k_HLLC}, the volume integral of the non-conservative term $\mathit{\Upsilon}$ in the phasic-total-energy equation can be obtained without the direct interpolation of the gradient of the phasic pressure.

Using these approximations, the volume integrals of the non-conservative terms in the six-equation model can be computed without the need for interpolating spatial derivatives. Consequently, combined with the volume integral of the flux divergence, the time derivative of the numerical solution of the homogeneous part can be obtained from the semi-discrete form given by \eqref{eq:six-eq_semi-discrete}. This completes the computational operator for the hyperbolic homogeneous part $\mathcal{L}_\mathrm{hyp}^{\Delta t}$.

\subsection{Relaxation}

The relaxation terms $\bm{\psi}_{\mathrm{p}}(\bm{q})$, $\bm{\psi}_{\mathrm{T}}(\bm{q})$, and $\bm{\psi}_{\mathrm{G}}(\bm{q})$ in the six-equation model are responsible for representing mechanical, thermal, and chemical equilibrium, respectively. After solving the hyperbolic homogeneous part of the six-equation model, these relaxation terms are solved to ensure that pressure, temperature, and Gibbs energy reach equilibrium. Here, it is assumed that the relaxation speeds are in the order of pressure, temperature, and then Gibbs energy, and the following relaxation equations are considered:
\begin{subequations}
    \begin{alignat}{2}
        &\dv{\bm{q}}{t} = \bm{\psi}_{\mathrm{p}}(\bm{q}), & & \text{(mechanical relaxation)}, \label{eq:mechanical_relaxation} \\
        &\dv{\bm{q}}{t} = \bm{\psi}_{\mathrm{p}}(\bm{q})+\bm{\psi}_{\mathrm{T}}(\bm{q}), & & \text{(mechanical and thermal relaxation)}, \label{eq:mechanical_thermal_relaxation} \\
        &\dv{\bm{q}}{t} = \bm{\psi}_{\mathrm{p}}(\bm{q})+\bm{\psi}_{\mathrm{T}}(\bm{q})+\bm{\psi}_{\mathrm{G}}(\bm{q}), \quad & & \text{(mechanical, thermal and chemical relaxation)}. \label{eq:mechanical_thermal_chemical_relaxation}
    \end{alignat}
\end{subequations}
Moreover, we assume that all of these relaxation rates are much faster than the characteristic time of the flow dynamics, treating the relaxation rates $\mu$, $\theta$, and $\nu$ as infinite. The calculation operators for solving each relaxation equation are denoted as $\mathcal{L}_{\mathrm{p}}$, $\mathcal{L}_{\mathrm{pT}}$, and $\mathcal{L}_{\mathrm{pTG}}$, respectively. Considering the possible combinations of these operators, the following four cases are taken into account \cite{Pelanti2014}:
\begin{subequations}
    \begin{alignat}{5}
        & \mathcal{L}_\mathrm{relax}&=& & & & &\mathcal{L}_\mathrm{p} , & & \text{(p-relaxation)}, \label{eq:p-relaxation} \\
        & \mathcal{L}_\mathrm{relax}&=& & &\mathcal{L}_\mathrm{pT}& &\mathcal{L}_\mathrm{p} , & & \text{(p-pT-relaxation)}, \label{eq:p-pT-relaxation} \\
        & \mathcal{L}_\mathrm{relax}&=&\mathcal{L}_\mathrm{pTG}& & & &\mathcal{L}_\mathrm{p} , & & \text{(p-pTG-relaxation)}, \label{eq:p-pTG-relaxation} \\
        & \mathcal{L}_\mathrm{relax}&=&\mathcal{L}_\mathrm{pTG}& &\mathcal{L}_\mathrm{pT}& &\mathcal{L}_\mathrm{p} , \quad & & \text{(p-pT-pTG-relaxation)}. \label{eq:p-pT-pTG-relaxation}
    \end{alignat}
    \label{eq:relaxation_operator}
\end{subequations}
For example, if the relaxation operator is $\mathrm{p}$-relaxation, the calculation corresponds to a model where pressure equilibrium is achieved while temperature and Gibbs energy are not in equilibrium. On the other hand, if the relaxation operator is $\mathrm{p}$-$\mathrm{pT}$-$\mathrm{pTG}$-relaxation, the calculation corresponds to a model where pressure, temperature, and Gibbs energy are all in equilibrium. The $\mathrm{p}$-$\mathrm{pTG}$-relaxation and $\mathrm{p}$-$\mathrm{pT}$-$\mathrm{pTG}$-relaxation cases account for phase change phenomena due to Gibbs energy equilibrium. Since the detailed procedure for each calculation operator $\mathcal{L}_{\mathrm{p}}$, $\mathcal{L}_{\mathrm{pT}}$, and $\mathcal{L}_{\mathrm{pTG}}$ is the same as in the case of structured grids, it will not be explained here. Please refer to \cite{Pelanti2014, Wakimura2024High-resolutionEvaporation} for more information.

\subsection{Time evolution}

The time evolution is conducted using the three-step third-order Runge-Kutta method \cite{Gottlieb2001}. Using the computational operators of the homogeneous hyperbolic part $\mathcal{L}_\mathrm{hyp}^{\Delta t}$ and the relaxation part $\mathcal{L}_\mathrm{relax}$, the three-step time evolution for the six-equation model is expressed as,
\begin{align}
    \begin{aligned}
        &\overline{\bm{q}}_i^{*}=\mathcal{L}_\mathrm{relax} \qty(\overline{\bm{q}}_i^{n} + \mathcal{L}_\mathrm{hyp}^{\Delta t}(\overline{\bm{q}}_i^{n})\Delta t), \\
        &\overline{\bm{q}}_i^{**}=\mathcal{L}_\mathrm{relax} \qty(\frac{3}{4}\overline{\bm{q}}_i^{n} + \frac{1}{4}\overline{\bm{q}}_i^{*} + \frac{1}{4}\mathcal{L}_\mathrm{hyp}^{\Delta t}(\overline{\bm{q}}_i^{*})\Delta t), \\
        &\overline{\bm{q}}_i^{n+1}=\mathcal{L}_\mathrm{relax} \qty(\frac{1}{3}\overline{\bm{q}}_i^{n} + \frac{2}{3}\overline{\bm{q}}_i^{**} + \frac{2}{3}\mathcal{L}_\mathrm{hyp}^{\Delta t}(\overline{\bm{q}}_i^{**})\Delta t).
    \end{aligned}
\end{align}
The time-step length $\Delta t$ is calculated as \cite{Chen2022RevisitRobustness},
\begin{align}
    \Delta t=\mathrm{CFL}\times \underset{i}{\min}\qty(\frac{|\Omega_i|}{\frac{1}{2}\sum_j|\bm{u}_{ij}\vdot\bm{n}_{ij} \pm c_{ij}||\Gamma_{ij}|}),
\end{align}
where the CFL number is set to $0.5$ unless otherwise specified.


\section{Numerical results}

\subsection{Water cavitation tube problem}
\label{sec:watercav}
In this problem, liquid initially fills a tube, and as the liquid moves toward both sides, the pressure at the center of the computational domain decreases, leading to cavitation. Although this problem is originally one-dimensional \cite{Saurel2008, Pelanti2014}, in this study instead of the exact one-dimensional calculation, it will be calculated on a two-dimensional computational domain which can not be configured rigorously as 1D. The computational domain is $x \in [0, 1\ \mathrm{m}] \times y \in [0, 0.01\ \mathrm{m}]$, the grid length is $h \approx 1/1000\ \mathrm{m}$, and the mesh number is 23992. Fig. \ref{fig:rectangle_mesh} shows a part of the computational domain and the triangle mesh, where the mesh is not aligned with the main flow direction. The computational results obtained with such a mesh generally differ from those of a purely one-dimensional cavitation tube problem. However, we assess the two-dimensionality of the results and confirm that the computed results are almost one-dimensional.
\iffigure
\begin{figure}[htbp]
    \centering
    \includegraphics[width=10cm]{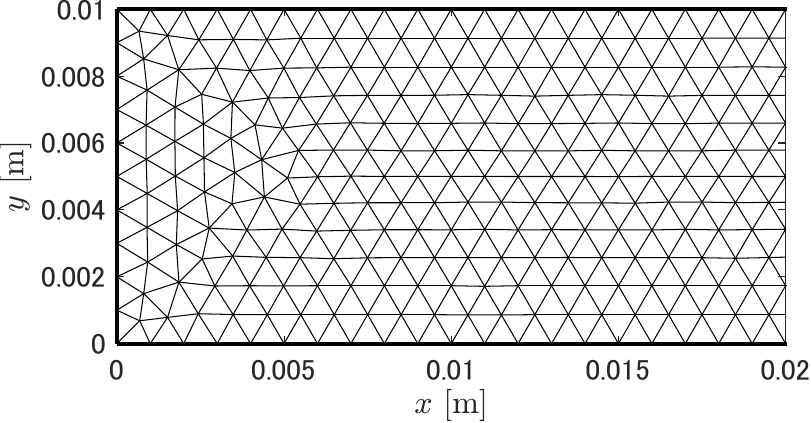}
    \caption{A part of the computational domain and the triangle mesh for the water cavitation tube problem and the water shock tube problem.}
    \label{fig:rectangle_mesh}
\end{figure}
\fi

The initial condition in terms of the primitive variables are shown as,
\begin{align}
    &\qty(\alpha_1,\rho_1\ \mathrm{[kg/m^3]},\rho_2\ \mathrm{[kg/m^3]},u\ \mathrm{[m/s]},v\ \mathrm{[m/s]},p\ \mathrm{[Pa]})_{t=0}= \nonumber \\
    &\quad
    \begin{dcases}
        \qty(0.99,1150,0.63,-2,0,10^{5}) & \text{for} \quad 0\ \mathrm{m}<x<0.5\ \mathrm{m},\\
        \qty(0.99,1150,0.63,2,0,10^{5}) & \text{for} \quad 0.5\ \mathrm{m}<x<1\ \mathrm{m}.
    \end{dcases}
\end{align}
The boundary condition is a zero-gradient condition for the left- and right-boundary, and a periodic condition for the top- and bottom-boundary. The material-dependent parameters in the SGEOS are based on the physical properties of water as listed in Table \ref{tab:SGEOS_param}.

\iffigure
\begin{figure}[htbp]
    \centering
    \includegraphics[width=17cm]{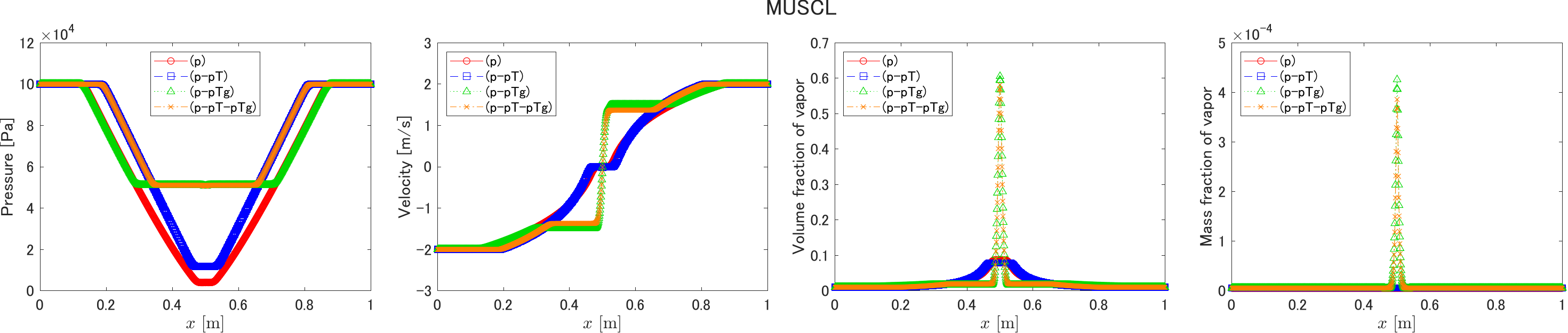} \\
    \includegraphics[width=17cm]{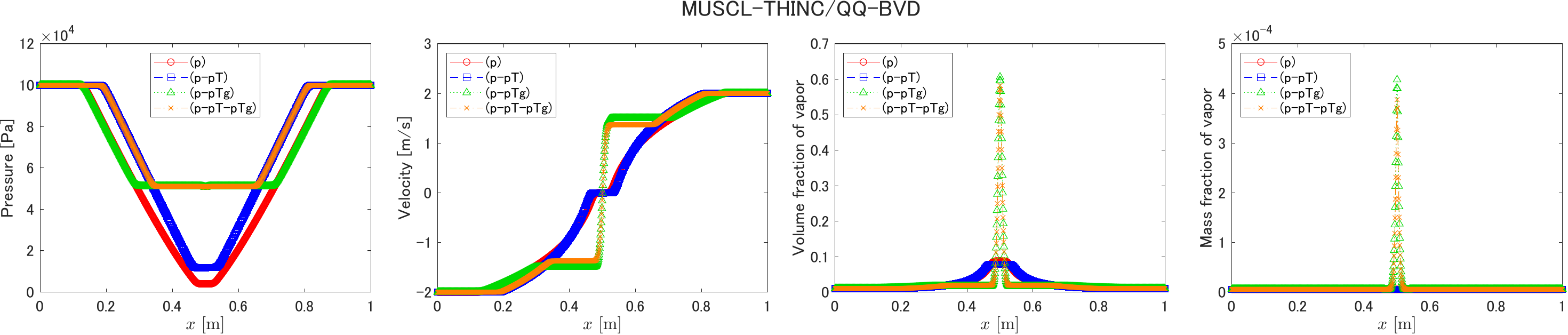} 
    \caption{Numerical results of pressure, velocity, volume fraction, and mass fraction of vapor at $y=0.005\ \mathrm{m}$ at $t=3.2\ \mathrm{ms}$ in water cavitation tube problem. The top and bottom rows show the numerical results of MUSCL and MUSCL-THINC/QQ-BVD schemes respectively.}
    \label{fig:watercav}
\end{figure}
\fi
The numerical results at $y=0.005\ \mathrm{m}$ are shown in Fig. \ref{fig:watercav}. When examining the results for the four relaxation patterns, it is observed that in the p-relaxation and p-pT-relaxation cases, the pressure at the center of the computational domain drops significantly. In contrast, in the p-pTG and p-pT-pTG relaxation cases, the pressure decrease stops at a certain value, and phase transition from liquid to gas is observed. These numerical results are consistent with those in reference \cite{Saurel2008, Pelanti2014}, indicating that the computations on unstructured grids can be performed similarly to those on the one-dimensional structured grid case. Furthermore, when comparing the results of the MUSCL and MUSCL-THINC/QQ-BVD schemes, the results are nearly identical. In fact, the maximum difference in the vapor mass fraction between the two schemes is on the order of $10^{-6}$. In this benchmark test, there are no pronounced shock waves or contact discontinuities, and only smooth expansion waves are generated. This confirms that the MUSCL-THINC/QQ-BVD scheme correctly selects the MUSCL reconstruction function for the smooth solutions. Furthermore, we evaluate the two-dimensionality of numerical results by assessing the ratio of the maximum velocity components in the $y$-direction to the maximum velocity components in the $x$-direction. Table \ref{tab:two-dimensionality_watercav} shows the two-dimensionality of each numerical result in the water cavitation tube problem. In each numerical result, the maximum value of the velocity component in the $y$-direction is only a few percent of the main flow. Therefore, the numerical results can be considered nearly one-dimensional.
\begin{table}[htbp]
    \centering
    \begin{tabular}{lcc}\hline
        & MUSCL & MUSCL-THINC/QQ-BVD \\ \hline
        p-relaxation & $3.46 \times 10^{-2}$ & $3.46 \times 10^{-2}$ \\
        p-pT-relaxation & $3.50 \times 10^{-2}$ & $3.50 \times 10^{-2}$ \\
        p-pTG-relaxation & $2.58 \times 10^{-2}$ & $2.56 \times 10^{-2}$ \\
        p-pT-pTG-relaxation & $2.72 \times 10^{-2}$ & $2.71 \times 10^{-2}$ \\ \hline
    \end{tabular}
    \caption{Two-dimensionality of each numerical result in the water cavitation tube problem.}
    \label{tab:two-dimensionality_watercav}
\end{table}

\subsection{Water shock tube problem}
This benchmark test is a shock tube problem using water \cite{Chiapolino2017}. The tube is filled with mainly water vapor ($Y_2=0.8$). The initial velocity is $0\ \mathrm{m/s}$, and the initial pressure is $2 \times 10^5 \mathrm{Pa}$ on left side and $10^5 \mathrm{Pa}$ on right side. The temperature is set to be the saturation temperature of a given pressure. The initial condition in terms of primitive variables can be derived as,
\begin{align}
    &\qty(\alpha_1,\rho_1\ \mathrm{[kg/m^3]},\rho_2\ \mathrm{[kg/m^3]},u\ \mathrm{[m/s]},v\ \mathrm{[m/s]},p\ \mathrm{[Pa]})_{t=0}= \nonumber \\
    &\quad
    \begin{dcases}
        \qty(2.7399 \times 10^{-4}, 1034.8, 1.1344, 0, 2 \times 10^{5}) & \text{for} \quad 0\ \mathrm{m}<x<0.5\ \mathrm{m},\\
        \qty(1.3702 \times 10^{-4}, 1094, 0.59969, 0, 10^{5}) & \text{for} \quad 0.5\ \mathrm{m}<x<1\ \mathrm{m}.
    \end{dcases}
\end{align}
Since this problem is also originally one-dimensional, the computation grid and the boundary condition are set as similar to the water cavitation tube problem. The material-dependent parameters in the SGEOS are based on the physical properties of water as listed in Table \ref{tab:SGEOS_param}.

\iffigure
\begin{figure}[htbp]
    \centering
    \includegraphics[width=17cm]{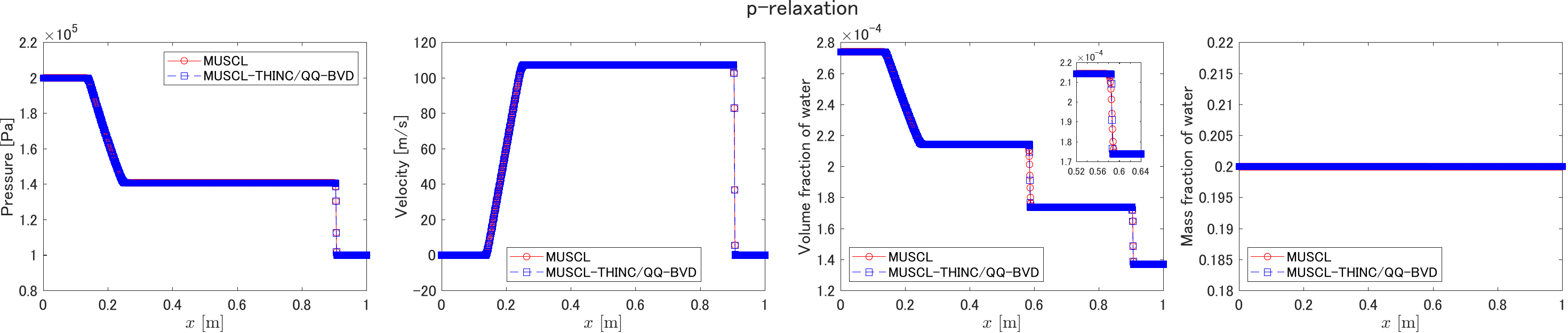} \\
    \includegraphics[width=17cm]{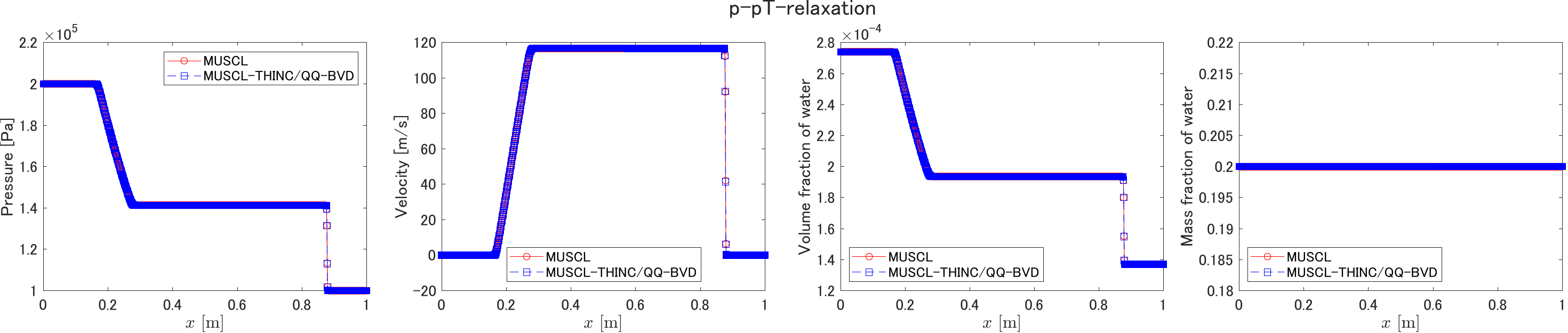} \\
    \includegraphics[width=17cm]{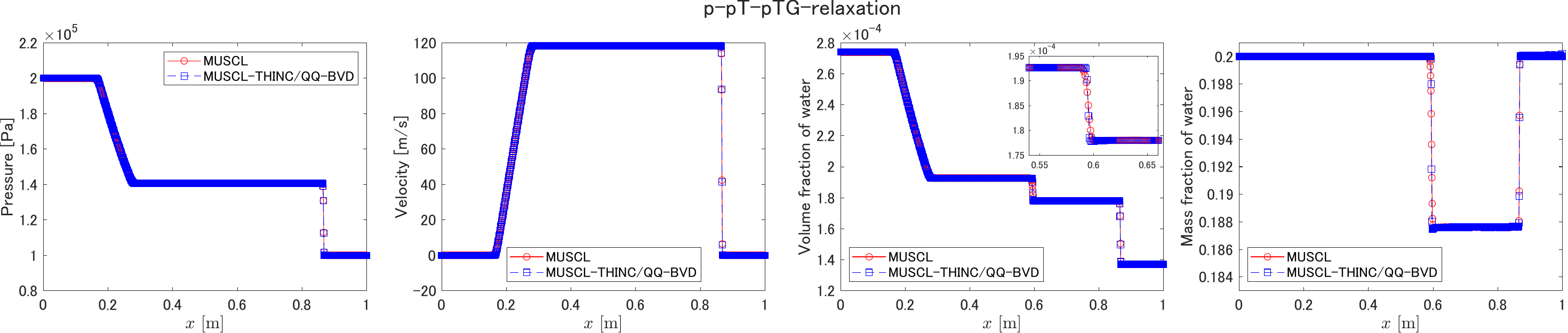}
    \caption{Numerical results of pressure, velocity, volume fraction of water, and mass fraction of water at $y=0.005\ \mathrm{m}$ at $t=0.8\ \mathrm{ms}$ in water shock tube problem. The top, middle, and bottom panels show the results of p-relaxation, p-pT-relaxation, and p-pT-pTG-relaxation respectively.}
    \label{fig:watershock}
\end{figure}
\fi
The numerical results are shown in Fig. \ref{fig:watershock}. The results of the p-pT-relaxation are consistent with the numerical results of the four-equation model (without phase change) \cite{Chiapolino2017}, which assumes pressure and temperature equilibrium. In the results of the p-relaxation and p-pT-pTG-relaxation, a difference between the MUSCL and MUSCL-THINC/QQ-BVD schemes is observed in the contact discontinuity near $x \approx 0.6\ \mathrm{m}$. The MUSCL-THINC/QQ-BVD scheme captures the discontinuity with lower dissipation compared to the MUSCL scheme. This difference is also observed in the distribution of mass fractions in the p-pT-pTG-relaxation. It is noted that the pTG-relaxation is activated only when the liquid temperature exceeds the saturation temperature of given pressure as in \cite{Pelanti2014, Wakimura2024High-resolutionEvaporation}. It became clear that the MUSCL-THINC/QQ-BVD scheme can capture the dynamically generated discontinuity due to phase change with high resolution. Additionally, Table \ref{tab:two-dimensionality_watershock} evaluates the two-dimensionality of the numerical results in the water shock tube problem similar to the water cavitation tube problem. The maximum value of the velocity component in the $y$-direction is a few percent of the maximum value of the velocity component in the $x$-direction, so the two-dimensionality of the numerical results can be neglected.
\begin{table}[htbp]
    \centering
    \begin{tabular}{lcc}\hline
        & MUSCL & MUSCL-THINC/QQ-BVD \\ \hline
        p-relaxation & $3.02 \times 10^{-2}$ & $3.03 \times 10^{-2}$ \\
        p-pT-relaxation & $3.06 \times 10^{-2}$ & $3.13 \times 10^{-2}$ \\
        p-pT-pTG-relaxation & $3.07 \times 10^{-2}$ & $3.16 \times 10^{-2}$ \\ \hline
    \end{tabular}
    \caption{Two-dimensionality of each numerical result in the water shock tube problem.}
    \label{tab:two-dimensionality_watershock}
\end{table}

\subsection{Vapor bubble compression problem}
This problem is a benchmark test where a vapor bubble of dodecane is placed at the center of a square region filled with liquid dodecane, enclosed by walls, and then compressed by a piston. According to \cite{Wakimura2024High-resolutionEvaporation}, at $ t = 1.4\ \mathrm{ms}$, the compressed liquid interferes with the shock wave, leading to a low-pressure region where cavitation occurs. Such dynamically generated gas-liquid interfaces are highly sensitive to the effects of numerical dissipation errors in the numerical scheme, making this test suitable for evaluating the numerical dissipation error of the numerical scheme. The computational domain is $x \in [0, 1\ \mathrm{m}] \times y \in [0, 1\ \mathrm{m}]$, the grid length is $h \approx 1/200\ \mathrm{m}$, and the mesh number is 90804. The initial condition in terms of the primitive variables are shown as,
\begin{align}
    &\qty(\alpha_1,\rho_1\ \mathrm{[kg/m^3]},\rho_2\ \mathrm{[kg/m^3]},u\ \mathrm{[m/s]},v\ \mathrm{[m/s]},p\ \mathrm{[Pa]})_{t=0}= \nonumber \\
    &\quad
    \begin{dcases}
        \qty(1-10^{-6}, 458.338, 3.408, 0, 0, 10^{5}) & \text{for} \quad \sqrt{(x-0.5)^2+(y-0.5)^2}>0.2\ \mathrm{m},\\
        \qty(10^{-6}, 458.338, 3.408, 0, 0, 10^{5}) & \text{otherwise}.
    \end{dcases}
\end{align}
The boundary condition is a moving piston with a constant speed $100\ \mathrm{m/s}$ for the left boundary and reflective condition for the other boundaries. The material-dependent parameters in the SGEOS are based on the physical properties of dodecane as listed in Table \ref{tab:SGEOS_param}.

\iffigure
\begin{figure}[htbp]
    \centering
    \includegraphics[width=12cm]{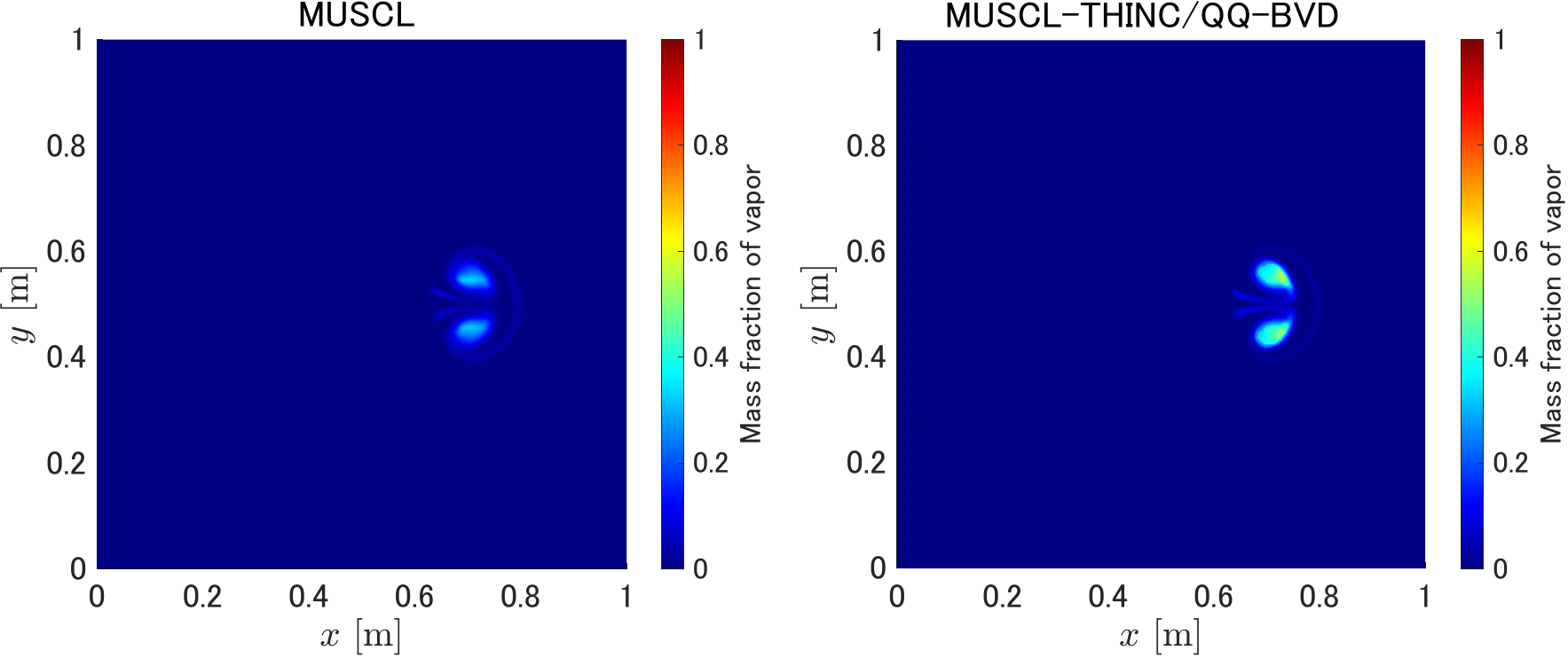}
    \includegraphics[width=12cm]{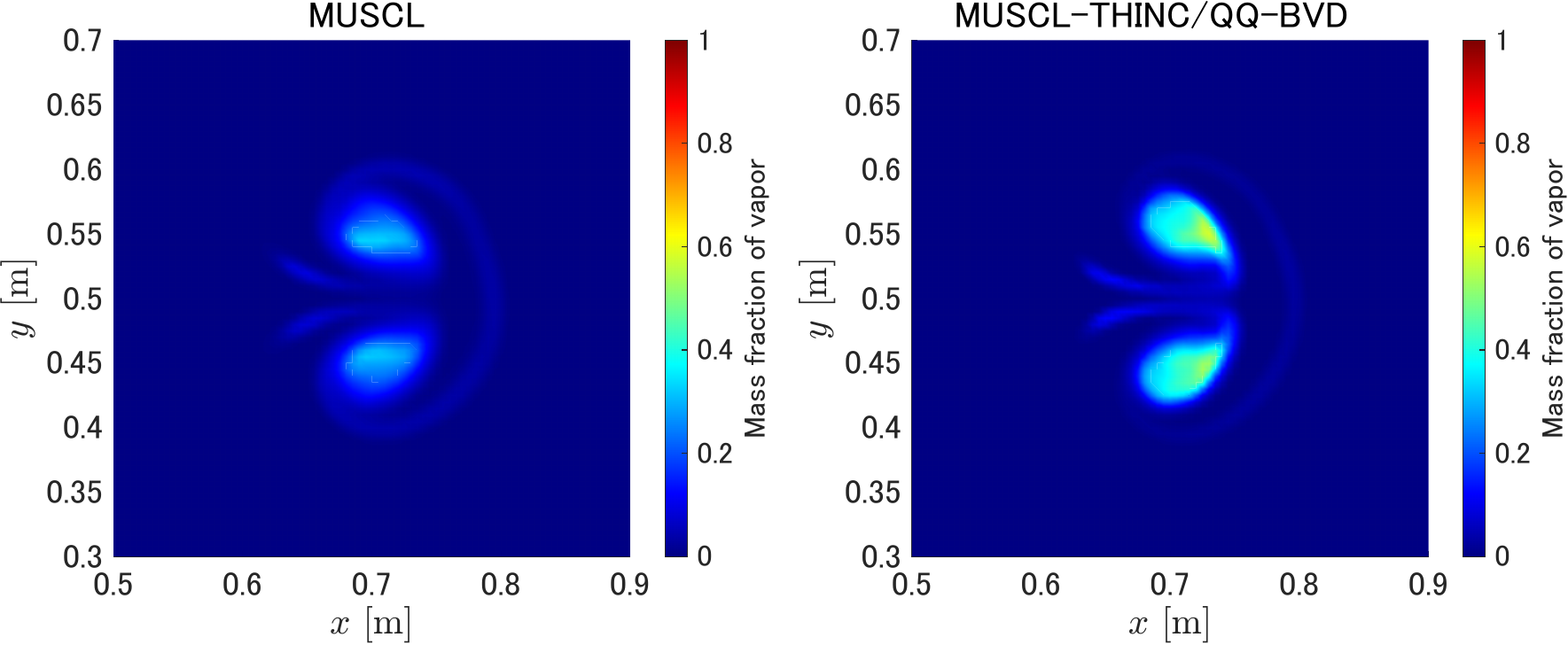}
    \caption{Pseudo-color plots of the mass fraction of vapor at $t=1.4\ \mathrm{ms}$ in vapor-bubble compression problem without phase change (p-relaxation). The left and right panels show the numerical results of MUSCL and MUSCL-THINC/QQ-BVD schemes respectively. The bottom row shows enlarged views of vapor bubbles.}
    \label{fig:2Dbubblecomp_Yg_t=1.4_allSchemes_noPhaseChange}
\end{figure}
\begin{figure}[htbp]
    \centering
    \includegraphics[width=12cm]{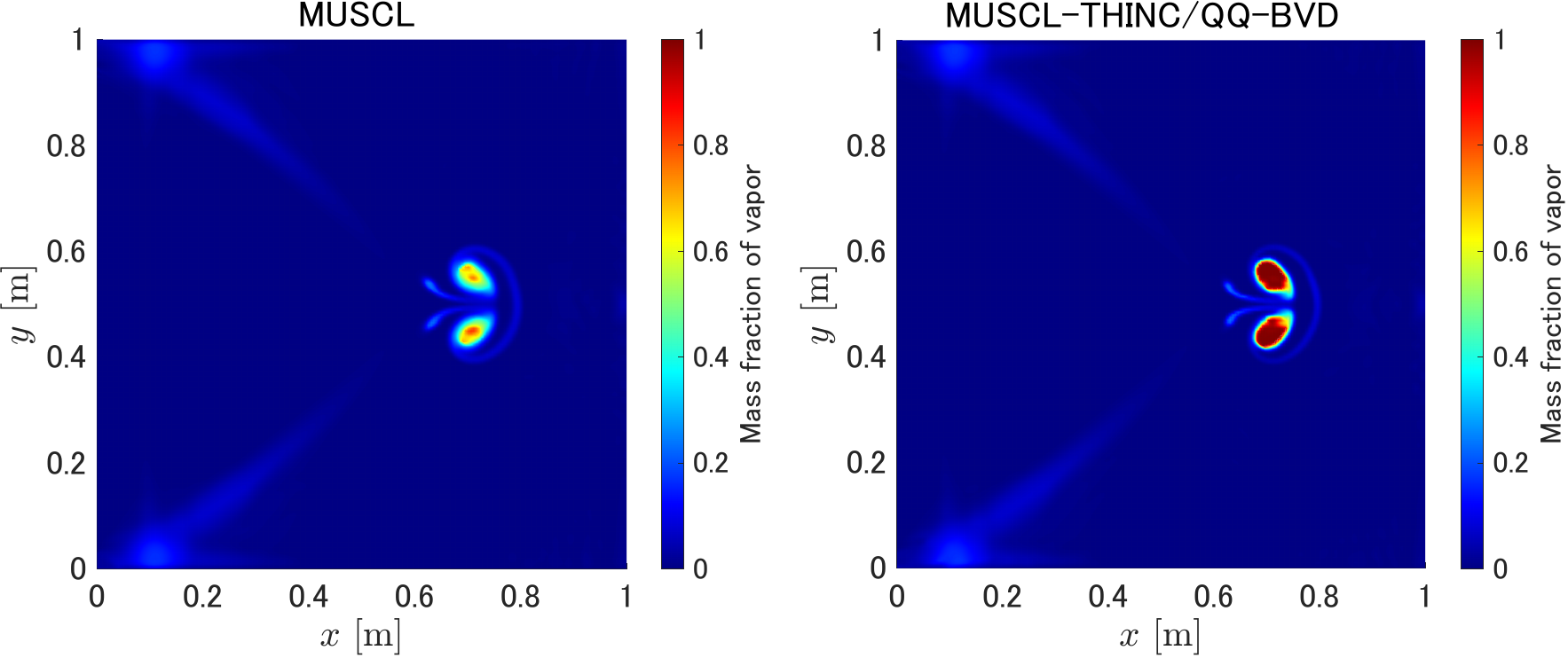}
    \includegraphics[width=12cm]{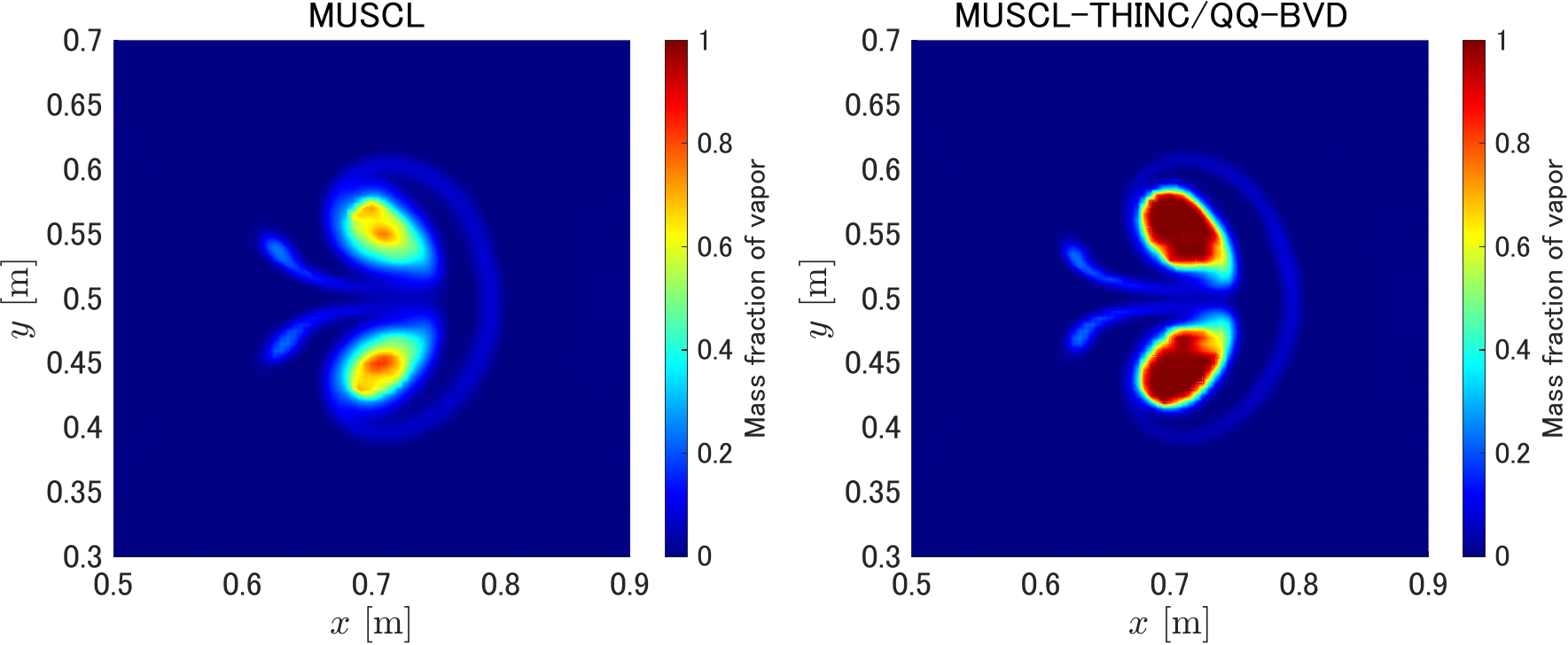}
    \caption{Same as Fig. \ref{fig:2Dbubblecomp_Yg_t=1.4_allSchemes_noPhaseChange}, but with phase change (p-pT-pTG-relaxation).}
    \label{fig:2Dbubblecomp_Yg_t=1.4_allSchemes_PhaseChange}
\end{figure}
\fi
The numerical results are shown in Figs. \ref{fig:2Dbubblecomp_Yg_t=1.4_allSchemes_noPhaseChange} and \ref{fig:2Dbubblecomp_Yg_t=1.4_allSchemes_PhaseChange}. Without phase change, the MUSCL-THINC/QQ-BVD scheme shows a slightly higher vapor mass fraction than the MUSCL scheme, but the difference is not significant. However, in the case with phase change, the MUSCL-THINC/QQ-BVD scheme clearly captures the bubbles generated by phase change, whereas the MUSCL scheme shows a blurred gas-liquid interface. The MUSCL scheme struggles to accurately represent the dynamically generated gas-liquid interface due to the excessive numerical dissipation errors. In contrast, the BVD scheme incorporating the THINC/QQ reconstruction function effectively suppresses numerical dissipation errors and captures the dynamically generated gas-liquid interface with high resolution.

\iffigure
\begin{figure}[htbp]
    \centering
    \includegraphics[width=8cm]{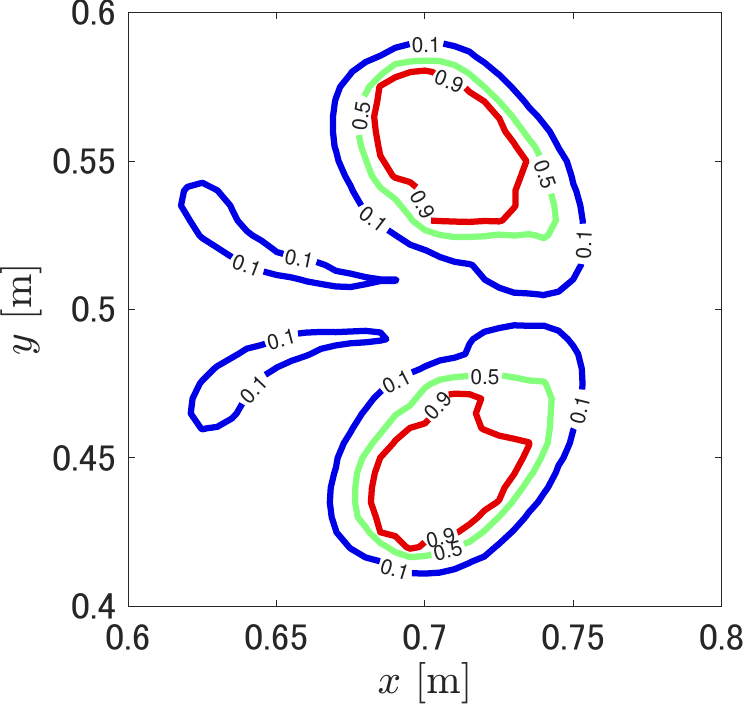}
    \includegraphics[width=8cm]{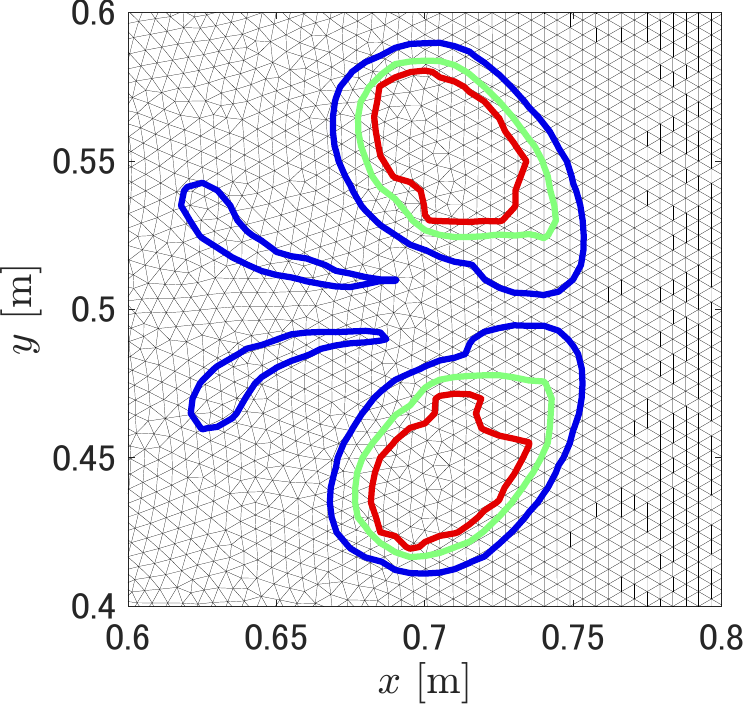}
    \caption{Contour plots of the mass fraction of vapor at $t=1.4\ \mathrm{ms}$ in vapor-bubble compression problem with phase change calculated with the MUSCL-THINC/QQ-BVD scheme. The left panel shows contour lines and their labels. The right panel shows contour lines and triangle mesh.}
    \label{fig:2Dbubblecomp_Yg_t=1.4_mesh_PhaseChange}
\end{figure}
\fi
In Fig. \ref{fig:2Dbubblecomp_Yg_t=1.4_mesh_PhaseChange}, the computational results of the MUSCL-THINC/QQ-BVD scheme from Fig. \ref{fig:2Dbubblecomp_Yg_t=1.4_allSchemes_PhaseChange} are presented as contour plots. As shown in the left panel, the red line represents a vapor mass fraction of 0.9, the green line 0.5, and the blue line 0.1. The right panel displays the contours with the triangular mesh. It can be observed that the distance between the red and blue lines is about 3 to 4 cells wide in the narrowest regions. This indicates that the MUSCL-THINC/QQ-BVD scheme is effective in achieving high-resolution calculations of dynamically generated gas-liquid interfaces with a relatively small number of grid cells.

\subsection{High-pressure fuel injector}
In this problem \cite{Saurel2008, Pelanti2014}, we simulate the high-pressure liquid fuel (dodecane) injection from a tank into an atmospheric pressure chamber. The nozzle has a length of 10 cm and a height of 4 cm, with a throat width of 1.2 cm. The inclination angles of the converging and diverging chambers with respect to the horizontal direction are 45 degrees and 10 degrees, respectively. The shape of the computational domain is shown in Fig. \ref{fig:mesh_injector_projectile}.
\iffigure
\begin{figure}[htbp]
    \centering
    \includegraphics[width=8cm]{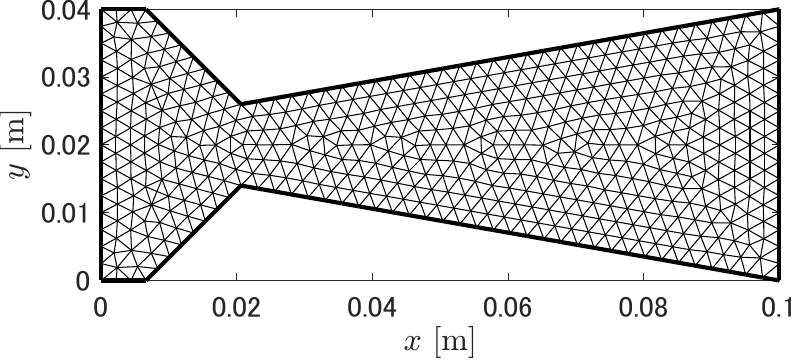}
    \includegraphics[width=8cm]{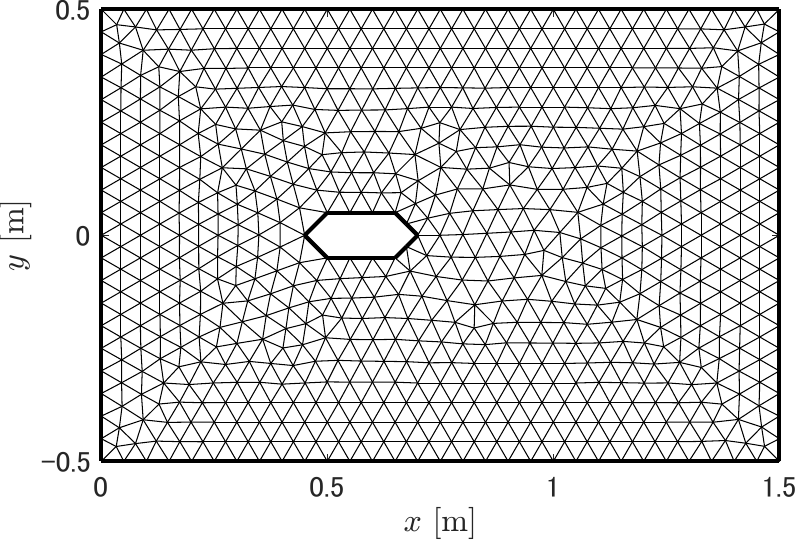}
    \caption{The computational domain and the triangle meshes for the high-pressure fuel injector problem (left) and the high-velocity underwater projectile problem (right). To improve visibility, the grid size is displayed at 10 times the size of the actual grid used in the calculations.}
    \label{fig:mesh_injector_projectile}
\end{figure}
\fi
The grid length is $h \approx 1/4000 \, \mathrm{m}$. The inside of the tank ($x<0.6 \, \mathrm{cm}$) is filled with liquid dodecane ($\alpha_1 = 1-10^{-4}$), with pressure and temperature set to $10^8 \, \mathrm{Pa}$ and $640 \, \mathrm{K}$, respectively. Our calculations show that the corresponding liquid density of dodecane is approximately $537 \, \mathrm{kg/m^3}$. The vapor density is set such that the vapor temperature also reaches $640 \, \mathrm{K}$. The pressure and temperature of vapor dodecane outside the tank are set to $10^5 \, \mathrm{Pa}$ and $300 \, \mathrm{K}$, respectively.
The initial condition in terms of the primitive variables are set as,
\begin{align}
    &\qty(\alpha_1,\rho_1\ \mathrm{[kg/m^3]},\rho_2\ \mathrm{[kg/m^3]},u\ \mathrm{[m/s]},v\ \mathrm{[m/s]},p\ \mathrm{[Pa]})_{t=0}= \nonumber \\
    &\quad
    \begin{dcases}
        \qty(1-10^{-4}, 536.99, 3.1946 \times 10^3, 0, 0, 10^{8}) & \text{for} \quad x<6 \times 10^{-3}\ \mathrm{m},\\
        \qty(10^{-4}, 916.68, 6.8151, 0, 0, 10^{5}) & \text{otherwise}.
    \end{dcases}
\end{align}
The boundary condition is an inlet boundary for the left boundary, an outlet boundary for the right boundary, and reflective conditions for the other boundaries. The material-dependent parameters in the SGEOS are based on the physical properties of dodecane as listed in Table \ref{tab:SGEOS_param}. In this problem, the CFL number is set to $0.2$ for computational stability.

\iffigure
\begin{figure}[htbp]
    \centering
    \includegraphics[width=15cm]{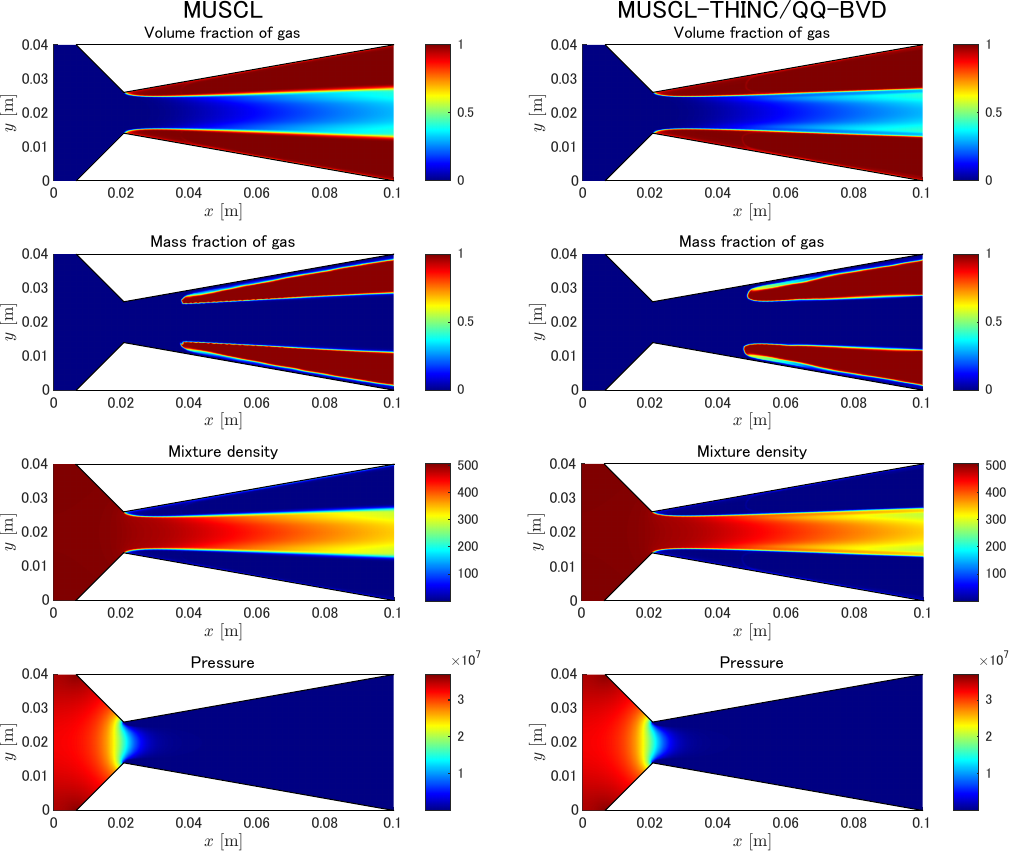}
    \caption{Pseudo-color plots of volume fraction of gas, mass fraction of gas, mixture density, and pressure at $t=\SI{1400}{\micro s}$ in high-pressure fuel injector problem without phase change (p-relaxation). The left and right panels show the numerical results of MUSCL and MUSCL-THINC/QQ-BVD schemes respectively.}
    \label{fig:injector_allSchemes_noPhaseChange}
\end{figure}
\begin{figure}[htbp]
    \centering
    \includegraphics[width=15cm]{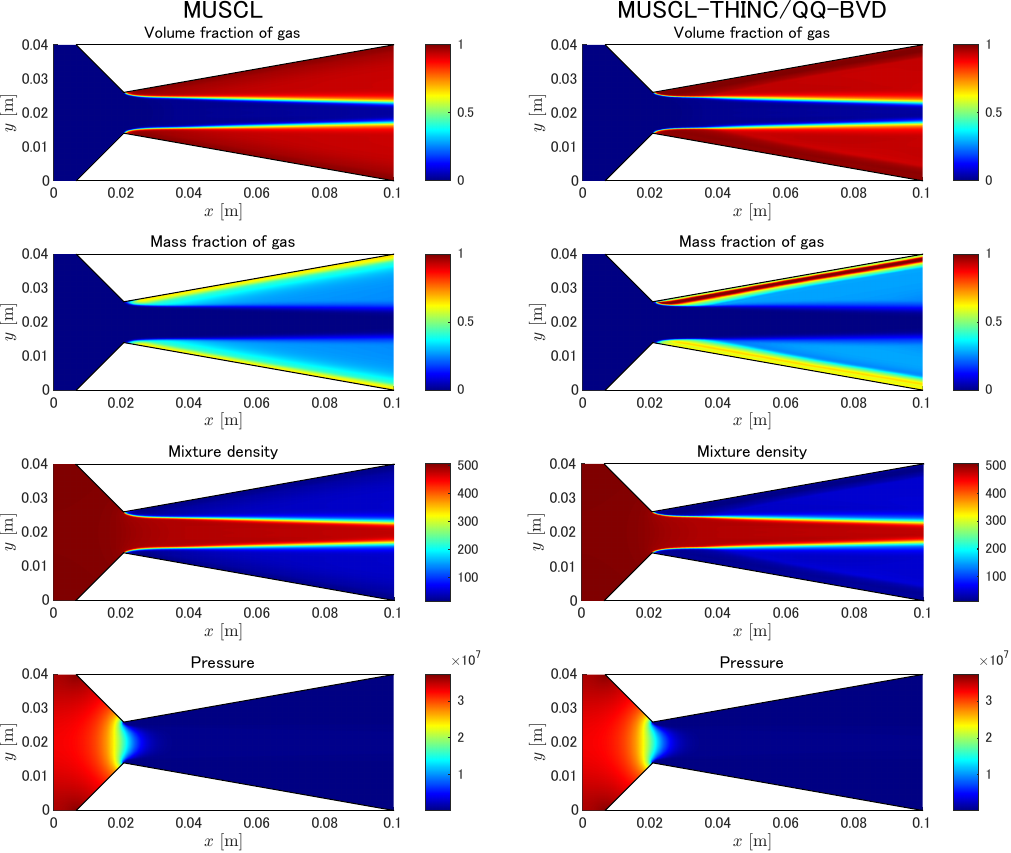}
    \caption{Same as Fig. \ref{fig:injector_allSchemes_noPhaseChange}, but with phase change (p-pT-pTG-relaxation) at $t=\SI{800}{\micro s}$.}
    \label{fig:injector_allSchemes_PhaseChange}
\end{figure}
\fi
Figs. \ref{fig:injector_allSchemes_noPhaseChange} and \ref{fig:injector_allSchemes_PhaseChange} show the numerical results without and with phase change respectively. In the case without phase change, significant differences between the results of MUSCL and MUSCL-THINC/QQ-BVD schemes are not observed. The results are generally consistent with those presented in \cite{Saurel2008, Pelanti2014}. In the case with phase change, a difference between the two schemes is observed in the results of the gas mass fraction. In the MUSCL-THINC/QQ-BVD scheme, the gas mass fraction near the upper walls of the diverging chamber approaches a value close to 1, whereas, in the MUSCL scheme, the gas mass fraction remains around 0.6, a significantly lower value. The increase in the gas mass fraction is considered to be caused by cavitation, which occurs as the liquid jet passing through the throat region (around $x \approx 6.6 \, \text{cm}$) experiences low pressure. The MUSCL scheme, due to numerical dissipation, fails to adequately capture this cavitation phenomenon, while the MUSCL-THINC/QQ-BVD scheme maintains a sharp gas-liquid interface, particularly near the upper boundary.

In this problem, spatial symmetry is present with respect to the axis at $ y = 0.02\ \mathrm{m} $. However, in the numerical results of the MUSCL-THINC/QQ-BVD scheme, an asymmetric distribution was observed. This result is attributed to the asymmetry of the grid and the low dissipation nature of the scheme. In reference \cite{Wakimura2024High-resolutionEvaporation}, it was noted that for structured grids, careful implementation of a computation sequence that avoids asymmetric rounding errors can yield a fully symmetric solution, emphasizing that rounding errors can depend on the order of operations involving three or more numbers. On the other hand, for unstructured grids with inherent asymmetry, maintaining spatial symmetry through such computation sequence adjustments is nearly impossible. Additionally, it is known that the lower the dissipation of a scheme, the more pronounced the spatial symmetry error becomes \cite{Fleischmann2019, Wakimura2021a}. Therefore, the spatial asymmetry observed in the numerical results of the MUSCL-THINC/QQ-BVD scheme seems to be a natural outcome.

Fig. \ref{fig:injector_timeev} shows the time evolution of the gas mass fraction obtained by the MUSCL and MUSCL-THINC/QQ-BVD schemes. In the results of the MUSCL-THINC/QQ-BVD scheme, asymmetric flow can already be observed at $ t = \SI{104}{\micro s} $. At $ t = \SI{160}{\micro s} $, the gas mass fraction reaches a value of 1 in the upper region of the throat, indicating the occurrence of cavitation. Due to small differences in the distribution of physical quantities, such as pressure and temperature, between the upper and lower regions of the throat, the application of chemical relaxation can vary, potentially leading to cavitation occurring in the upper part but not in the lower part due to the asymmetry in the flow. After $ t = \SI{240}{\micro s} $, the dynamically generated gas continues to form in the upper region of the throat due to phase change, and a layer of gas extends toward the rear of the diverging chamber. In contrast, the results of the MUSCL scheme show that the mass fraction distribution is more diffused compared to the MUSCL-THINC/QQ-BVD scheme, and no cavitation is observed near the throat. This difference highlights the advantage of the MUSCL-THINC/QQ-BVD scheme in capturing the sharp interfaces and dynamic behavior of the flow, including the cavitation phenomenon.

\iffigure
\begin{figure}[htbp]
    \centering
    \includegraphics[width=15cm]{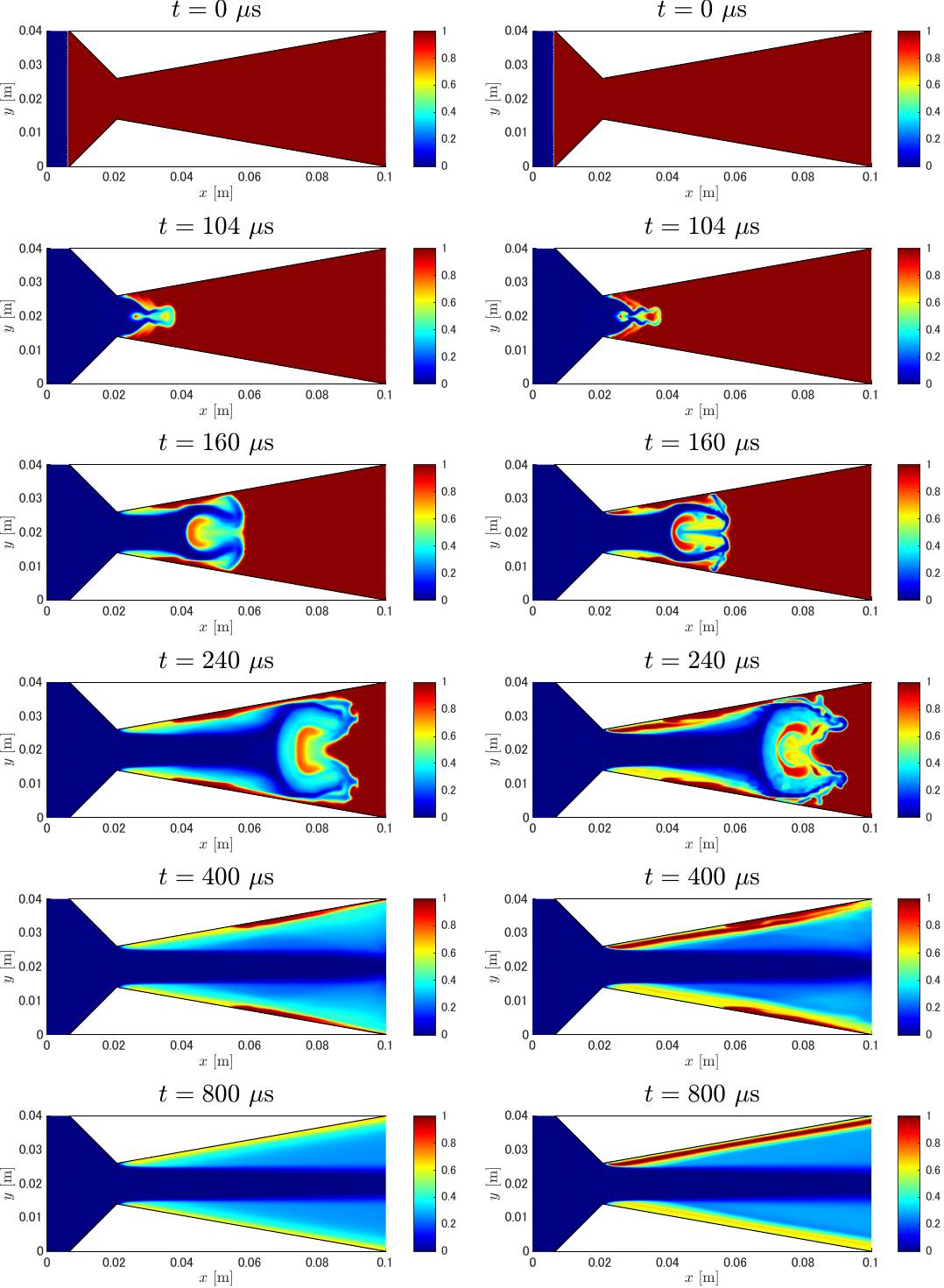}
    \caption{Pseudo-color plots of mass fraction of gas at different time steps in high-pressure fuel injector problem with phase change (p-pT-pTG-relaxation). The left and right panels show the numerical results of MUSCL and MUSCL-THINC/QQ-BVD schemes respectively.}
    \label{fig:injector_timeev}
\end{figure}
\fi


\subsection{High-velocity underwater projectile}
In this problem \cite{Saurel2008, Pelanti2014}, a hexagonal immersed obstacle is placed in a high-velocity water flow. Behind the obstacle, a region called a cavitation pocket is formed, where the pressure drops significantly. In the case with phase change, the cavitation phenomenon is observed in this region. The liquid flow velocity is set to $600 \, \mathrm{m/s}$, with an atmospheric pressure of $10^5 \, \mathrm{Pa}$, and a liquid density of $1150 \, \mathrm{kg/m^3}$. Under these conditions, the liquid temperature becomes approximately $355 \, \mathrm{K}$, and the corresponding gas density, where the gas temperature matches the liquid temperature, is calculated to be $0.6304 \, \mathrm{kg/m^3}$. The liquid contains a small amount of vapor ($\alpha_2=10^{-3}$). In summary, the initial condition is written as,
\begin{align}
    &\qty(\alpha_1,\rho_1\ \mathrm{[kg/m^3]},\rho_2\ \mathrm{[kg/m^3]},u\ \mathrm{[m/s]},v\ \mathrm{[m/s]},p\ \mathrm{[Pa]})_{t=0} \nonumber \\
    &=\qty(1-10^{-3}, 1150, 0.6304, 600, 0, 10^{5}) .
\end{align}
The shape of the computational domain is shown in Fig. \ref{fig:mesh_injector_projectile}. The grid length is $h \approx 1/200 \, \mathrm{m}$. The boundary condition is an inlet boundary for the left boundary, outlet boundaries for the top, bottom, and right boundaries, and reflective conditions for the obstacle boundaries.

\iffigure
\begin{figure}[htbp]
    \centering
    \includegraphics[width=15cm]{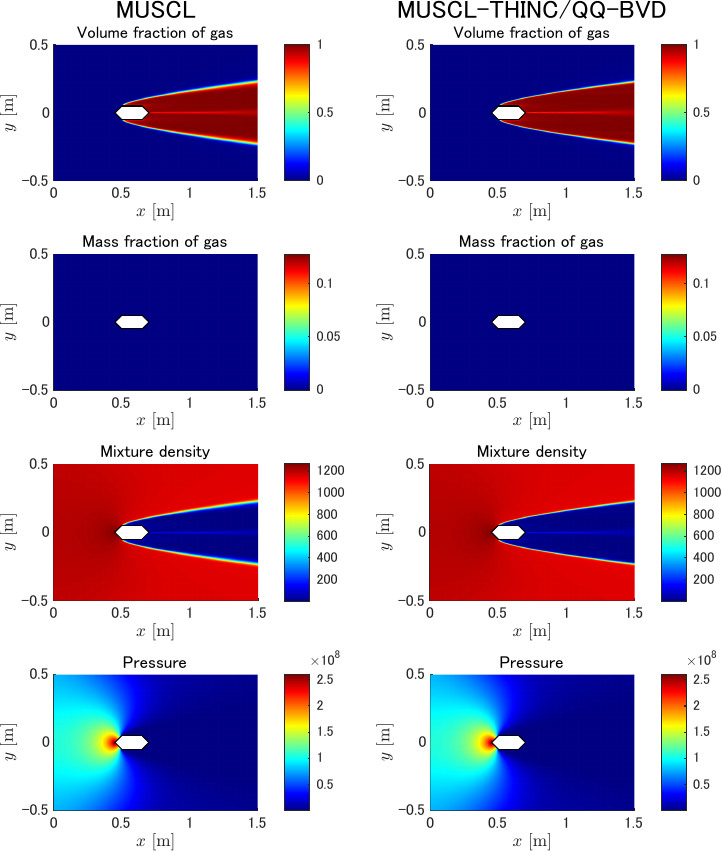}
    \caption{Pseudo-color plots of volume fraction of gas, mass fraction of gas, mixture density, and pressure at $t=10\ \mathrm{ms}$ in high-velocity underwater projectile problem without phase change (p-relaxation). The left and right panels show the numerical results of MUSCL and MUSCL-THINC/QQ-BVD schemes respectively.}
    \label{fig:projectile_allSchemes_noPhaseChange}
\end{figure}
\begin{figure}[htbp]
    \centering
    \includegraphics[width=15cm]{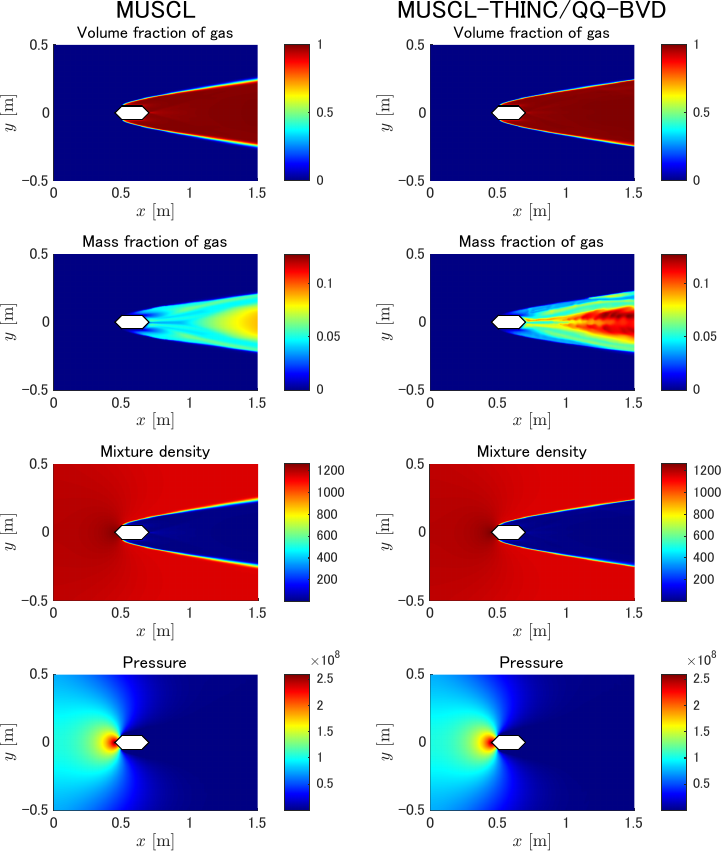}
    \caption{Same as Fig. \ref{fig:projectile_allSchemes_noPhaseChange}, but with phase change (p-pT-pTG-relaxation).}
    \label{fig:projectile_allSchemes_PhaseChange}
\end{figure}
\fi
Figs. \ref{fig:projectile_allSchemes_noPhaseChange} and \ref{fig:projectile_allSchemes_PhaseChange} show the numerical results without and with phase change respectively. The general flow structure of these numerical results is consistent with the findings of \cite{Saurel2008, Pelanti2014}. When comparing the results of the MUSCL and MUSCL-THINC/QQ-BVD schemes in the absence of phase change, the MUSCL-THINC/QQ-BVD scheme captures the discontinuous surfaces of gas volume fraction and mixture density slightly more sharply. Despite the pressure difference across these discontinuities, the MUSCL scheme exhibits some diffusion due to numerical dissipation error, which smears the discontinuous interfaces. 

The most noticeable difference between the two schemes emerges in the results when phase change is included. Behind the obstacle, both schemes show an increase in the gas mass fraction, which corresponds to cavitation caused by the pressure drop. However, the gas mass fraction is higher in the MUSCL-THINC/QQ-BVD scheme than in the MUSCL scheme. This result is considered to be attributed to the low-dissipation nature of the MUSCL-THINC/QQ-BVD scheme. While the dynamically generated liquid-gas interface becomes blurred due to numerical dissipation in the MUSCL scheme, the MUSCL-THINC/QQ-BVD scheme captures it with high resolution.

Thus, the low-dissipation MUSCL-THINC/QQ-BVD scheme is more suitable for accurately reproducing phase change phenomena compared to existing schemes.

\section{Conclusion}

Compressible gas-liquid two-phase flow, including phase changes, was computed on unstructured grids using the MUSCL-THINC/QQ-BVD scheme. The six-equation model was adopted as the governing equation, and the detailed computational procedure for the non-conservative terms that appear in the volume fraction and phasic total energy equations on unstructured grids was clarified. By defining and calculating the non-conservative terms on cell boundaries using second-order approximations, interpolation of spatial derivatives was avoided, and it was confirmed that the numerical results for the Riemann problems were consistent with those on structured grids.

For the reconstruction functions, in addition to the MUSCL method with the existing MLP limiter, the THINC/QQ method, originally proposed for moving interface capturing, was incorporated as a candidate interpolation function. The appropriate reconstruction function was selected for each cell based on the BVD principle. The MUSCL-THINC/QQ-BVD scheme is designed to select the MUSCL method for smooth solutions and the THINC/QQ method near discontinuities. Compared to the MUSCL scheme, the numerical results of the MUSCL-THINC/QQ-BVD scheme demonstrated that it captures discontinuities, such as contact discontinuities and gas-liquid interfaces, with low dissipation. 

Furthermore, the MUSCL-THINC/QQ-BVD scheme accurately captured dynamically generated gas-liquid interfaces due to phase changes. Thanks to its low dissipation property for discontinuous solutions, the MUSCL-THINC/QQ-BVD scheme is well-suited for high-fidelity calculations of phase change phenomena, such as cavitation.

\section*{Acknowledgements}

This work was supported in part by the fund from JSPS (Japan Society for the Promotion of Science) under Grant Nos. 18H01366, 19H05613, 22J21252 and 22KJ1331.

\bibliography{references}

\begin{thebibliography}{83}%
\makeatletter
\providecommand \@ifxundefined [1]{%
 \@ifx{#1\undefined}
}%
\providecommand \@ifnum [1]{%
 \ifnum #1\expandafter \@firstoftwo
 \else \expandafter \@secondoftwo
 \fi
}%
\providecommand \@ifx [1]{%
 \ifx #1\expandafter \@firstoftwo
 \else \expandafter \@secondoftwo
 \fi
}%
\providecommand \natexlab [1]{#1}%
\providecommand \enquote  [1]{``#1''}%
\providecommand \bibnamefont  [1]{#1}%
\providecommand \bibfnamefont [1]{#1}%
\providecommand \citenamefont [1]{#1}%
\providecommand \href@noop [0]{\@secondoftwo}%
\providecommand \href [0]{\begingroup \@sanitize@url \@href}%
\providecommand \@href[1]{\@@startlink{#1}\@@href}%
\providecommand \@@href[1]{\endgroup#1\@@endlink}%
\providecommand \@sanitize@url [0]{\catcode `\\12\catcode `\$12\catcode `\&12\catcode `\#12\catcode `\^12\catcode `\_12\catcode `\%12\relax}%
\providecommand \@@startlink[1]{}%
\providecommand \@@endlink[0]{}%
\providecommand \url  [0]{\begingroup\@sanitize@url \@url }%
\providecommand \@url [1]{\endgroup\@href {#1}{\urlprefix }}%
\providecommand \urlprefix  [0]{URL }%
\providecommand \Eprint [0]{\href }%
\providecommand \doibase [0]{http://dx.doi.org/}%
\providecommand \selectlanguage [0]{\@gobble}%
\providecommand \bibinfo  [0]{\@secondoftwo}%
\providecommand \bibfield  [0]{\@secondoftwo}%
\providecommand \translation [1]{[#1]}%
\providecommand \BibitemOpen [0]{}%
\providecommand \bibitemStop [0]{}%
\providecommand \bibitemNoStop [0]{.\EOS\space}%
\providecommand \EOS [0]{\spacefactor3000\relax}%
\providecommand \BibitemShut  [1]{\csname bibitem#1\endcsname}%
\let\auto@bib@innerbib\@empty
\bibitem [{\citenamefont {Hirt}\ and\ \citenamefont {Nichols}(1981)}]{Hirt1981}%
  \BibitemOpen
  \bibfield  {author} {\bibinfo {author} {\bibfnamefont {C.~W.}\ \bibnamefont {Hirt}}\ and\ \bibinfo {author} {\bibfnamefont {B.~D.}\ \bibnamefont {Nichols}},\ }\bibfield  {title} {\enquote {\bibinfo {title} {{Volume of Fluid (VOF) Method for the Dynamics of Free Boundaries}},}\ }\href@noop {} {\bibfield  {journal} {\bibinfo  {journal} {Journal of Computational Physics}\ }\textbf {\bibinfo {volume} {39}},\ \bibinfo {pages} {201--225} (\bibinfo {year} {1981})}\BibitemShut {NoStop}%
\bibitem [{\citenamefont {Kothe}\ and\ \citenamefont {Rider}(1998)}]{Kothe1998ReconstructingTracking}%
  \BibitemOpen
  \bibfield  {author} {\bibinfo {author} {\bibfnamefont {D.~B.}\ \bibnamefont {Kothe}}\ and\ \bibinfo {author} {\bibfnamefont {W.~J.}\ \bibnamefont {Rider}},\ }\bibfield  {title} {\enquote {\bibinfo {title} {{Reconstructing Volume Tracking}},}\ }\href@noop {} {\bibfield  {journal} {\bibinfo  {journal} {Journal of Computational Physics}\ }\textbf {\bibinfo {volume} {141}},\ \bibinfo {pages} {112–152} (\bibinfo {year} {1998})}\BibitemShut {NoStop}%
\bibitem [{\citenamefont {Welch}\ and\ \citenamefont {Wilson}(2000)}]{Welch2000}%
  \BibitemOpen
  \bibfield  {author} {\bibinfo {author} {\bibfnamefont {S.~W.}\ \bibnamefont {Welch}}\ and\ \bibinfo {author} {\bibfnamefont {J.}~\bibnamefont {Wilson}},\ }\bibfield  {title} {\enquote {\bibinfo {title} {{A Volume of Fluid Based Method for Fluid Flows with Phase Change}},}\ }\href {\doibase 10.1006/jcph.2000.6481} {\bibfield  {journal} {\bibinfo  {journal} {Journal of Computational Physics}\ }\textbf {\bibinfo {volume} {160}},\ \bibinfo {pages} {662--682} (\bibinfo {year} {2000})}\BibitemShut {NoStop}%
\bibitem [{\citenamefont {Mulder}, \citenamefont {Osher},\ and\ \citenamefont {Sethian}(1992)}]{Mulder1992}%
  \BibitemOpen
  \bibfield  {author} {\bibinfo {author} {\bibfnamefont {W.}~\bibnamefont {Mulder}}, \bibinfo {author} {\bibfnamefont {S.}~\bibnamefont {Osher}}, \ and\ \bibinfo {author} {\bibfnamefont {J.~A.}\ \bibnamefont {Sethian}},\ }\bibfield  {title} {\enquote {\bibinfo {title} {{Computing interface motion in compressible gas dynamics}},}\ }\href {\doibase 10.1016/0021-9991(92)90229-R} {\bibfield  {journal} {\bibinfo  {journal} {Journal of Computational Physics}\ }\textbf {\bibinfo {volume} {100}},\ \bibinfo {pages} {209--228} (\bibinfo {year} {1992})}\BibitemShut {NoStop}%
\bibitem [{\citenamefont {Sussman}(1994)}]{Sussman1994}%
  \BibitemOpen
  \bibfield  {author} {\bibinfo {author} {\bibfnamefont {M.}~\bibnamefont {Sussman}},\ }\bibfield  {title} {\enquote {\bibinfo {title} {{A level set approach for computing solutions to incompressible two-phase flow}},}\ }\href {\doibase 10.1006/jcph.1994.1155} {\bibfield  {journal} {\bibinfo  {journal} {Journal of Computational Physics}\ }\textbf {\bibinfo {volume} {114}},\ \bibinfo {pages} {146--159} (\bibinfo {year} {1994})}\BibitemShut {NoStop}%
\bibitem [{\citenamefont {Unverdi}\ and\ \citenamefont {Tryggvason}(1992)}]{Unverdi1992AFlows}%
  \BibitemOpen
  \bibfield  {author} {\bibinfo {author} {\bibfnamefont {S.~O.}\ \bibnamefont {Unverdi}}\ and\ \bibinfo {author} {\bibfnamefont {G.}~\bibnamefont {Tryggvason}},\ }\bibfield  {title} {\enquote {\bibinfo {title} {{A front-tracking method for viscous, incompressible, multi-fluid flows}},}\ }\href {\doibase 10.1016/0021-9991(92)90307-K} {\bibfield  {journal} {\bibinfo  {journal} {Journal of Computational Physics}\ }\textbf {\bibinfo {volume} {100}},\ \bibinfo {pages} {25--37} (\bibinfo {year} {1992})}\BibitemShut {NoStop}%
\bibitem [{\citenamefont {Tryggvason}\ \emph {et~al.}(2001)\citenamefont {Tryggvason}, \citenamefont {Bunner}, \citenamefont {Esmaeeli}, \citenamefont {Juric}, \citenamefont {Al-Rawahi}, \citenamefont {Tauber}, \citenamefont {Han}, \citenamefont {Nas},\ and\ \citenamefont {Jan}}]{Tryggvason2001AFlow}%
  \BibitemOpen
  \bibfield  {author} {\bibinfo {author} {\bibfnamefont {G.}~\bibnamefont {Tryggvason}}, \bibinfo {author} {\bibfnamefont {B.}~\bibnamefont {Bunner}}, \bibinfo {author} {\bibfnamefont {A.}~\bibnamefont {Esmaeeli}}, \bibinfo {author} {\bibfnamefont {D.}~\bibnamefont {Juric}}, \bibinfo {author} {\bibfnamefont {N.}~\bibnamefont {Al-Rawahi}}, \bibinfo {author} {\bibfnamefont {W.}~\bibnamefont {Tauber}}, \bibinfo {author} {\bibfnamefont {J.}~\bibnamefont {Han}}, \bibinfo {author} {\bibfnamefont {S.}~\bibnamefont {Nas}}, \ and\ \bibinfo {author} {\bibfnamefont {Y.~J.}\ \bibnamefont {Jan}},\ }\bibfield  {title} {\enquote {\bibinfo {title} {{A Front-Tracking Method for the Computations of Multiphase Flow}},}\ }\href {\doibase 10.1006/jcph.2001.6726} {\bibfield  {journal} {\bibinfo  {journal} {Journal of Computational Physics}\ }\textbf {\bibinfo {volume} {169}},\ \bibinfo {pages} {708--759} (\bibinfo {year} {2001})}\BibitemShut {NoStop}%
\bibitem [{\citenamefont {Baer}\ and\ \citenamefont {Nunziato}(1986)}]{Baer1986}%
  \BibitemOpen
  \bibfield  {author} {\bibinfo {author} {\bibfnamefont {M.~R.}\ \bibnamefont {Baer}}\ and\ \bibinfo {author} {\bibfnamefont {J.~W.}\ \bibnamefont {Nunziato}},\ }\bibfield  {title} {\enquote {\bibinfo {title} {{A two-phase mixture theory for the deflagration-to-detonation transition (ddt) in reactive granular materials}},}\ }\href {\doibase 10.1016/0301-9322(86)90033-9} {\bibfield  {journal} {\bibinfo  {journal} {International Journal of Multiphase Flow}\ }\textbf {\bibinfo {volume} {12}},\ \bibinfo {pages} {861--889} (\bibinfo {year} {1986})}\BibitemShut {NoStop}%
\bibitem [{\citenamefont {Saurel}\ and\ \citenamefont {Abgrall}(1999)}]{Saurel1999}%
  \BibitemOpen
  \bibfield  {author} {\bibinfo {author} {\bibfnamefont {R.}~\bibnamefont {Saurel}}\ and\ \bibinfo {author} {\bibfnamefont {R.}~\bibnamefont {Abgrall}},\ }\bibfield  {title} {\enquote {\bibinfo {title} {{A Multiphase Godunov Method for Compressible Multifluid and Multiphase Flows}},}\ }\href {\doibase 10.1006/jcph.1999.6187} {\bibfield  {journal} {\bibinfo  {journal} {Journal of Computational Physics}\ }\textbf {\bibinfo {volume} {150}},\ \bibinfo {pages} {425--467} (\bibinfo {year} {1999})}\BibitemShut {NoStop}%
\bibitem [{\citenamefont {Kapila}\ \emph {et~al.}(2001)\citenamefont {Kapila}, \citenamefont {Menikoff}, \citenamefont {Bdzil}, \citenamefont {Son},\ and\ \citenamefont {Stewart}}]{Kapila2001}%
  \BibitemOpen
  \bibfield  {author} {\bibinfo {author} {\bibfnamefont {A.~K.}\ \bibnamefont {Kapila}}, \bibinfo {author} {\bibfnamefont {R.}~\bibnamefont {Menikoff}}, \bibinfo {author} {\bibfnamefont {J.~B.}\ \bibnamefont {Bdzil}}, \bibinfo {author} {\bibfnamefont {S.~F.}\ \bibnamefont {Son}}, \ and\ \bibinfo {author} {\bibfnamefont {D.~S.}\ \bibnamefont {Stewart}},\ }\bibfield  {title} {\enquote {\bibinfo {title} {{Two-phase modeling of deflagration-to-detonation transition in granular materials: Reduced equations}},}\ }\href {\doibase 10.1063/1.1398042} {\bibfield  {journal} {\bibinfo  {journal} {Physics of Fluids}\ }\textbf {\bibinfo {volume} {13}},\ \bibinfo {pages} {3002--3024} (\bibinfo {year} {2001})}\BibitemShut {NoStop}%
\bibitem [{\citenamefont {{Richard Saurel}}, \citenamefont {Petitpas},\ and\ \citenamefont {Berry}(2009)}]{RichardSaurel2009}%
  \BibitemOpen
  \bibfield  {author} {\bibinfo {author} {\bibnamefont {{Richard Saurel}}}, \bibinfo {author} {\bibfnamefont {F.}~\bibnamefont {Petitpas}}, \ and\ \bibinfo {author} {\bibfnamefont {R.~A.}\ \bibnamefont {Berry}},\ }\bibfield  {title} {\enquote {\bibinfo {title} {{Simple and efficient relaxation methods for interfaces separating compressible fluids, cavitating flows and shocks in multiphase mixtures}},}\ }\href {\doibase 10.1016/j.jcp.2008.11.002} {\bibfield  {journal} {\bibinfo  {journal} {Journal of Computational Physics}\ }\textbf {\bibinfo {volume} {228}},\ \bibinfo {pages} {1678--1712} (\bibinfo {year} {2009})}\BibitemShut {NoStop}%
\bibitem [{\citenamefont {Picard}\ and\ \citenamefont {Bishnoi}(1987)}]{Picard1987}%
  \BibitemOpen
  \bibfield  {author} {\bibinfo {author} {\bibfnamefont {D.~J.}\ \bibnamefont {Picard}}\ and\ \bibinfo {author} {\bibfnamefont {P.~R.}\ \bibnamefont {Bishnoi}},\ }\bibfield  {title} {\enquote {\bibinfo {title} {{Calculation of the thermodynamic sound velocity in two-phase multicomponent fluids}},}\ }\href {\doibase 10.1016/0301-9322(87)90050-4} {\bibfield  {journal} {\bibinfo  {journal} {International Journal of Multiphase Flow}\ }\textbf {\bibinfo {volume} {13}},\ \bibinfo {pages} {295--308} (\bibinfo {year} {1987})}\BibitemShut {NoStop}%
\bibitem [{\citenamefont {Fl{\aa}tten}\ and\ \citenamefont {Lund}(2011)}]{Flatten2011}%
  \BibitemOpen
  \bibfield  {author} {\bibinfo {author} {\bibfnamefont {T.}~\bibnamefont {Fl{\aa}tten}}\ and\ \bibinfo {author} {\bibfnamefont {H.}~\bibnamefont {Lund}},\ }\bibfield  {title} {\enquote {\bibinfo {title} {{Relaxation two-phase flow models and the subcharacteristic condition}},}\ }\href {\doibase 10.1142/S0218202511005775} {\bibfield  {journal} {\bibinfo  {journal} {Mathematical Models and Methods in Applied Sciences}\ }\textbf {\bibinfo {volume} {21}},\ \bibinfo {pages} {2379--2407} (\bibinfo {year} {2011})}\BibitemShut {NoStop}%
\bibitem [{\citenamefont {Pelanti}\ and\ \citenamefont {Shyue}(2014)}]{Pelanti2014}%
  \BibitemOpen
  \bibfield  {author} {\bibinfo {author} {\bibfnamefont {M.}~\bibnamefont {Pelanti}}\ and\ \bibinfo {author} {\bibfnamefont {K.~M.}\ \bibnamefont {Shyue}},\ }\bibfield  {title} {\enquote {\bibinfo {title} {{A mixture-energy-consistent six-equation two-phase numerical model for fluids with interfaces, cavitation and evaporation waves}},}\ }\href {\doibase 10.1016/j.jcp.2013.12.003} {\bibfield  {journal} {\bibinfo  {journal} {Journal of Computational Physics}\ }\textbf {\bibinfo {volume} {259}},\ \bibinfo {pages} {331--357} (\bibinfo {year} {2014})}\BibitemShut {NoStop}%
\bibitem [{\citenamefont {Cockburn}\ and\ \citenamefont {Shu}(1998)}]{Cockburn1998TheV}%
  \BibitemOpen
  \bibfield  {author} {\bibinfo {author} {\bibfnamefont {B.}~\bibnamefont {Cockburn}}\ and\ \bibinfo {author} {\bibfnamefont {C.~W.}\ \bibnamefont {Shu}},\ }\bibfield  {title} {\enquote {\bibinfo {title} {{The Runge–Kutta Discontinuous Galerkin Method for Conservation Laws V}},}\ }\href@noop {} {\bibfield  {journal} {\bibinfo  {journal} {Journal of Computational Physics}\ }\textbf {\bibinfo {volume} {141}},\ \bibinfo {pages} {199--224} (\bibinfo {year} {1998})}\BibitemShut {NoStop}%
\bibitem [{\citenamefont {Krivodonova}\ \emph {et~al.}(2004)\citenamefont {Krivodonova}, \citenamefont {Xin}, \citenamefont {Remacle}, \citenamefont {Chevaugeon},\ and\ \citenamefont {Flaherty}}]{Krivodonova2004ShockLaws}%
  \BibitemOpen
  \bibfield  {author} {\bibinfo {author} {\bibfnamefont {L.}~\bibnamefont {Krivodonova}}, \bibinfo {author} {\bibfnamefont {J.}~\bibnamefont {Xin}}, \bibinfo {author} {\bibfnamefont {J.~F.}\ \bibnamefont {Remacle}}, \bibinfo {author} {\bibfnamefont {N.}~\bibnamefont {Chevaugeon}}, \ and\ \bibinfo {author} {\bibfnamefont {J.~E.}\ \bibnamefont {Flaherty}},\ }\bibfield  {title} {\enquote {\bibinfo {title} {{Shock detection and limiting with discontinuous Galerkin methods for hyperbolic conservation laws}},}\ }\href {\doibase 10.1016/j.apnum.2003.11.002} {\bibfield  {journal} {\bibinfo  {journal} {Applied Numerical Mathematics}\ }\textbf {\bibinfo {volume} {48}},\ \bibinfo {pages} {323--338} (\bibinfo {year} {2004})}\BibitemShut {NoStop}%
\bibitem [{\citenamefont {Dumbser}\ \emph {et~al.}(2008)\citenamefont {Dumbser}, \citenamefont {Balsara}, \citenamefont {Toro},\ and\ \citenamefont {Munz}}]{Dumbser2008AMeshes}%
  \BibitemOpen
  \bibfield  {author} {\bibinfo {author} {\bibfnamefont {M.}~\bibnamefont {Dumbser}}, \bibinfo {author} {\bibfnamefont {D.~S.}\ \bibnamefont {Balsara}}, \bibinfo {author} {\bibfnamefont {E.~F.}\ \bibnamefont {Toro}}, \ and\ \bibinfo {author} {\bibfnamefont {C.~D.}\ \bibnamefont {Munz}},\ }\bibfield  {title} {\enquote {\bibinfo {title} {{A unified framework for the construction of one-step finite volume and discontinuous Galerkin schemes on unstructured meshes}},}\ }\href {\doibase 10.1016/j.jcp.2008.05.025} {\bibfield  {journal} {\bibinfo  {journal} {Journal of Computational Physics}\ }\textbf {\bibinfo {volume} {227}},\ \bibinfo {pages} {8209--8253} (\bibinfo {year} {2008})}\BibitemShut {NoStop}%
\bibitem [{\citenamefont {Zhu}\ \emph {et~al.}(2013)\citenamefont {Zhu}, \citenamefont {Zhong}, \citenamefont {Shu},\ and\ \citenamefont {Qiu}}]{Zhu2013Runge-KuttaMeshes}%
  \BibitemOpen
  \bibfield  {author} {\bibinfo {author} {\bibfnamefont {J.}~\bibnamefont {Zhu}}, \bibinfo {author} {\bibfnamefont {X.}~\bibnamefont {Zhong}}, \bibinfo {author} {\bibfnamefont {C.~W.}\ \bibnamefont {Shu}}, \ and\ \bibinfo {author} {\bibfnamefont {J.}~\bibnamefont {Qiu}},\ }\bibfield  {title} {\enquote {\bibinfo {title} {{Runge-Kutta discontinuous Galerkin method using a new type of WENO limiters on unstructured meshes}},}\ }\href {\doibase 10.1016/j.jcp.2013.04.012} {\bibfield  {journal} {\bibinfo  {journal} {Journal of Computational Physics}\ }\textbf {\bibinfo {volume} {248}},\ \bibinfo {pages} {200--220} (\bibinfo {year} {2013})}\BibitemShut {NoStop}%
\bibitem [{\citenamefont {Wang}(2002)}]{Wang2002SpectralFormulation}%
  \BibitemOpen
  \bibfield  {author} {\bibinfo {author} {\bibfnamefont {Z.~J.}\ \bibnamefont {Wang}},\ }\bibfield  {title} {\enquote {\bibinfo {title} {{Spectral (finite) volume method for conservation laws on unstructured grids. Basic Formulation}},}\ }\href {\doibase 10.1006/jcph.2002.7041} {\bibfield  {journal} {\bibinfo  {journal} {Journal of Computational Physics}\ }\textbf {\bibinfo {volume} {178}},\ \bibinfo {pages} {210--251} (\bibinfo {year} {2002})}\BibitemShut {NoStop}%
\bibitem [{\citenamefont {Wang}\ \emph {et~al.}(2007)\citenamefont {Wang}, \citenamefont {Liu}, \citenamefont {May},\ and\ \citenamefont {Jameson}}]{Wang2007SpectralEquations}%
  \BibitemOpen
  \bibfield  {author} {\bibinfo {author} {\bibfnamefont {Z.~J.}\ \bibnamefont {Wang}}, \bibinfo {author} {\bibfnamefont {Y.}~\bibnamefont {Liu}}, \bibinfo {author} {\bibfnamefont {G.}~\bibnamefont {May}}, \ and\ \bibinfo {author} {\bibfnamefont {A.}~\bibnamefont {Jameson}},\ }\bibfield  {title} {\enquote {\bibinfo {title} {{Spectral difference method for unstructured grids II: Extension to the Euler equations}},}\ }\href {\doibase 10.1007/s10915-006-9113-9} {\bibfield  {journal} {\bibinfo  {journal} {Journal of Scientific Computing}\ }\textbf {\bibinfo {volume} {32}},\ \bibinfo {pages} {45--71} (\bibinfo {year} {2007})}\BibitemShut {NoStop}%
\bibitem [{\citenamefont {Huynh}(2007)}]{Huynh2007AMethods}%
  \BibitemOpen
  \bibfield  {author} {\bibinfo {author} {\bibfnamefont {H.~T.}\ \bibnamefont {Huynh}},\ }\bibfield  {title} {\enquote {\bibinfo {title} {{A Flux Reconstruction Approach to High-Order Schemes Including Discontinuous Galerkin Methods}},}\ }in\ \href {\doibase 10.2514/6.2007-4079} {\emph {\bibinfo {booktitle} {18th AIAA Computational Fluid Dynamics Conference}}}\ (\bibinfo {year} {2007})\BibitemShut {NoStop}%
\bibitem [{\citenamefont {Witherden}, \citenamefont {Farrington},\ and\ \citenamefont {Vincent}(2014)}]{Witherden2014PyFR:Approach}%
  \BibitemOpen
  \bibfield  {author} {\bibinfo {author} {\bibfnamefont {F.~D.}\ \bibnamefont {Witherden}}, \bibinfo {author} {\bibfnamefont {A.~M.}\ \bibnamefont {Farrington}}, \ and\ \bibinfo {author} {\bibfnamefont {P.~E.}\ \bibnamefont {Vincent}},\ }\bibfield  {title} {\enquote {\bibinfo {title} {{PyFR: An open source framework for solving advection-diffusion type problems on streaming architectures using the flux reconstruction approach}},}\ }\href {\doibase 10.1016/j.cpc.2014.07.011} {\bibfield  {journal} {\bibinfo  {journal} {Computer Physics Communications}\ }\textbf {\bibinfo {volume} {185}},\ \bibinfo {pages} {3028--3040} (\bibinfo {year} {2014})}\BibitemShut {NoStop}%
\bibitem [{\citenamefont {Haga}\ and\ \citenamefont {Kawai}(2019)}]{Haga2019OnMethod}%
  \BibitemOpen
  \bibfield  {author} {\bibinfo {author} {\bibfnamefont {T.}~\bibnamefont {Haga}}\ and\ \bibinfo {author} {\bibfnamefont {S.}~\bibnamefont {Kawai}},\ }\bibfield  {title} {\enquote {\bibinfo {title} {{On a robust and accurate localized artificial diffusivity scheme for the high-order flux-reconstruction method}},}\ }\href {\doibase 10.1016/j.jcp.2018.09.052} {\bibfield  {journal} {\bibinfo  {journal} {Journal of Computational Physics}\ }\textbf {\bibinfo {volume} {376}},\ \bibinfo {pages} {534--563} (\bibinfo {year} {2019})}\BibitemShut {NoStop}%
\bibitem [{\citenamefont {Abe}, \citenamefont {Sun},\ and\ \citenamefont {Xiao}(2020)}]{Abe2020}%
  \BibitemOpen
  \bibfield  {author} {\bibinfo {author} {\bibfnamefont {Y.}~\bibnamefont {Abe}}, \bibinfo {author} {\bibfnamefont {Z.}~\bibnamefont {Sun}}, \ and\ \bibinfo {author} {\bibfnamefont {F.}~\bibnamefont {Xiao}},\ }\bibfield  {title} {\enquote {\bibinfo {title} {{Boundary variation diminishing algorithm for high-order local polynomial-based schemes}},}\ }\href {\doibase 10.1002/fld.4899} {\bibfield  {journal} {\bibinfo  {journal} {International Journal for Numerical Methods in Fluids}\ }\textbf {\bibinfo {volume} {93}},\ \bibinfo {pages} {892--907} (\bibinfo {year} {2020})}\BibitemShut {NoStop}%
\bibitem [{\citenamefont {Ii}\ and\ \citenamefont {Xiao}(2009)}]{Ii2009HighFormulation}%
  \BibitemOpen
  \bibfield  {author} {\bibinfo {author} {\bibfnamefont {S.}~\bibnamefont {Ii}}\ and\ \bibinfo {author} {\bibfnamefont {F.}~\bibnamefont {Xiao}},\ }\bibfield  {title} {\enquote {\bibinfo {title} {{High order multi-moment constrained finite volume method. Part I: Basic formulation}},}\ }\href {\doibase 10.1016/j.jcp.2009.02.009} {\bibfield  {journal} {\bibinfo  {journal} {Journal of Computational Physics}\ }\textbf {\bibinfo {volume} {228}},\ \bibinfo {pages} {3669--3707} (\bibinfo {year} {2009})}\BibitemShut {NoStop}%
\bibitem [{\citenamefont {Xiao}\ \emph {et~al.}(2013)\citenamefont {Xiao}, \citenamefont {Ii}, \citenamefont {Chen},\ and\ \citenamefont {Li}}]{Xiao2013ASchemes}%
  \BibitemOpen
  \bibfield  {author} {\bibinfo {author} {\bibfnamefont {F.}~\bibnamefont {Xiao}}, \bibinfo {author} {\bibfnamefont {S.}~\bibnamefont {Ii}}, \bibinfo {author} {\bibfnamefont {C.}~\bibnamefont {Chen}}, \ and\ \bibinfo {author} {\bibfnamefont {X.}~\bibnamefont {Li}},\ }\bibfield  {title} {\enquote {\bibinfo {title} {{A note on the general multi-moment constrained flux reconstruction formulation for high order schemes}},}\ }\href {\doibase 10.1016/j.apm.2012.10.050} {\bibfield  {journal} {\bibinfo  {journal} {Applied Mathematical Modelling}\ }\textbf {\bibinfo {volume} {37}},\ \bibinfo {pages} {5092--5108} (\bibinfo {year} {2013})}\BibitemShut {NoStop}%
\bibitem [{\citenamefont {Van~Leer}(1977)}]{VanLeer1977}%
  \BibitemOpen
  \bibfield  {author} {\bibinfo {author} {\bibfnamefont {B.}~\bibnamefont {Van~Leer}},\ }\bibfield  {title} {\enquote {\bibinfo {title} {{Towards the ultimate conservative difference scheme. IV. A new approach to numerical convection}},}\ }\href {\doibase 10.1016/0021-9991(77)90095-X} {\bibfield  {journal} {\bibinfo  {journal} {Journal of Computational Physics}\ }\textbf {\bibinfo {volume} {23}},\ \bibinfo {pages} {276--299} (\bibinfo {year} {1977})}\BibitemShut {NoStop}%
\bibitem [{\citenamefont {Liu}, \citenamefont {Osher},\ and\ \citenamefont {Chan}(1994)}]{Liu1994}%
  \BibitemOpen
  \bibfield  {author} {\bibinfo {author} {\bibfnamefont {X.-D.}\ \bibnamefont {Liu}}, \bibinfo {author} {\bibfnamefont {S.}~\bibnamefont {Osher}}, \ and\ \bibinfo {author} {\bibfnamefont {T.}~\bibnamefont {Chan}},\ }\bibfield  {title} {\enquote {\bibinfo {title} {{Weighted Essentially Non-oscillatory Schemes}},}\ }\href {\doibase 10.1002/fld.3889} {\bibfield  {journal} {\bibinfo  {journal} {Journal of Computational Physics}\ }\textbf {\bibinfo {volume} {115}},\ \bibinfo {pages} {200--212} (\bibinfo {year} {1994})}\BibitemShut {NoStop}%
\bibitem [{\citenamefont {Jiang}\ and\ \citenamefont {Shu}(1996)}]{Jiang1996}%
  \BibitemOpen
  \bibfield  {author} {\bibinfo {author} {\bibfnamefont {G.-S.}\ \bibnamefont {Jiang}}\ and\ \bibinfo {author} {\bibfnamefont {C.-W.}\ \bibnamefont {Shu}},\ }\bibfield  {title} {\enquote {\bibinfo {title} {{Efficient implementation of weighted ENO schemes}},}\ }\href@noop {} {\bibfield  {journal} {\bibinfo  {journal} {Journal of Computational Physics}\ }\textbf {\bibinfo {volume} {126}},\ \bibinfo {pages} {202--228} (\bibinfo {year} {1996})}\BibitemShut {NoStop}%
\bibitem [{\citenamefont {Harten}(1983)}]{Harten1983}%
  \BibitemOpen
  \bibfield  {author} {\bibinfo {author} {\bibfnamefont {A.}~\bibnamefont {Harten}},\ }\bibfield  {title} {\enquote {\bibinfo {title} {{High Resolution Schemes for Hyperbolic Conservation Laws}},}\ }\href {\doibase 10.1002/(SICI)1097-0363(19960830)23:4<309::AID-FLD410>3.0.CO;2-Z} {\bibfield  {journal} {\bibinfo  {journal} {International Journal for Numerical Methods in Fluids}\ }\textbf {\bibinfo {volume} {23}},\ \bibinfo {pages} {309--323} (\bibinfo {year} {1983})}\BibitemShut {NoStop}%
\bibitem [{\citenamefont {Venkatakrishnan}(1995)}]{Venkatakrishnan1995ConvergenceLimiters}%
  \BibitemOpen
  \bibfield  {author} {\bibinfo {author} {\bibfnamefont {V.}~\bibnamefont {Venkatakrishnan}},\ }\bibfield  {title} {\enquote {\bibinfo {title} {{Convergence to Steady State Solutions of the Euler Equations on Unstructured Grids with Limiters}},}\ }\href {\doibase 10.1006/jcph.1995.1084} {\bibfield  {journal} {\bibinfo  {journal} {Journal of Computational Physics}\ }\textbf {\bibinfo {volume} {118}},\ \bibinfo {pages} {120--130} (\bibinfo {year} {1995})}\BibitemShut {NoStop}%
\bibitem [{\citenamefont {Hubbard}(1999)}]{Hubbard1999MultidimensionalGrids}%
  \BibitemOpen
  \bibfield  {author} {\bibinfo {author} {\bibfnamefont {M.~E.}\ \bibnamefont {Hubbard}},\ }\bibfield  {title} {\enquote {\bibinfo {title} {{Multidimensional Slope Limiters for MUSCL-Type Finite Volume Schemes on Unstructured Grids}},}\ }\href {\doibase 10.1006/jcph.1999.6329} {\bibfield  {journal} {\bibinfo  {journal} {Journal of Computational Physics}\ }\textbf {\bibinfo {volume} {155}},\ \bibinfo {pages} {54--74} (\bibinfo {year} {1999})}\BibitemShut {NoStop}%
\bibitem [{\citenamefont {Darwish}\ and\ \citenamefont {Moukalled}(2003)}]{Darwish2003TVDGrids}%
  \BibitemOpen
  \bibfield  {author} {\bibinfo {author} {\bibfnamefont {M.~S.}\ \bibnamefont {Darwish}}\ and\ \bibinfo {author} {\bibfnamefont {F.}~\bibnamefont {Moukalled}},\ }\bibfield  {title} {\enquote {\bibinfo {title} {{TVD schemes for unstructured grids}},}\ }\href {\doibase 10.1016/S0017-9310(02)00330-7} {\bibfield  {journal} {\bibinfo  {journal} {International Journal of Heat and Mass Transfer}\ }\textbf {\bibinfo {volume} {46}},\ \bibinfo {pages} {599--611} (\bibinfo {year} {2003})}\BibitemShut {NoStop}%
\bibitem [{\citenamefont {Li}, \citenamefont {Liao},\ and\ \citenamefont {Qi}(2008)}]{Li2008AnSchemes}%
  \BibitemOpen
  \bibfield  {author} {\bibinfo {author} {\bibfnamefont {L.~x.}\ \bibnamefont {Li}}, \bibinfo {author} {\bibfnamefont {H.~s.}\ \bibnamefont {Liao}}, \ and\ \bibinfo {author} {\bibfnamefont {L.~j.}\ \bibnamefont {Qi}},\ }\bibfield  {title} {\enquote {\bibinfo {title} {{An improved r-factor algorithm for TVD schemes}},}\ }\href {\doibase 10.1016/j.ijheatmasstransfer.2007.04.051} {\bibfield  {journal} {\bibinfo  {journal} {International Journal of Heat and Mass Transfer}\ }\textbf {\bibinfo {volume} {51}},\ \bibinfo {pages} {610--617} (\bibinfo {year} {2008})}\BibitemShut {NoStop}%
\bibitem [{\citenamefont {Castro~D{\'{i}}az}\ \emph {et~al.}(2009)\citenamefont {Castro~D{\'{i}}az}, \citenamefont {Fern{\'{a}}ndez-Nieto}, \citenamefont {Ferreiro},\ and\ \citenamefont {Par{\'{e}}s}}]{CastroDiaz2009Two-dimensionalMeshes}%
  \BibitemOpen
  \bibfield  {author} {\bibinfo {author} {\bibfnamefont {M.~J.}\ \bibnamefont {Castro~D{\'{i}}az}}, \bibinfo {author} {\bibfnamefont {E.~D.}\ \bibnamefont {Fern{\'{a}}ndez-Nieto}}, \bibinfo {author} {\bibfnamefont {A.~M.}\ \bibnamefont {Ferreiro}}, \ and\ \bibinfo {author} {\bibfnamefont {C.}~\bibnamefont {Par{\'{e}}s}},\ }\bibfield  {title} {\enquote {\bibinfo {title} {{Two-dimensional sediment transport models in shallow water equations. A second order finite volume approach on unstructured meshes}},}\ }\href {\doibase 10.1016/j.cma.2009.03.001} {\bibfield  {journal} {\bibinfo  {journal} {Computer Methods in Applied Mechanics and Engineering}\ }\textbf {\bibinfo {volume} {198}},\ \bibinfo {pages} {2520--2538} (\bibinfo {year} {2009})}\BibitemShut {NoStop}%
\bibitem [{\citenamefont {Park}, \citenamefont {Yoon},\ and\ \citenamefont {Kim}(2010)}]{Park2010Multi-dimensionalGrids}%
  \BibitemOpen
  \bibfield  {author} {\bibinfo {author} {\bibfnamefont {J.~S.}\ \bibnamefont {Park}}, \bibinfo {author} {\bibfnamefont {S.~H.}\ \bibnamefont {Yoon}}, \ and\ \bibinfo {author} {\bibfnamefont {C.}~\bibnamefont {Kim}},\ }\bibfield  {title} {\enquote {\bibinfo {title} {{Multi-dimensional limiting process for hyperbolic conservation laws on unstructured grids}},}\ }\href {\doibase 10.1016/j.jcp.2009.10.011} {\bibfield  {journal} {\bibinfo  {journal} {Journal of Computational Physics}\ }\textbf {\bibinfo {volume} {229}},\ \bibinfo {pages} {788--812} (\bibinfo {year} {2010})}\BibitemShut {NoStop}%
\bibitem [{\citenamefont {Park}\ and\ \citenamefont {Kim}(2012)}]{Park2012Multi-dimensionalGrids}%
  \BibitemOpen
  \bibfield  {author} {\bibinfo {author} {\bibfnamefont {J.~S.}\ \bibnamefont {Park}}\ and\ \bibinfo {author} {\bibfnamefont {C.}~\bibnamefont {Kim}},\ }\bibfield  {title} {\enquote {\bibinfo {title} {{Multi-dimensional limiting process for finite volume methods on unstructured grids}},}\ }\href {\doibase 10.1016/j.compfluid.2012.04.015} {\bibfield  {journal} {\bibinfo  {journal} {Computers {\&} Fluids}\ }\textbf {\bibinfo {volume} {65}},\ \bibinfo {pages} {8--24} (\bibinfo {year} {2012})}\BibitemShut {NoStop}%
\bibitem [{\citenamefont {Zhang}\ \emph {et~al.}(2016)\citenamefont {Zhang}, \citenamefont {Jiang}, \citenamefont {Cheng},\ and\ \citenamefont {Liang}}]{Zhang2016AMeshes}%
  \BibitemOpen
  \bibfield  {author} {\bibinfo {author} {\bibfnamefont {D.}~\bibnamefont {Zhang}}, \bibinfo {author} {\bibfnamefont {C.}~\bibnamefont {Jiang}}, \bibinfo {author} {\bibfnamefont {L.}~\bibnamefont {Cheng}}, \ and\ \bibinfo {author} {\bibfnamefont {D.}~\bibnamefont {Liang}},\ }\bibfield  {title} {\enquote {\bibinfo {title} {{A refined r-factor algorithm for TVD schemes on arbitrary unstructured meshes}},}\ }\href {\doibase 10.1002/fld.4073} {\bibfield  {journal} {\bibinfo  {journal} {International Journal for Numerical Methods in Fluids}\ }\textbf {\bibinfo {volume} {80}},\ \bibinfo {pages} {105--139} (\bibinfo {year} {2016})}\BibitemShut {NoStop}%
\bibitem [{\citenamefont {Li}\ \emph {et~al.}(2021)\citenamefont {Li}, \citenamefont {Zhang}, \citenamefont {Zhai}, \citenamefont {Sun},\ and\ \citenamefont {Liao}}]{Li2021AnGrids}%
  \BibitemOpen
  \bibfield  {author} {\bibinfo {author} {\bibfnamefont {J.}~\bibnamefont {Li}}, \bibinfo {author} {\bibfnamefont {Q.}~\bibnamefont {Zhang}}, \bibinfo {author} {\bibfnamefont {Z.~Q.}\ \bibnamefont {Zhai}}, \bibinfo {author} {\bibfnamefont {X.}~\bibnamefont {Sun}}, \ and\ \bibinfo {author} {\bibfnamefont {S.}~\bibnamefont {Liao}},\ }\bibfield  {title} {\enquote {\bibinfo {title} {{An improved r-factor algorithm for total variational diminishing (TVD) schemes on two-dimension non-uniform unstructured grids}},}\ }\href {\doibase 10.1002/fld.4937} {\bibfield  {journal} {\bibinfo  {journal} {International Journal for Numerical Methods in Fluids}\ }\textbf {\bibinfo {volume} {93}},\ \bibinfo {pages} {1446--1467} (\bibinfo {year} {2021})}\BibitemShut {NoStop}%
\bibitem [{\citenamefont {Hu}\ and\ \citenamefont {Shu}(1999)}]{Hu1999WeightedMeshes}%
  \BibitemOpen
  \bibfield  {author} {\bibinfo {author} {\bibfnamefont {C.}~\bibnamefont {Hu}}\ and\ \bibinfo {author} {\bibfnamefont {C.-W.}\ \bibnamefont {Shu}},\ }\bibfield  {title} {\enquote {\bibinfo {title} {{Weighted Essentially Non-oscillatory Schemes on Triangular Meshes}},}\ }\href {\doibase 10.1006/jcph.1998.6165} {\bibfield  {journal} {\bibinfo  {journal} {Journal of Computational Physics}\ }\textbf {\bibinfo {volume} {150}},\ \bibinfo {pages} {97--127} (\bibinfo {year} {1999})}\BibitemShut {NoStop}%
\bibitem [{\citenamefont {Dumbser}\ \emph {et~al.}(2007)\citenamefont {Dumbser}, \citenamefont {K{\"{a}}ser}, \citenamefont {Titarev},\ and\ \citenamefont {Toro}}]{Dumbser2007Quadrature-freeSystems}%
  \BibitemOpen
  \bibfield  {author} {\bibinfo {author} {\bibfnamefont {M.}~\bibnamefont {Dumbser}}, \bibinfo {author} {\bibfnamefont {M.}~\bibnamefont {K{\"{a}}ser}}, \bibinfo {author} {\bibfnamefont {V.~A.}\ \bibnamefont {Titarev}}, \ and\ \bibinfo {author} {\bibfnamefont {E.~F.}\ \bibnamefont {Toro}},\ }\bibfield  {title} {\enquote {\bibinfo {title} {{Quadrature-free non-oscillatory finite volume schemes on unstructured meshes for nonlinear hyperbolic systems}},}\ }\href {\doibase 10.1016/j.jcp.2007.04.004} {\bibfield  {journal} {\bibinfo  {journal} {Journal of Computational Physics}\ }\textbf {\bibinfo {volume} {226}},\ \bibinfo {pages} {204--243} (\bibinfo {year} {2007})}\BibitemShut {NoStop}%
\bibitem [{\citenamefont {Dumbser}\ and\ \citenamefont {K{\"{a}}ser}(2007)}]{Dumbser2007ArbitrarySystems}%
  \BibitemOpen
  \bibfield  {author} {\bibinfo {author} {\bibfnamefont {M.}~\bibnamefont {Dumbser}}\ and\ \bibinfo {author} {\bibfnamefont {M.}~\bibnamefont {K{\"{a}}ser}},\ }\bibfield  {title} {\enquote {\bibinfo {title} {{Arbitrary high order non-oscillatory finite volume schemes on unstructured meshes for linear hyperbolic systems}},}\ }\href {\doibase 10.1016/j.jcp.2006.06.043} {\bibfield  {journal} {\bibinfo  {journal} {Journal of Computational Physics}\ }\textbf {\bibinfo {volume} {221}},\ \bibinfo {pages} {693--723} (\bibinfo {year} {2007})}\BibitemShut {NoStop}%
\bibitem [{\citenamefont {Zhang}\ and\ \citenamefont {Shu}(2009)}]{Zhang2009ThirdMeshes}%
  \BibitemOpen
  \bibfield  {author} {\bibinfo {author} {\bibfnamefont {Y.~T.}\ \bibnamefont {Zhang}}\ and\ \bibinfo {author} {\bibfnamefont {C.~W.}\ \bibnamefont {Shu}},\ }\bibfield  {title} {\enquote {\bibinfo {title} {{Third order WENO scheme on three dimensional tetrahedral meshes}},}\ }\href@noop {} {\bibfield  {journal} {\bibinfo  {journal} {Communications in Computational Physics}\ }\textbf {\bibinfo {volume} {5}},\ \bibinfo {pages} {836--848} (\bibinfo {year} {2009})}\BibitemShut {NoStop}%
\bibitem [{\citenamefont {Li}\ and\ \citenamefont {Ren}(2012)}]{Li2012HighorderGrids}%
  \BibitemOpen
  \bibfield  {author} {\bibinfo {author} {\bibfnamefont {W.}~\bibnamefont {Li}}\ and\ \bibinfo {author} {\bibfnamefont {Y.-X.}\ \bibnamefont {Ren}},\ }\bibfield  {title} {\enquote {\bibinfo {title} {{High‐order k ‐exact WENO finite volume schemes for solving gas dynamic Euler equations on unstructured grids}},}\ }\href {\doibase 10.1002/fld.2710} {\bibfield  {journal} {\bibinfo  {journal} {International Journal for Numerical Methods in Fluids}\ }\textbf {\bibinfo {volume} {70}},\ \bibinfo {pages} {742--763} (\bibinfo {year} {2012})}\BibitemShut {NoStop}%
\bibitem [{\citenamefont {Liu}\ and\ \citenamefont {Zhang}(2013)}]{Liu2013ASchemes}%
  \BibitemOpen
  \bibfield  {author} {\bibinfo {author} {\bibfnamefont {Y.}~\bibnamefont {Liu}}\ and\ \bibinfo {author} {\bibfnamefont {Y.~T.}\ \bibnamefont {Zhang}},\ }\bibfield  {title} {\enquote {\bibinfo {title} {{A robust reconstruction for unstructured WENO schemes}},}\ }\href {\doibase 10.1007/s10915-012-9598-3} {\bibfield  {journal} {\bibinfo  {journal} {Journal of Scientific Computing}\ }\textbf {\bibinfo {volume} {54}},\ \bibinfo {pages} {603--621} (\bibinfo {year} {2013})}\BibitemShut {NoStop}%
\bibitem [{\citenamefont {Christlieb}\ \emph {et~al.}(2015)\citenamefont {Christlieb}, \citenamefont {Liu}, \citenamefont {Tang},\ and\ \citenamefont {Xu}}]{Christlieb2015HighMeshes}%
  \BibitemOpen
  \bibfield  {author} {\bibinfo {author} {\bibfnamefont {A.~J.}\ \bibnamefont {Christlieb}}, \bibinfo {author} {\bibfnamefont {Y.}~\bibnamefont {Liu}}, \bibinfo {author} {\bibfnamefont {Q.}~\bibnamefont {Tang}}, \ and\ \bibinfo {author} {\bibfnamefont {Z.}~\bibnamefont {Xu}},\ }\bibfield  {title} {\enquote {\bibinfo {title} {{High order parametrized maximum-principle-preserving and positivity-preserving WENO schemes on unstructured meshes}},}\ }\href {\doibase 10.1016/j.jcp.2014.10.029} {\bibfield  {journal} {\bibinfo  {journal} {Journal of Computational Physics}\ }\textbf {\bibinfo {volume} {281}},\ \bibinfo {pages} {334--351} (\bibinfo {year} {2015})}\BibitemShut {NoStop}%
\bibitem [{\citenamefont {Ji}, \citenamefont {Liang},\ and\ \citenamefont {Fu}(2022)}]{Ji2022AMeshes}%
  \BibitemOpen
  \bibfield  {author} {\bibinfo {author} {\bibfnamefont {Z.}~\bibnamefont {Ji}}, \bibinfo {author} {\bibfnamefont {T.}~\bibnamefont {Liang}}, \ and\ \bibinfo {author} {\bibfnamefont {L.}~\bibnamefont {Fu}},\ }\bibfield  {title} {\enquote {\bibinfo {title} {{A Class of New High-order Finite-Volume TENO Schemes for Hyperbolic Conservation Laws with Unstructured Meshes}},}\ }\href {\doibase 10.1007/s10915-022-01925-5} {\bibfield  {journal} {\bibinfo  {journal} {Journal of Scientific Computing}\ }\textbf {\bibinfo {volume} {92}},\ \bibinfo {pages} {1--39} (\bibinfo {year} {2022})}\BibitemShut {NoStop}%
\bibitem [{\citenamefont {Ji}, \citenamefont {Liang},\ and\ \citenamefont {Fu}(2023)}]{Ji2023High-OrderMeshes}%
  \BibitemOpen
  \bibfield  {author} {\bibinfo {author} {\bibfnamefont {Z.}~\bibnamefont {Ji}}, \bibinfo {author} {\bibfnamefont {T.}~\bibnamefont {Liang}}, \ and\ \bibinfo {author} {\bibfnamefont {L.}~\bibnamefont {Fu}},\ }\bibfield  {title} {\enquote {\bibinfo {title} {{High-Order Finite-Volume TENO Schemes with Dual ENO-Like Stencil Selection for Unstructured Meshes}},}\ }\href {\doibase 10.1007/s10915-023-02199-1} {\bibfield  {journal} {\bibinfo  {journal} {Journal of Scientific Computing}\ }\textbf {\bibinfo {volume} {95}},\ \bibinfo {pages} {76} (\bibinfo {year} {2023})}\BibitemShut {NoStop}%
\bibitem [{\citenamefont {Chiapolino}, \citenamefont {Saurel},\ and\ \citenamefont {Nkonga}(2017)}]{Chiapolino2017b}%
  \BibitemOpen
  \bibfield  {author} {\bibinfo {author} {\bibfnamefont {A.}~\bibnamefont {Chiapolino}}, \bibinfo {author} {\bibfnamefont {R.}~\bibnamefont {Saurel}}, \ and\ \bibinfo {author} {\bibfnamefont {B.}~\bibnamefont {Nkonga}},\ }\bibfield  {title} {\enquote {\bibinfo {title} {{Sharpening diffuse interfaces with compressible fluids on unstructured meshes}},}\ }\href {\doibase 10.1016/j.jcp.2017.03.042} {\bibfield  {journal} {\bibinfo  {journal} {Journal of Computational Physics}\ }\textbf {\bibinfo {volume} {340}},\ \bibinfo {pages} {389--417} (\bibinfo {year} {2017})}\BibitemShut {NoStop}%
\bibitem [{\citenamefont {Tsoutsanis}\ \emph {et~al.}(2021)\citenamefont {Tsoutsanis}, \citenamefont {Adebayo}, \citenamefont {Merino}, \citenamefont {Arjona},\ and\ \citenamefont {Skote}}]{Tsoutsanis2021CWENOMeshes}%
  \BibitemOpen
  \bibfield  {author} {\bibinfo {author} {\bibfnamefont {P.}~\bibnamefont {Tsoutsanis}}, \bibinfo {author} {\bibfnamefont {E.~M.}\ \bibnamefont {Adebayo}}, \bibinfo {author} {\bibfnamefont {A.~C.}\ \bibnamefont {Merino}}, \bibinfo {author} {\bibfnamefont {A.~P.}\ \bibnamefont {Arjona}}, \ and\ \bibinfo {author} {\bibfnamefont {M.}~\bibnamefont {Skote}},\ }\bibfield  {title} {\enquote {\bibinfo {title} {{CWENO Finite-Volume Interface Capturing Schemes for Multicomponent Flows Using Unstructured Meshes}},}\ }\href {\doibase 10.1007/s10915-021-01673-y} {\bibfield  {journal} {\bibinfo  {journal} {Journal of Scientific Computing}\ }\textbf {\bibinfo {volume} {89}},\ \bibinfo {pages} {1--27} (\bibinfo {year} {2021})}\BibitemShut {NoStop}%
\bibitem [{\citenamefont {Maltsev}, \citenamefont {Skote},\ and\ \citenamefont {Tsoutsanis}(2024)}]{Maltsev2024High-orderMeshes}%
  \BibitemOpen
  \bibfield  {author} {\bibinfo {author} {\bibfnamefont {V.}~\bibnamefont {Maltsev}}, \bibinfo {author} {\bibfnamefont {M.}~\bibnamefont {Skote}}, \ and\ \bibinfo {author} {\bibfnamefont {P.}~\bibnamefont {Tsoutsanis}},\ }\bibfield  {title} {\enquote {\bibinfo {title} {{High-order hybrid DG-FV framework for compressible multi-fluid problems on unstructured meshes}},}\ }\href {\doibase 10.1016/j.jcp.2024.112819} {\bibfield  {journal} {\bibinfo  {journal} {Journal of Computational Physics}\ }\textbf {\bibinfo {volume} {502}},\ \bibinfo {pages} {112819} (\bibinfo {year} {2024})}\BibitemShut {NoStop}%
\bibitem [{\citenamefont {Sun}, \citenamefont {Inaba},\ and\ \citenamefont {Xiao}(2016)}]{Sun2016}%
  \BibitemOpen
  \bibfield  {author} {\bibinfo {author} {\bibfnamefont {Z.}~\bibnamefont {Sun}}, \bibinfo {author} {\bibfnamefont {S.}~\bibnamefont {Inaba}}, \ and\ \bibinfo {author} {\bibfnamefont {F.}~\bibnamefont {Xiao}},\ }\bibfield  {title} {\enquote {\bibinfo {title} {{Boundary Variation Diminishing (BVD) reconstruction: A new approach to improve Godunov schemes}},}\ }\href {\doibase 10.1016/j.jcp.2016.06.051} {\bibfield  {journal} {\bibinfo  {journal} {Journal of Computational Physics}\ }\textbf {\bibinfo {volume} {322}},\ \bibinfo {pages} {309--325} (\bibinfo {year} {2016})}\BibitemShut {NoStop}%
\bibitem [{\citenamefont {Xie}\ \emph {et~al.}(2017)\citenamefont {Xie}, \citenamefont {Deng}, \citenamefont {Sun},\ and\ \citenamefont {Xiao}}]{Xie2017a}%
  \BibitemOpen
  \bibfield  {author} {\bibinfo {author} {\bibfnamefont {B.}~\bibnamefont {Xie}}, \bibinfo {author} {\bibfnamefont {X.}~\bibnamefont {Deng}}, \bibinfo {author} {\bibfnamefont {Z.}~\bibnamefont {Sun}}, \ and\ \bibinfo {author} {\bibfnamefont {F.}~\bibnamefont {Xiao}},\ }\bibfield  {title} {\enquote {\bibinfo {title} {{A hybrid pressure–density-based Mach uniform algorithm for 2D Euler equations on unstructured grids by using multi-moment finite volume method}},}\ }\href {\doibase 10.1016/j.jcp.2017.01.043} {\bibfield  {journal} {\bibinfo  {journal} {Journal of Computational Physics}\ }\textbf {\bibinfo {volume} {335}},\ \bibinfo {pages} {637--663} (\bibinfo {year} {2017})}\BibitemShut {NoStop}%
\bibitem [{\citenamefont {Deng}, \citenamefont {Xie},\ and\ \citenamefont {Xiao}(2017)}]{Deng2017}%
  \BibitemOpen
  \bibfield  {author} {\bibinfo {author} {\bibfnamefont {X.}~\bibnamefont {Deng}}, \bibinfo {author} {\bibfnamefont {B.}~\bibnamefont {Xie}}, \ and\ \bibinfo {author} {\bibfnamefont {F.}~\bibnamefont {Xiao}},\ }\bibfield  {title} {\enquote {\bibinfo {title} {{A finite volume multi-moment method with boundary variation diminishing principle for Euler equation on three-dimensional hybrid unstructured grids}},}\ }\href {\doibase 10.1016/j.compfluid.2017.05.007} {\bibfield  {journal} {\bibinfo  {journal} {Computers and Fluids}\ }\textbf {\bibinfo {volume} {153}},\ \bibinfo {pages} {85--101} (\bibinfo {year} {2017})}\BibitemShut {NoStop}%
\bibitem [{\citenamefont {Deng}\ \emph {et~al.}(2018{\natexlab{a}})\citenamefont {Deng}, \citenamefont {Xie}, \citenamefont {Loub{\`{e}}re}, \citenamefont {Shimizu},\ and\ \citenamefont {Xiao}}]{Deng2018}%
  \BibitemOpen
  \bibfield  {author} {\bibinfo {author} {\bibfnamefont {X.}~\bibnamefont {Deng}}, \bibinfo {author} {\bibfnamefont {B.}~\bibnamefont {Xie}}, \bibinfo {author} {\bibfnamefont {R.}~\bibnamefont {Loub{\`{e}}re}}, \bibinfo {author} {\bibfnamefont {Y.}~\bibnamefont {Shimizu}}, \ and\ \bibinfo {author} {\bibfnamefont {F.}~\bibnamefont {Xiao}},\ }\bibfield  {title} {\enquote {\bibinfo {title} {{Limiter-free discontinuity-capturing scheme for compressible gas dynamics with reactive fronts}},}\ }\href {\doibase 10.1016/j.compfluid.2018.05.015} {\bibfield  {journal} {\bibinfo  {journal} {Computers and Fluids}\ }\textbf {\bibinfo {volume} {171}},\ \bibinfo {pages} {1--14} (\bibinfo {year} {2018}{\natexlab{a}})}\BibitemShut {NoStop}%
\bibitem [{\citenamefont {Deng}\ \emph {et~al.}(2018{\natexlab{b}})\citenamefont {Deng}, \citenamefont {Inaba}, \citenamefont {Xie}, \citenamefont {Shyue},\ and\ \citenamefont {Xiao}}]{Deng2018a}%
  \BibitemOpen
  \bibfield  {author} {\bibinfo {author} {\bibfnamefont {X.}~\bibnamefont {Deng}}, \bibinfo {author} {\bibfnamefont {S.}~\bibnamefont {Inaba}}, \bibinfo {author} {\bibfnamefont {B.}~\bibnamefont {Xie}}, \bibinfo {author} {\bibfnamefont {K.~M.}\ \bibnamefont {Shyue}}, \ and\ \bibinfo {author} {\bibfnamefont {F.}~\bibnamefont {Xiao}},\ }\bibfield  {title} {\enquote {\bibinfo {title} {{High fidelity discontinuity-resolving reconstruction for compressible multiphase flows with moving interfaces}},}\ }\href {\doibase 10.1016/j.jcp.2018.03.036} {\bibfield  {journal} {\bibinfo  {journal} {Journal of Computational Physics}\ }\textbf {\bibinfo {volume} {371}},\ \bibinfo {pages} {945--966} (\bibinfo {year} {2018}{\natexlab{b}})}\BibitemShut {NoStop}%
\bibitem [{\citenamefont {Deng}, \citenamefont {Shimizu},\ and\ \citenamefont {Xiao}(2019)}]{Deng2019}%
  \BibitemOpen
  \bibfield  {author} {\bibinfo {author} {\bibfnamefont {X.}~\bibnamefont {Deng}}, \bibinfo {author} {\bibfnamefont {Y.}~\bibnamefont {Shimizu}}, \ and\ \bibinfo {author} {\bibfnamefont {F.}~\bibnamefont {Xiao}},\ }\bibfield  {title} {\enquote {\bibinfo {title} {{A fifth-order shock capturing scheme with two-stage boundary variation diminishing algorithm}},}\ }\href {\doibase 10.1016/j.jcp.2019.02.024} {\bibfield  {journal} {\bibinfo  {journal} {Journal of Computational Physics}\ }\textbf {\bibinfo {volume} {386}},\ \bibinfo {pages} {323--349} (\bibinfo {year} {2019})}\BibitemShut {NoStop}%
\bibitem [{\citenamefont {Xie}\ \emph {et~al.}(2019)\citenamefont {Xie}, \citenamefont {Deng}, \citenamefont {Liao},\ and\ \citenamefont {Xiao}}]{Xie2019High-orderFlow}%
  \BibitemOpen
  \bibfield  {author} {\bibinfo {author} {\bibfnamefont {B.}~\bibnamefont {Xie}}, \bibinfo {author} {\bibfnamefont {X.}~\bibnamefont {Deng}}, \bibinfo {author} {\bibfnamefont {S.~J.}\ \bibnamefont {Liao}}, \ and\ \bibinfo {author} {\bibfnamefont {F.}~\bibnamefont {Xiao}},\ }\bibfield  {title} {\enquote {\bibinfo {title} {{High-order multi-moment finite volume method with smoothness adaptive fitting reconstruction for compressible viscous flow}},}\ }\href {\doibase 10.1016/j.jcp.2019.06.002} {\bibfield  {journal} {\bibinfo  {journal} {Journal of Computational Physics}\ }\textbf {\bibinfo {volume} {394}},\ \bibinfo {pages} {559--593} (\bibinfo {year} {2019})}\BibitemShut {NoStop}%
\bibitem [{\citenamefont {Deng}\ \emph {et~al.}(2020)\citenamefont {Deng}, \citenamefont {Shimizu}, \citenamefont {Xie},\ and\ \citenamefont {Xiao}}]{Deng2020}%
  \BibitemOpen
  \bibfield  {author} {\bibinfo {author} {\bibfnamefont {X.}~\bibnamefont {Deng}}, \bibinfo {author} {\bibfnamefont {Y.}~\bibnamefont {Shimizu}}, \bibinfo {author} {\bibfnamefont {B.}~\bibnamefont {Xie}}, \ and\ \bibinfo {author} {\bibfnamefont {F.}~\bibnamefont {Xiao}},\ }\bibfield  {title} {\enquote {\bibinfo {title} {{Constructing higher order discontinuity-capturing schemes with upwind-biased interpolations and boundary variation diminishing algorithm}},}\ }\href {\doibase 10.1016/j.compfluid.2020.104433} {\bibfield  {journal} {\bibinfo  {journal} {Computers and Fluids}\ }\textbf {\bibinfo {volume} {200}},\ \bibinfo {pages} {104433} (\bibinfo {year} {2020})}\BibitemShut {NoStop}%
\bibitem [{\citenamefont {Jiang}\ \emph {et~al.}(2021)\citenamefont {Jiang}, \citenamefont {Deng}, \citenamefont {Xiao}, \citenamefont {Yan}, \citenamefont {Yu},\ and\ \citenamefont {Lou}}]{Jiang2021HybridFlows}%
  \BibitemOpen
  \bibfield  {author} {\bibinfo {author} {\bibfnamefont {Z.-H.}\ \bibnamefont {Jiang}}, \bibinfo {author} {\bibfnamefont {X.}~\bibnamefont {Deng}}, \bibinfo {author} {\bibfnamefont {F.}~\bibnamefont {Xiao}}, \bibinfo {author} {\bibfnamefont {C.}~\bibnamefont {Yan}}, \bibinfo {author} {\bibfnamefont {J.}~\bibnamefont {Yu}}, \ and\ \bibinfo {author} {\bibfnamefont {S.}~\bibnamefont {Lou}},\ }\bibfield  {title} {\enquote {\bibinfo {title} {{Hybrid Discontinuous Galerkin/Finite Volume Method with Subcell Resolution for Shocked Flows}},}\ }\href {\doibase 10.2514/1.j059763} {\bibfield  {journal} {\bibinfo  {journal} {AIAA Journal}\ }\textbf {\bibinfo {volume} {59}},\ \bibinfo {pages} {2027--2044} (\bibinfo {year} {2021})}\BibitemShut {NoStop}%
\bibitem [{\citenamefont {Cheng}\ \emph {et~al.}(2021)\citenamefont {Cheng}, \citenamefont {Deng}, \citenamefont {Xie}, \citenamefont {Jiang},\ and\ \citenamefont {Xiao}}]{Cheng2021}%
  \BibitemOpen
  \bibfield  {author} {\bibinfo {author} {\bibfnamefont {L.}~\bibnamefont {Cheng}}, \bibinfo {author} {\bibfnamefont {X.}~\bibnamefont {Deng}}, \bibinfo {author} {\bibfnamefont {B.}~\bibnamefont {Xie}}, \bibinfo {author} {\bibfnamefont {Y.}~\bibnamefont {Jiang}}, \ and\ \bibinfo {author} {\bibfnamefont {F.}~\bibnamefont {Xiao}},\ }\bibfield  {title} {\enquote {\bibinfo {title} {{Low-dissipation BVD schemes for single and multi-phase compressible flows on unstructured grids}},}\ }\href {\doibase 10.1016/j.jcp.2020.110088} {\bibfield  {journal} {\bibinfo  {journal} {Journal of Computational Physics}\ }\textbf {\bibinfo {volume} {428}},\ \bibinfo {pages} {110088} (\bibinfo {year} {2021})}\BibitemShut {NoStop}%
\bibitem [{\citenamefont {Wakimura}, \citenamefont {Takagi},\ and\ \citenamefont {Xiao}(2022)}]{Wakimura2021a}%
  \BibitemOpen
  \bibfield  {author} {\bibinfo {author} {\bibfnamefont {H.}~\bibnamefont {Wakimura}}, \bibinfo {author} {\bibfnamefont {S.}~\bibnamefont {Takagi}}, \ and\ \bibinfo {author} {\bibfnamefont {F.}~\bibnamefont {Xiao}},\ }\bibfield  {title} {\enquote {\bibinfo {title} {{Symmetry-preserving enforcement of low-dissipation method based on boundary variation diminishing principle}},}\ }\href {\doibase 10.1016/j.compfluid.2021.105227} {\bibfield  {journal} {\bibinfo  {journal} {Computers {\&} Fluids}\ }\textbf {\bibinfo {volume} {233}},\ \bibinfo {pages} {105227} (\bibinfo {year} {2022})}\BibitemShut {NoStop}%
\bibitem [{\citenamefont {Majima}\ \emph {et~al.}(2023)\citenamefont {Majima}, \citenamefont {Wakimura}, \citenamefont {Aoki},\ and\ \citenamefont {Xiao}}]{Majima2023AFlows}%
  \BibitemOpen
  \bibfield  {author} {\bibinfo {author} {\bibfnamefont {Y.}~\bibnamefont {Majima}}, \bibinfo {author} {\bibfnamefont {H.}~\bibnamefont {Wakimura}}, \bibinfo {author} {\bibfnamefont {T.}~\bibnamefont {Aoki}}, \ and\ \bibinfo {author} {\bibfnamefont {F.}~\bibnamefont {Xiao}},\ }\bibfield  {title} {\enquote {\bibinfo {title} {{A new high-fidelity Total Variation Diminishing scheme based on the Boundary Variation Diminishing principle for compressible flows}},}\ }\href {\doibase 10.1016/j.compfluid.2023.106070} {\bibfield  {journal} {\bibinfo  {journal} {Computers and Fluids}\ }\textbf {\bibinfo {volume} {266}},\ \bibinfo {pages} {106070} (\bibinfo {year} {2023})}\BibitemShut {NoStop}%
\bibitem [{\citenamefont {Li}, \citenamefont {Liu},\ and\ \citenamefont {Li}(2023)}]{Li2023AFlows}%
  \BibitemOpen
  \bibfield  {author} {\bibinfo {author} {\bibfnamefont {J.}~\bibnamefont {Li}}, \bibinfo {author} {\bibfnamefont {C.}~\bibnamefont {Liu}}, \ and\ \bibinfo {author} {\bibfnamefont {Z.}~\bibnamefont {Li}},\ }\bibfield  {title} {\enquote {\bibinfo {title} {{A revised WENO‐THINC scheme for the general structured mesh and applications in the direct numerical simulation of compressible turbulent flows}},}\ }\href {\doibase 10.1002/fld.5196} {\bibfield  {journal} {\bibinfo  {journal} {International Journal for Numerical Methods in Fluids}\ }\textbf {\bibinfo {volume} {95}},\ \bibinfo {pages} {1372--1403} (\bibinfo {year} {2023})}\BibitemShut {NoStop}%
\bibitem [{\citenamefont {Wakimura}\ \emph {et~al.}(2024)\citenamefont {Wakimura}, \citenamefont {Li}, \citenamefont {Shyue}, \citenamefont {Aoki},\ and\ \citenamefont {Xiao}}]{Wakimura2024High-resolutionEvaporation}%
  \BibitemOpen
  \bibfield  {author} {\bibinfo {author} {\bibfnamefont {H.}~\bibnamefont {Wakimura}}, \bibinfo {author} {\bibfnamefont {T.}~\bibnamefont {Li}}, \bibinfo {author} {\bibfnamefont {K.~M.}\ \bibnamefont {Shyue}}, \bibinfo {author} {\bibfnamefont {T.}~\bibnamefont {Aoki}}, \ and\ \bibinfo {author} {\bibfnamefont {F.}~\bibnamefont {Xiao}},\ }\bibfield  {title} {\enquote {\bibinfo {title} {{High-resolution boundary variation diminishing scheme for two-phase compressible flow with cavitation and evaporation}},}\ }\href {\doibase 10.1016/j.jcp.2024.113164} {\bibfield  {journal} {\bibinfo  {journal} {Journal of Computational Physics}\ }\textbf {\bibinfo {volume} {513}},\ \bibinfo {pages} {113164} (\bibinfo {year} {2024})}\BibitemShut {NoStop}%
\bibitem [{\citenamefont {Xiao}, \citenamefont {Honma},\ and\ \citenamefont {Kono}(2005)}]{Xiao2005}%
  \BibitemOpen
  \bibfield  {author} {\bibinfo {author} {\bibfnamefont {F.}~\bibnamefont {Xiao}}, \bibinfo {author} {\bibfnamefont {Y.}~\bibnamefont {Honma}}, \ and\ \bibinfo {author} {\bibfnamefont {T.}~\bibnamefont {Kono}},\ }\bibfield  {title} {\enquote {\bibinfo {title} {{A simple algebraic interface capturing scheme using hyperbolic tangent function}},}\ }\href {\doibase 10.1002/fld.975} {\bibfield  {journal} {\bibinfo  {journal} {International Journal for Numerical Methods in Fluids}\ }\textbf {\bibinfo {volume} {48}},\ \bibinfo {pages} {1023--1040} (\bibinfo {year} {2005})}\BibitemShut {NoStop}%
\bibitem [{\citenamefont {Xiao}, \citenamefont {Ii},\ and\ \citenamefont {Chen}(2011)}]{Xiao2011}%
  \BibitemOpen
  \bibfield  {author} {\bibinfo {author} {\bibfnamefont {F.}~\bibnamefont {Xiao}}, \bibinfo {author} {\bibfnamefont {S.}~\bibnamefont {Ii}}, \ and\ \bibinfo {author} {\bibfnamefont {C.}~\bibnamefont {Chen}},\ }\bibfield  {title} {\enquote {\bibinfo {title} {{Revisit to the THINC scheme: A simple algebraic VOF algorithm}},}\ }\href {\doibase 10.1016/j.jcp.2011.06.012} {\bibfield  {journal} {\bibinfo  {journal} {Journal of Computational Physics}\ }\textbf {\bibinfo {volume} {230}},\ \bibinfo {pages} {7086--7092} (\bibinfo {year} {2011})}\BibitemShut {NoStop}%
\bibitem [{\citenamefont {Xie}\ and\ \citenamefont {Xiao}(2017)}]{Xie2017}%
  \BibitemOpen
  \bibfield  {author} {\bibinfo {author} {\bibfnamefont {B.}~\bibnamefont {Xie}}\ and\ \bibinfo {author} {\bibfnamefont {F.}~\bibnamefont {Xiao}},\ }\bibfield  {title} {\enquote {\bibinfo {title} {{Toward efficient and accurate interface capturing on arbitrary hybrid unstructured grids: The THINC method with quadratic surface representation and Gaussian quadrature}},}\ }\href {\doibase 10.1016/j.jcp.2017.08.028} {\bibfield  {journal} {\bibinfo  {journal} {Journal of Computational Physics}\ }\textbf {\bibinfo {volume} {349}},\ \bibinfo {pages} {415--440} (\bibinfo {year} {2017})}\BibitemShut {NoStop}%
\bibitem [{\citenamefont {Saurel}, \citenamefont {Petitpas},\ and\ \citenamefont {Abgrall}(2008)}]{Saurel2008}%
  \BibitemOpen
  \bibfield  {author} {\bibinfo {author} {\bibfnamefont {R.}~\bibnamefont {Saurel}}, \bibinfo {author} {\bibfnamefont {F.}~\bibnamefont {Petitpas}}, \ and\ \bibinfo {author} {\bibfnamefont {R.}~\bibnamefont {Abgrall}},\ }\bibfield  {title} {\enquote {\bibinfo {title} {{Modelling phase transition in metastable liquids: application to cavitating and flashing flows}},}\ }\href {\doibase 10.1017/S0022112008002061} {\bibfield  {journal} {\bibinfo  {journal} {Journal of Fluid Mechanics}\ }\textbf {\bibinfo {volume} {607}},\ \bibinfo {pages} {313--350} (\bibinfo {year} {2008})}\BibitemShut {NoStop}%
\bibitem [{\citenamefont {LeVeque}(1997)}]{LeVeque1997}%
  \BibitemOpen
  \bibfield  {author} {\bibinfo {author} {\bibfnamefont {R.~J.}\ \bibnamefont {LeVeque}},\ }\bibfield  {title} {\enquote {\bibinfo {title} {{Wave propagation algorithms for multidimensional hyperbolic systems}},}\ }\href {\doibase 10.1006/jcph.1996.5603} {\bibfield  {journal} {\bibinfo  {journal} {Journal of Computational Physics}\ }\textbf {\bibinfo {volume} {131}},\ \bibinfo {pages} {327--353} (\bibinfo {year} {1997})}\BibitemShut {NoStop}%
\bibitem [{\citenamefont {Ketcheson}, \citenamefont {Parsani},\ and\ \citenamefont {LeVeque}(2013)}]{Ketcheson2013High-OrderSystems}%
  \BibitemOpen
  \bibfield  {author} {\bibinfo {author} {\bibfnamefont {D.~I.}\ \bibnamefont {Ketcheson}}, \bibinfo {author} {\bibfnamefont {M.}~\bibnamefont {Parsani}}, \ and\ \bibinfo {author} {\bibfnamefont {R.~J.}\ \bibnamefont {LeVeque}},\ }\bibfield  {title} {\enquote {\bibinfo {title} {{High-Order Wave Propagation Algorithms for Hyperbolic Systems}},}\ }\href {\doibase 10.1137/110830320} {\bibfield  {journal} {\bibinfo  {journal} {SIAM Journal on Scientific Computing}\ }\textbf {\bibinfo {volume} {35}},\ \bibinfo {pages} {A351--A377} (\bibinfo {year} {2013})}\BibitemShut {NoStop}%
\bibitem [{\citenamefont {Dumbser}\ \emph {et~al.}(2010)\citenamefont {Dumbser}, \citenamefont {Hidalgo}, \citenamefont {Castro}, \citenamefont {Par{\'{e}}s},\ and\ \citenamefont {Toro}}]{Dumbser2010FORCESystems}%
  \BibitemOpen
  \bibfield  {author} {\bibinfo {author} {\bibfnamefont {M.}~\bibnamefont {Dumbser}}, \bibinfo {author} {\bibfnamefont {A.}~\bibnamefont {Hidalgo}}, \bibinfo {author} {\bibfnamefont {M.}~\bibnamefont {Castro}}, \bibinfo {author} {\bibfnamefont {C.}~\bibnamefont {Par{\'{e}}s}}, \ and\ \bibinfo {author} {\bibfnamefont {E.~F.}\ \bibnamefont {Toro}},\ }\bibfield  {title} {\enquote {\bibinfo {title} {{FORCE schemes on unstructured meshes II: Non-conservative hyperbolic systems}},}\ }\href {\doibase 10.1016/j.cma.2009.10.016} {\bibfield  {journal} {\bibinfo  {journal} {Computer Methods in Applied Mechanics and Engineering}\ }\textbf {\bibinfo {volume} {199}},\ \bibinfo {pages} {625--647} (\bibinfo {year} {2010})}\BibitemShut {NoStop}%
\bibitem [{\citenamefont {Tokareva}\ and\ \citenamefont {Toro}(2010)}]{Tokareva2010}%
  \BibitemOpen
  \bibfield  {author} {\bibinfo {author} {\bibfnamefont {S.~A.}\ \bibnamefont {Tokareva}}\ and\ \bibinfo {author} {\bibfnamefont {E.~F.}\ \bibnamefont {Toro}},\ }\bibfield  {title} {\enquote {\bibinfo {title} {{HLLC-type Riemann solver for the Baer-Nunziato equations of compressible two-phase flow}},}\ }\href {\doibase 10.1016/j.jcp.2010.01.016} {\bibfield  {journal} {\bibinfo  {journal} {Journal of Computational Physics}\ }\textbf {\bibinfo {volume} {229}},\ \bibinfo {pages} {3573--3604} (\bibinfo {year} {2010})}\BibitemShut {NoStop}%
\bibitem [{\citenamefont {Nguyen}\ and\ \citenamefont {Dumbser}(2015)}]{Nguyen2015ATension}%
  \BibitemOpen
  \bibfield  {author} {\bibinfo {author} {\bibfnamefont {N.~T.}\ \bibnamefont {Nguyen}}\ and\ \bibinfo {author} {\bibfnamefont {M.}~\bibnamefont {Dumbser}},\ }\bibfield  {title} {\enquote {\bibinfo {title} {{A path-conservative finite volume scheme for compressible multi-phase flows with surface tension}},}\ }\href {\doibase 10.1016/j.amc.2015.09.026} {\bibfield  {journal} {\bibinfo  {journal} {Applied Mathematics and Computation}\ }\textbf {\bibinfo {volume} {271}},\ \bibinfo {pages} {959--978} (\bibinfo {year} {2015})}\BibitemShut {NoStop}%
\bibitem [{\citenamefont {Chen}, \citenamefont {Xie},\ and\ \citenamefont {Xiao}(2022)}]{Chen2022RevisitRobustness}%
  \BibitemOpen
  \bibfield  {author} {\bibinfo {author} {\bibfnamefont {D.}~\bibnamefont {Chen}}, \bibinfo {author} {\bibfnamefont {B.}~\bibnamefont {Xie}}, \ and\ \bibinfo {author} {\bibfnamefont {F.}~\bibnamefont {Xiao}},\ }\bibfield  {title} {\enquote {\bibinfo {title} {{Revisit to the THINC/QQ scheme: Recent progress to improve accuracy and robustness}},}\ }\href {\doibase 10.1002/fld.5072} {\bibfield  {journal} {\bibinfo  {journal} {International Journal for Numerical Methods in Fluids}\ }\textbf {\bibinfo {volume} {94}},\ \bibinfo {pages} {719--755} (\bibinfo {year} {2022})}\BibitemShut {NoStop}%
\bibitem [{\citenamefont {Kumar}\ \emph {et~al.}(2021)\citenamefont {Kumar}, \citenamefont {Cheng}, \citenamefont {Xiong}, \citenamefont {Xie}, \citenamefont {Abgrall},\ and\ \citenamefont {Xiao}}]{Kumar2021THINCSchemes}%
  \BibitemOpen
  \bibfield  {author} {\bibinfo {author} {\bibfnamefont {R.}~\bibnamefont {Kumar}}, \bibinfo {author} {\bibfnamefont {L.}~\bibnamefont {Cheng}}, \bibinfo {author} {\bibfnamefont {Y.}~\bibnamefont {Xiong}}, \bibinfo {author} {\bibfnamefont {B.}~\bibnamefont {Xie}}, \bibinfo {author} {\bibfnamefont {R.}~\bibnamefont {Abgrall}}, \ and\ \bibinfo {author} {\bibfnamefont {F.}~\bibnamefont {Xiao}},\ }\bibfield  {title} {\enquote {\bibinfo {title} {{THINC scaling method that bridges VOF and level set schemes}},}\ }\href {\doibase 10.1016/j.jcp.2021.110323} {\bibfield  {journal} {\bibinfo  {journal} {Journal of Computational Physics}\ }\textbf {\bibinfo {volume} {436}},\ \bibinfo {pages} {110323} (\bibinfo {year} {2021})}\BibitemShut {NoStop}%
\bibitem [{\citenamefont {Toro}(2009)}]{Toro2009}%
  \BibitemOpen
  \bibfield  {author} {\bibinfo {author} {\bibfnamefont {E.~F.}\ \bibnamefont {Toro}},\ }\href {\doibase 10.1007/b79761} {\emph {\bibinfo {title} {{Riemann Solvers and Numerical Methods for Fluid Dynamics}}}}\ (\bibinfo  {publisher} {Springer},\ \bibinfo {year} {2009})\BibitemShut {NoStop}%
\bibitem [{\citenamefont {Zein}, \citenamefont {Hantke},\ and\ \citenamefont {Warnecke}(2010)}]{Zein2010}%
  \BibitemOpen
  \bibfield  {author} {\bibinfo {author} {\bibfnamefont {A.}~\bibnamefont {Zein}}, \bibinfo {author} {\bibfnamefont {M.}~\bibnamefont {Hantke}}, \ and\ \bibinfo {author} {\bibfnamefont {G.}~\bibnamefont {Warnecke}},\ }\bibfield  {title} {\enquote {\bibinfo {title} {{Modeling phase transition for compressible two-phase flows applied to metastable liquids}},}\ }\href {\doibase 10.1016/j.jcp.2009.12.026} {\bibfield  {journal} {\bibinfo  {journal} {Journal of Computational Physics}\ }\textbf {\bibinfo {volume} {229}},\ \bibinfo {pages} {2964--2998} (\bibinfo {year} {2010})}\BibitemShut {NoStop}%
\bibitem [{\citenamefont {De~Lorenzo}, \citenamefont {Pelanti},\ and\ \citenamefont {Lafon}(2018)}]{DeLorenzo2018}%
  \BibitemOpen
  \bibfield  {author} {\bibinfo {author} {\bibfnamefont {M.}~\bibnamefont {De~Lorenzo}}, \bibinfo {author} {\bibfnamefont {M.}~\bibnamefont {Pelanti}}, \ and\ \bibinfo {author} {\bibfnamefont {P.}~\bibnamefont {Lafon}},\ }\bibfield  {title} {\enquote {\bibinfo {title} {{HLLC-type and path-conservative schemes for a single-velocity six-equation two-phase flow model: A comparative study}},}\ }\href {\doibase 10.1016/j.amc.2018.03.092} {\bibfield  {journal} {\bibinfo  {journal} {Applied Mathematics and Computation}\ }\textbf {\bibinfo {volume} {333}},\ \bibinfo {pages} {95--117} (\bibinfo {year} {2018})}\BibitemShut {NoStop}%
\bibitem [{\citenamefont {Johnsen}\ and\ \citenamefont {Colonius}(2006)}]{Johnsen2006}%
  \BibitemOpen
  \bibfield  {author} {\bibinfo {author} {\bibfnamefont {E.}~\bibnamefont {Johnsen}}\ and\ \bibinfo {author} {\bibfnamefont {T.}~\bibnamefont {Colonius}},\ }\bibfield  {title} {\enquote {\bibinfo {title} {{Implementation of WENO schemes in compressible multicomponent flow problems}},}\ }\href {\doibase 10.1016/j.jcp.2006.04.018} {\bibfield  {journal} {\bibinfo  {journal} {Journal of Computational Physics}\ }\textbf {\bibinfo {volume} {219}},\ \bibinfo {pages} {715--732} (\bibinfo {year} {2006})}\BibitemShut {NoStop}%
\bibitem [{\citenamefont {Gottlieb}, \citenamefont {Shu},\ and\ \citenamefont {Tadmor}(2001)}]{Gottlieb2001}%
  \BibitemOpen
  \bibfield  {author} {\bibinfo {author} {\bibfnamefont {S.}~\bibnamefont {Gottlieb}}, \bibinfo {author} {\bibfnamefont {C.-w.}\ \bibnamefont {Shu}}, \ and\ \bibinfo {author} {\bibfnamefont {E.}~\bibnamefont {Tadmor}},\ }\bibfield  {title} {\enquote {\bibinfo {title} {{Strong Stability-Preserving High-Order Time Discretization Methods}},}\ }\href {\doibase 10.1137/S003614450036757X} {\bibfield  {journal} {\bibinfo  {journal} {SIAM Review}\ }\textbf {\bibinfo {volume} {43}},\ \bibinfo {pages} {89--112} (\bibinfo {year} {2001})}\BibitemShut {NoStop}%
\bibitem [{\citenamefont {Chiapolino}, \citenamefont {Boivin},\ and\ \citenamefont {Saurel}(2017)}]{Chiapolino2017}%
  \BibitemOpen
  \bibfield  {author} {\bibinfo {author} {\bibfnamefont {A.}~\bibnamefont {Chiapolino}}, \bibinfo {author} {\bibfnamefont {P.}~\bibnamefont {Boivin}}, \ and\ \bibinfo {author} {\bibfnamefont {R.}~\bibnamefont {Saurel}},\ }\bibfield  {title} {\enquote {\bibinfo {title} {{A simple phase transition relaxation solver for liquid–vapor flows}},}\ }\href {\doibase 10.1002/fld.4282} {\bibfield  {journal} {\bibinfo  {journal} {International Journal for Numerical Methods in Fluids}\ }\textbf {\bibinfo {volume} {83}},\ \bibinfo {pages} {583--605} (\bibinfo {year} {2017})}\BibitemShut {NoStop}%
\bibitem [{\citenamefont {Fleischmann}, \citenamefont {Adami},\ and\ \citenamefont {Adams}(2019)}]{Fleischmann2019}%
  \BibitemOpen
  \bibfield  {author} {\bibinfo {author} {\bibfnamefont {N.}~\bibnamefont {Fleischmann}}, \bibinfo {author} {\bibfnamefont {S.}~\bibnamefont {Adami}}, \ and\ \bibinfo {author} {\bibfnamefont {N.~A.}\ \bibnamefont {Adams}},\ }\bibfield  {title} {\enquote {\bibinfo {title} {{Numerical symmetry-preserving techniques for low-dissipation shock-capturing schemes}},}\ }\href {\doibase 10.1016/j.compfluid.2019.04.004} {\bibfield  {journal} {\bibinfo  {journal} {Computers and Fluids}\ }\textbf {\bibinfo {volume} {189}},\ \bibinfo {pages} {94--107} (\bibinfo {year} {2019})}\BibitemShut {NoStop}%
\end{thebibliography}%

\end{document}